\documentclass{aastex631}
\usepackage{tabularx}
\usepackage{amsmath}

\newcommand{\rin}{{r_\mathrm{in}}}
\newcommand{\rout}{{r_\mathrm{out}}}

\newcommand{\cm}{{\,\mathrm{cm}}}
\newcommand{\yr}{{\,\mathrm{yr}}}
\newcommand{\rJ}{r_{\!\!\raisebox{-0.5ex}{$\scriptstyle J$}}}

\shorttitle{Poisson solvers for Astaroth}
\shortauthors{Krasnopolsky et al.}

\begin{document}

\title{Iterative Poisson Solvers for Self-gravity with the GPU Code Astaroth}

\author[0000-0001-5557-5387]{Ruben Krasnopolsky}
\affiliation{Institute of Astronomy and Astrophysics, Academia Sinica, Taipei 106319, Taiwan}

\author[0009-0008-8632-0385]{Touko Puro}
\affiliation{High-Performance Computing Lab, Department of Computer Science, Aalto University, Finland}

\author[0009-0008-2938-8260]{Wei-Wen Li}
\affiliation{Institute of Astronomy and Astrophysics, Academia Sinica, Taipei 106319, Taiwan}

\author[0000-0001-8385-9838]{Hsien Shang}
\affiliation{Institute of Astronomy and Astrophysics, Academia Sinica, Taipei 106319, Taiwan}

\author[0000-0002-8782-4664]{Miikka S. V\"ais\"al\"a}
\affiliation{Institute of Astronomy and Astrophysics, Academia Sinica, Taipei 106319, Taiwan}
\affiliation{Faculty of Information Technology and Electrical Engineering, University of Oulu, P.O. Box 8000, FI-90014 University of Oulu, Finland}

\author[0000-0003-0064-4060]{Mordecai-Mark Mac Low}
\affiliation{Department of Astrophysics, American Museum of Natural History, New York, NY 10024, USA}

\author[0000-0001-9840-5986]{Matthias Rheinhardt}
\affiliation{High-Performance Computing Lab, Department of Computer Science, Aalto University, Finland}

\author[0000-0002-9614-2200]{Maarit Korpi-Lagg}
\affiliation{High-Performance Computing Lab, Department of Computer Science, Aalto University, Finland}

\correspondingauthor{\newline{}Ruben Krasnopolsky, Hsien Shang}
\email{ruben@asiaa.sinica.edu.tw, shang@asiaa.sinica.edu.tw}

\begin{abstract}
We present the development and benchmarking of Poisson solvers for graphics processing units (GPUs).  Implemented in the Astaroth platform, the solvers feature high computational efficiency.
We present novel combinations of discretizations and smoothers and document practical and performance-focused implementations aimed at reducing time-to-solution for self-gravitating systems.
We describe the solver architectures and validate their accuracy against known analytic solutions. We measure convergence and timing per iteration for various solver algorithms, including conjugate gradient, successive overrelaxation, and multigrid in Cartesian coordinates, along with biconjugate gradient stabilized in spherical coordinates. We also couple the solvers to the Astaroth hydrodynamics to simulate a classic time-dependent problem in star formation, measuring accuracy and time-to-solution, for self-gravity on three-dimensional structured grids.
Our results demonstrate that the solvers achieve performance similar to other algorithms implemented in Astaroth, and provide a solid foundation for integration into production-scale  astrophysical simulations. 
\end{abstract}


\section{Introduction} \label{sec:intro}
Poisson solvers are a performance bottleneck in self-gravitating simulations.
We present our implementation in Astaroth, a graphics processing unit (GPU) platform for stencil-based operations, with flexible implementation of operations enabled by the use of its own domain-specific language \citep{pekkila2019astaroth}.
Astaroth supports efficient magnetohydrodynamics (MHD) and hydrodynamics (HD) solvers, implementing the algorithms of the Pencil Code \citep{Pencil-JOSS}, with broad applications, especially in turbulence and dynamo theory.  Astaroth can run on both CUDA and HIP-based GPU machines.
It has been used for dynamos in turbulent flows \citep{vaisala2021}, for pseudodisk formation under the gravity field of a growing central mass \citep{vaisala2023}, and for galactic small-scale dynamos \citep{gent2025arXiv}.
Astaroth has been benchmarked at 0.3 billion zone updates per second on one GPU, with weak scaling speed of 0.26 in the same units up to 8192 GPUs. Further high performance benchmarks are reported in \citet{pekkila2025}.

Full application of Astaroth to problems in star formation requires a flexible implementation of self-gravity that can take full advantage of the stencil-based operations at the core of Astaroth's high speed and is compatible with its flexible grids, which include uniform and nonuniform grids in spherical and Cartesian coordinates.
In this work, we implement self-gravity via iterative, stencil-based Poisson solvers for the gravitational potential, with the open boundary of the potential implemented using multipolar expansion of user-adjustable order.

The efficient Poisson solver of \citet{moon2019} extensively uses Fourier transform methods, including the use of the \citet{James1977Poisson} algorithm for implementation of open boundaries. These methods can achieve high accuracy, but they are not stencil-based as required to give Astaroth its speed, and they require specific regularities of the grids that are not compatible with the general curvilinear and nonuniform grids
that Astaroth implements. Such grids are of great utility in some self-gravity problems (e.g., the nonuniform spherical grids used in \citealt{wang2025}), and therefore we need algorithms capable of handling that flexibility. Similar considerations limit the application of the Fourier methods developed for uniform 3D Cartesian grids by \citet{krasnopolsky2021} to Astaroth, which are based on James's algorithm throughout the 3D volume.
The convolution method for nested meshes of \citet{vorobyob2023} is also based on a Fourier transform and does not use stencil operations.

Iterative solvers on structured grids are, on the other hand, inherently stencil-based and compatible with grid flexibility.
Therefore, we implement such solvers for Astaroth.
Conjugate gradient (CG), successive overrelaxation (SOR), biconjugate gradient stabilized \citep[BICGSTAB;][]{saad2003iterative}, and multigrid \citep[MG;][]{trottenberg2001multigrid} are examples of iterative solvers that can be employed with the Astaroth stencil methods and can thus achieve Astaroth's efficiency of memory use, single-GPU parallelization, and multi-GPU use based on the Message Passing Interface \citep{pekkila2022}. Other iterative solvers can potentially retain these advantages, and could also be considered in extended applications.

The SOR method is based on an acceleration of the Gauss--Seidel method by overrelaxation.
It has been used in our self-gravity calculations in spherical coordinates using ZEUS-TW, from \citet{krasnopolsky2012} to \citet{wang2025}. Fastest convergence for the application at hand
is achieved for an optimally adjusted overrelaxation parameter $\omega$. 
We use a nonuniform $\omega$, following \citet{Clarke15}.

The MG method has recent applications in nested grid codes,
with that of \citet{wang2020} for Cartesian nested grids being close to our purposes as it uses stencil-based operations. Being MG, they also need methods to exchange information between different levels of refinement, although they do not explicitly describe their treatment of the outer boundary conditions for a general situation beyond their two static tests, in 2D and 3D\@.
\citet{tomida2023} present an MG method for Athena++, implemented for nested uniform Cartesian grids as they appear in adaptive mesh refinement (AMR) computations. They describe a boundary condition based on multipolar expansion up to order $\ell=4$ in Cartesian coordinates, comparable to the spherical adjustable boundary conditions for choosable $\ell_{\mathrm{max}}$ in Cartesian and spherical coordinates of our work presented here.

GPU implementations of Poisson solvers  that remain close to stencil- and structured-grid-based algorithms range from framework-provided solvers in AMR applications to reusable linear solver libraries and a smaller set of application codes with in-house implementations. 
Framework-based approaches, such as those available through AMReX-based applications \citep{Zhang2019}, are, e.g. \cite{Nyx2013, CASTRO2020, QUOKKA2022}.
There are also library-based approaches, such as PETSc \citep{petsc-efficient, petsc-user-ref, petsc-web-page}, hypre \citep{FALGOUT2021102840}, MueLu \citep{MueLu, MueLuURL}, AMGX \citep{AMGX2015}, and rocALUTION \citep{rocalution}, that define much of the current ecosystem.
However, more direct comparisons can be made with GPU codes such as 
GAMER \citep{schive2018gamer} and IDEFIX \citep{IDEFIX2023}, which illustrate that MG- and relaxation-based Poisson solvers can also be implemented as part of the application code itself. Here, IDEFIX utilizes the performance portability framework Kokkos \citep{Kokkos, Kokkos3} for GPU use. Indeed, the approach of Kokkos is perhaps closest to Astaroth, as  Astaroth also provides a language to interact with GPUs without explicitly writing CUDA.

The Poisson solvers presented here are not limited to self-gravity. The iterative methods (SOR, CG, BICGSTAB, MG) apply also to more general elliptic equations, including those arising in electrostatics, magnetostatics, and incompressible flows, with the added benefit of the large dynamic range in space afforded by the nonuniform and curvilinear grids these solvers allow.
These extensions to general elliptic problems need attention regarding symmetry and positivity of the coefficient matrix. Some also require implementation of physical boundary conditions, including but not limited to application of the multipolar expansions, which is a traditional approach in electrostatics and magnetostatics.

In this work, we present the implementation of these solvers in Astaroth and measure their efficiency and accuracy for the self-gravity problem.
The paper is organized as follows.
In \S\ref{sec:methods} we show the self-gravity implementations for Astaroth, describing the iterative methods in \S\ref{sec:methods:iterative}, and the boundary conditions in \S\ref{sec:methods:BC}.
In \S\ref{sec:results} we test the methods, comparing the Astaroth results for various simple models of astrophysical interest having known solutions. Results, timings, and convergence of the self-gravity solvers are shown in \S\ref{sec:static}.
In \S\ref{sec:dynamic} we apply the methods to a time evolving problem of astrophysical collapse driven by self-gravity.
In \S\ref{sec:conclusions} we discuss other applications of the Poisson solver, assess its integration with the other components of Astaroth, and conclude with our future directions of work.

\vspace{4mm}
\section{Methods} \label{sec:methods}
The Poisson equation for the gravitational potential $\Phi$ in Cartesian coordinates is
\begin{equation}
\nabla^2 \Phi(x, y, z) = \frac{\partial^2 \Phi}{\partial x^2} + \frac{\partial^2 \Phi}{\partial y^2} + \frac{\partial^2 \Phi}{\partial z^2} = 4\pi G \rho(x, y, z),
\end{equation}
while in spherical coordinates it is
\begin{equation}
\nabla^2 \Phi(r, \theta, \phi) = \frac{1}{r^2} \left[\frac{\partial}{\partial r} \left( r^2 \frac{\partial \Phi}{\partial r} \right)
+ \frac{1}{\sin\theta} \frac{\partial}{\partial \theta} \left( \sin\theta \frac{\partial \Phi}{\partial \theta} \right)
+ \frac{1}{\sin^2\theta} \frac{\partial^2 \Phi}{\partial \phi^2}\right]
= 4\pi G \rho(r, \theta, \phi)\ ,
\end{equation}
where $G$ is Newton's gravitational constant, and $\rho$ is the mass density.
We describe implementations of iterative solvers for these equations (\S\ref{sec:methods:iterative}) using boundary conditions of astrophysical interest (\S\ref{sec:methods:BC}).

\subsection{Iterative solvers} \label{sec:methods:iterative}
Iterative solvers for the Poisson equation are based on successive updates of an estimate of the gravitational potential.
The iteration steps are set to terminate when the residual (difference between the left and right hand side of the Poisson equation) is less than a tolerance threshold value. An appropriate warning message is produced if the threshold is not achieved within a maximum number of iterations.
In practice, the number of iterations is strongly dependent on the initial estimate for the potential. In time-dependent flows, a good initial estimate is given by the gravitational potential in the previous time step, often leading in practice to convergence within a modest number of iterations.

Iterative methods are extremely flexible, allowing solution of the Poisson equation on uniform or nonuniform grids in different coordinate systems
by computing the corresponding coefficients of the finite-difference form of the Laplacian operator. In the case of Astaroth, this task is made immediate because its domain specific language allows GPU-efficient expressions of all kinds of stencil operations, including the Laplacian operator for all the grids relevant to this work.

\subsubsection{Successive Overrelaxation of Gauss--Seidel} \label{sec:sor}
The Gauss--Seidel method for the Poisson equation updates the gravitational potential at a given grid point based on the values at neighboring points.
A red-black implementation \citep{numerical_recipes} is used for efficient parallelization (each grid point can be updated independently), in which
red and black cells of a 3D checkerboard pattern
are alternately updated in successive sweeps\footnote{A sweep is one complete round of unknown updating according to the iteration rule.} of the update algorithm.
With two colors only, independence of grid points is strictly obeyed only for stencil radius 1. However, it has been found that for bigger stencil radii, two-coloring still grants good convergence.

The SOR method is also based on local neighbors. It is an overrelaxation of the Gauss--Seidel corrections to the current iteration, based on a weight factor $\omega$. Each SOR update amplifies the corresponding Gauss--Seidel update by the factor $\omega$. If Gauss--Seidel can converge, and $0<\omega<2$, SOR converges to the same solution. If $\omega$ is chosen in an optimal range, it can converge far faster than Gauss--Seidel while for $\omega=1$, it coincides with Gauss--Seidel.
Again, we use the red-black implementation for efficient parallelization. The optimal value of the weight factor $\omega_\mathrm{opt}$ that achieves fastest convergence is estimated \citep[][Equation 20.5.19]{numerical_recipes} as
\begin{equation}\label{eq:omega_final}
    \omega_\mathrm{opt} = \frac{2}{1+\sqrt{1-\rJ^2}}
\end{equation}
where $\rJ$ is the spectral radius of the Jacobi iteration, which is derived from the matrix elements of the discretized Laplacian operator.
For this case, and stencil radius one, the spectral radius for a uniform Cartesian 3D grid can be estimated as
\begin{equation}\label{eq:rjac}
    \rJ = \frac{\left({\Delta x}\right)^{-2} \cos{(\pi/N_x})+\left({\Delta y}\right)^{-2} \cos{({\pi}/{N_y})}+\left({\Delta z}\right)^{-2} \cos{({\pi}/{N_z})}}{\left({\Delta x}\right)^{-2}+\left({\Delta y}\right)^{-2}+\left({\Delta z}\right)^{-2}} ,
\end{equation}
where $N_x$,$N_y$, $N_z$ are the gridpoint numbers in the $x$, $y$ and $z$ directions, respectively and $\Delta x$, $\Delta y$, and $\Delta z$ are the grid spacings.
For the case $N_x=N_y=N_z=N$ with $\Delta x=\Delta y=\Delta z$, this simplifies to $\rJ=\cos(\pi/N)$, leading to
$\omega_\mathrm{opt} = {2}/[1+\sin(\pi/N)]$. Following \citet[][Equation 20.5.21]{numerical_recipes} we can adopt for big enough $N$
\begin{equation}
    \omega_\mathrm{opt}\approx\omega_\mathrm{opt,a}=\frac{2}{1+\pi/N}\ .
\end{equation}

It has been experimentally observed that the first steps of the SOR iteration may be more accurate by using instead of $\omega=\omega_\mathrm{opt}$
a sequence $\omega_n\rightarrow\omega_\mathrm{opt}$, given by
\begin{equation}\label{eq:omega_n}
\omega_0=1,\quad\ \omega_1=1/(1-\rJ^2/2),\quad\ \omega_{n+1}=1/(1-\rJ^2\omega_n/4)
\end{equation}
as in \citet[][Equation 20.5.30]{numerical_recipes}. A variant of this sequence, which is used in \citet{Clarke15} and \citet{krasnopolsky2012}, starts the sequence from $\omega_1$, skipping $\omega_0=1$.

The Astaroth implementation of SOR for Cartesian uniform grids is straightforward using the domain specific language. For uniform Cartesian grids we use either the fixed optimal $\omega_\mathrm{opt}$, or the sequence $\omega_n$ starting from $n=1$.
For spherical coordinates, we use, as in \citet{Clarke15} and \citet{krasnopolsky2012}, a variant in which the values of $\omega_n$ are not uniform, but depend on position on the grid. This approach also converges to the same value as Gauss--Seidel, and it is experimentally observed to improve speed of convergence for nonuniform and curvilinear grids.
For this work we use the iteration sequence Equation \eqref{eq:omega_n} starting from $n=1$, and we generalize Equation \eqref{eq:rjac} to obtain a local $\rJ$ which now depends on position,
\begin{equation}\label{eq:local_rjac}
\rJ=\frac{\cos({\pi}/{N_r})\,({\Delta}r_i)^{-2}+ \cos({\pi}/{N_\theta})\,(r_i\,\Delta\theta_j)^{-2} +  \cos({\pi}/{N_\phi})\,(r_i\sin\theta_j\,\Delta\phi_k)^{-2}}
{({\Delta} r_i)^{-2}+(r\,\Delta\theta_j)^{-2} + (r_i\sin\theta_j\,\Delta\phi_k)^{-2}}
\end{equation}
where $(N_r, N_\theta, N_\phi)$ are the extents of the spherical grid  and $(i,j,k)$ are the grid indexes.
Equation \eqref{eq:local_rjac} can also be adapted to a nonuniform Cartesian grid by omitting the metric factors $r_i$ and $r_i\sin\theta_j$.

\subsubsection{Conjugate gradient and variants} \label{sec:cg}
In the following sections, we refer to
\begin{equation}
A x = b
\end{equation}
as the equation system we want to solve, with coefficient matrix $A$ and right-hand side $b$.

The CG method finds the approximate solution in the $m$th iteration, $\tilde{x}_m \in x_0 + \mathcal{K}_m(A,r_0)$, such that the residual $b-A\tilde{x}_m \perp \mathcal{K}_m(A,r_0)$, where $x_0$ is the initial guess, $r_0=b-A x_0$, and $\mathcal{K}_m(A,r_0) = \operatorname{span} \{r_0,Ar_0,A^2r_0,\ldots,A^{m-1}r_0 \}$ is the Krylov subspace of dimension $m$.

Due to $A$  being symmetric and positive definite in our case, $\tilde{x}_m$ minimizes the A-norm\footnote{The $A$-norm of a vector $x$ is defined as $\sqrt{x^{\rm T} A x}$.} of the error, $||x_{\star}-\tilde{x}_m||_A$, where $x_{\star}$ is the true solution. Importantly, CG needs only inner products, which map to reduction operations, and applications of $A$, which map to stencil operations. Also, it requires only a small amount of additional memory because there exists a three-term recurrence relation to calculate the needed basis vectors. This makes it an ideal method to be implemented in Astaroth.
For general non-symmetric matrices, such as for the direct finite difference discretization of the  Laplacian in spherical coordinates, we use the BICGSTAB method \citep{van1992bi}. BICGSTAB likewise finds the iterative solution in $x_0 + \mathcal{K}_m(A,r_0)$, but imposes that $b-A\tilde{x}_m \perp \mathcal{K}_m(A^{\rm T}\!,r_0)$. Importantly, in BICGSTAB, short-term recurrence formulae are derived that do not need applications of $A^{\rm T}$ even though it is used to define the Krylov subspace to which the residual should be orthogonal. Thus, it shares many properties of CG that make it ideal to be implemented with Astaroth. For further details on CG and BICGSTAB we refer the reader to \citet{saad2003iterative}, which our description follows closely. 

Applying an iterative method directly to $A$ rarely gives sufficiently good convergence. Instead, a better approach called \textit{preconditioning} is to solve a modified equation that is formed by multiplying $A$ by a preconditioner $B^{-1}$ such that the system with the combined matrix $AB^{-1}$ or $B^{-1}A$ is easier to solve numerically. A guiding rule for choosing $B$ is that $B^{-1}$ should approximate $A^{-1}$, but be less expensive to apply. It is worth pointing out that the matrix product is never explicitly formed. Instead, $B^{-1}$ is applied to a vector by solving the corresponding equation system, and then $A$ is applied to the result (or vice versa).

For our GPU-based implementation, we use multi-color Gauss--Seidel \citep{koester1994parallel} as a preconditioner, due to the inherent sequential nature of normal lexicographical Gauss--Seidel iteration. First, the grid is ``colored" with $n$ colors such that each grid point only depends on grid points of different colors. Two, three, and four colors are needed for the second-, fourth-, and sixth-order discretization, respectively. Thus, for example for the sixth-order discretization of the Laplacian using red, blue, green, and yellow, the matrix $A$ can be written as
\begin{equation}
A = 
\begin{bmatrix}
    D_R & N_{RB} & N_{RG} & N_{RY} \\
    N_{BR} & D_{B} & N_{BG} & N_{BY} \\
    N_{GR} & N_{GB} & D_{G} & N_{GY} \\
     N_{YR} & N_{YB} & N_{YG} & D_{Y}
\end{bmatrix},
\end{equation}
where $D_{c}$ are the diagonal coefficient matrices for color $c$ and $N_{c_ic_j}$ contains the couplings between colors $c_i$ and $c_j$. 
An easy to invert approximation to $A$ then becomes its lower triangular part 
\begin{equation}
B = 
\begin{bmatrix}
    D_R & 0 & 0 & 0 \\
    N_{BR} & D_{B} & 0 & 0 \\
    N_{GR} & N_{GB} & D_{G} & 0 \\
     N_{YR} & N_{YB} & N_{YG} & D_{Y}
\end{bmatrix}
\end{equation}
of $A$. The required systems can now be solved directly block-wise in a number of steps equal to the number of colors.

\subsubsection{Multigrid} \label{sec:MG}

We next describe the MG solver implemented in this work, which is at the moment limited to uniform Cartesian grids only. For aspects of MG solvers and the theory behind it not described in this work we refer the reader to \citet{trottenberg2001multigrid}. MG solvers work by considering the high and low frequency errors of the approximate solutions. High frequencies can be damped out  on a given discretization, while low frequency errors are transferred to coarser rediscretizations, where some of the low frequencies map to high frequency errors that  in turn can be damped out effectively. This is continued recursively until the system is solved on the coarsest level and a correction from that level is passed to the level before it. When each level is visited only once in each direction (forth and back), these iterations are called \textit{V-cycles}. This gives the following iteration operator for the error 
\begin{equation}
    E_i = S^{\nu_1}(I - PE_{i+1}R)A_iS^{\nu_2}, \quad i\in[0,N-1]
\end{equation}
with
\begin{equation}
    E_N = A_N^{-1},
\end{equation}
where $i=0$ refers to the finest level, $N$ is the number of coarse levels, $S$ is the applied smoother, $\nu_1$ and $\nu_2$ are the number of times the smoother is applied, $A_i$ is the operator to be applied on level $i$, $R$ is the restriction operator from the fine grid to the coarse grid and $P$ is the prolongation operator from the coarse grid to the fine grid. 

For MG to be an effective solver it is crucial that the high frequency errors can be damped out effectively, which is true for the standard finite difference discretization of the Laplacian. When this holds true, then the convergence of MG is independent of the grid spacing, so it is an optimal solver in the sense that achieving a given error reduction factor only requires $O(N)$ operations.

For the restriction operator we use the standard full weighting operator, which is given in the one-dimensional case as\footnote{For representing stencils by the $[.]$ notation, see \cite{saad2003iterative}.}
\begin{equation}
R_{1d}= \frac{1}{4}
\begin{bmatrix}
1 & 2 & 1\\
\end{bmatrix}
\end{equation}
and in three dimensions as $R = R_{1d} \otimes R_{1d} \otimes R_{1d}$, where $\otimes$ denotes the Kronecker product. In accordance with standard MG theory for the prolongation operator, we use $P = cR^{\rm T}$, where $c$ is adjusted such that $P$ preserves constant functions. On the coarse grids, we use the Galerkin operator given by 
\begin{equation}
  A_{i+1} = RA_iR^{\rm T} ,
\end{equation}
which has a more robust theory behind it in comparison to the simpler approach of rediscretizing the Laplacian on the coarse grids. For solving $A_N^{-1}$, we employ the CG method, since the Galerkin operators inherit the symmetric positive definite property from $A$.
Due to the symmetry of $A$, the Galerkin operators can be easily built with Astaroth by applying their definition to a grid point $(i,j,k)$ and reading the stencil coefficients from it and its neighbours. Importantly, if the original stencil for $A_0$ has a radius of one, then all the Galerkin operators will also have a radius of one.

The given restriction operator works only for odd-sized grids which limits the fine grid to be of the form of $2^Nk-1$, where $k$ is an arbitrary positive integer. Furthermore, different grid sizes require different amounts of redistribution of grid points among processes after restricting to a coarse level. In multiprocess applications, a good choice for the grid size would be $2^Np-1$, where $p$ is the number of processes in a given direction, since it does not require any redistribution of coarse grid points among the processes. This limitation is not severe since
subdomains used for computational efficiency are already naturally cubes with side lengths of powers of two and thus one needs only to reduce the subdomain length at the edges by at most one grid point in each direction.

We use sparse approximate inverse (SPAI) smoothers  \citep{broker2002sparse} due to their inherent parallelism, rather than following \citet{trottenberg2001multigrid}. The action of SPAI smoothers can be written in the following preconditioned Richardson iteration form:
\begin{equation}\label{eq:richardson}
    x_{n+1} = (I-\omega MA)x_n + \omega Mb\ ,
\end{equation}
where $M$ and $\omega$ are chosen to have the best damping properties and $I$ is the identity operator. The smoother is called approximate inverse since it is supposed to approximate $A^{-1}$ in the sense that clearly the inverse has the best damping property, due to making the error exactly zero. Being sparse means that $M$ is constructed to correspond to a stencil of small radius. Out of the possible forms $M$ can take, in this work we restrict it to be the largest symmetric stencil of radius one
\begin{equation}\label{eq:Smoother}
M(\alpha,\beta,\gamma,\delta) := h^2
\left\{
\begin{bmatrix}
\delta & \gamma & \delta \\
\gamma & \beta & \gamma \\
\delta & \gamma & \delta 
\end{bmatrix}
\begin{bmatrix}
\gamma & \beta & \gamma \\
\beta & \alpha & \beta \\
\gamma & \beta & \gamma
\end{bmatrix}
\begin{bmatrix}
\delta & \gamma & \delta \\
\gamma & \beta & \gamma \\
\delta & \gamma & \delta
\end{bmatrix}
\right\}\ ,
\end{equation}
where $h$ is the grid spacing and $\alpha$, $\beta$, $\gamma$, and $\delta$ are coefficients to be determined; we restrict the smoothers here to isotropic grids.
The smoothing properties of $S$ can be measured using local Fourier analysis. This is motivated by the fact that any stencil operator can be diagonalized by the Fourier modes $\varphi(\mathbf{k},\mathbf{x}) = e^{i\mathbf{k} \cdot \mathbf{x}/h}$, where $\mathbf{x}=(x,y,z)$, $\mathbf{k}=(k_x,k_y,k_z)$.

The damping efficiency of the smoother $S = I-\omega MA$ on high-frequency errors is measured by the smoothing factor $\mu$. For each Fourier mode with frequencies $\mathbf{k}$, a single smoothing step multiplies its amplitude by the eigenvalue $\tilde{S}(\mathbf{k})$.  The high-frequency modes to be considered are chosen to be in the set $T = [-\pi,\pi]^3 \setminus [-(\pi/2),(\pi/2)]^3$.  The smoothing factor 
\begin{equation}
\mu = \max_{\mathbf{k} \in T}\tilde{S}(\mathbf{k}),
\end{equation}
corresponds to the worst error reduction factor among the high-frequency modes for a single step. 
The coefficients of the smoother in Equation~(\ref{eq:Smoother}) were numerically optimized with the Hybrid Algorithm for Non-Smooth Optimization  \citep[HANSO][]{burke2005robust} to reduce $\mu$ as much as possible, following \citet{he2023optimized}. 

Possibly even more efficient smoothers could have been created by optimizing for the two-grid convergence factor of the error as in \citet{brown2021tuning}, but the employed smoothers already give sufficiently low $\mu$.

Due to the communication bottleneck, in applications of $A$ one is always interested in using stencils with as small a radius as possible. Using the form of the Poisson equation in Cartesian coordinates allows the use of compact stencils of radius unity even for fourth order, or, assuming isotropic spacing, even for sixth order discretizations of the Laplacian \citep{spotz1996high}. In comparison, the naive stencils for these orders require stencil radii of two and three, respectively. As an additional benefit of compact stencils, we could numerically find SPAI smoothers with low values of $\mu$, while for higher radius stencils, the straightforward application of HANSO was not able to produce such smoothers.

We are not aware of any work using SPAI smoothers in this context, although other compact discretizations have been combined with MG solvers previously \citep{ghaffar2020multigrid,nwankwo2020multigrid,gupta1997compact}. Thus the effective use of numerically optimized SPAI smoothers for the compact discretizations presents the main algorithmic novelty of this paper. We emphasize that our combination is not only attractive due to the computational expense being the same for higher orders, but also because the same smoothing procedure can be employed at up to sixth order effectively, which is not always true for the naive discretizations and smoothers. The coefficients of the optimized smoothers and their approximate smoothing factors are given in Table \ref{table:coeffs}.

\begin{table} 
\begin{center}
\caption{Optimized Coefficients and Smoothing Factors}
\hspace*{-2.3cm}
    \begin{tabular}{c c c c c c c c} 
 \hline
 Order & $\omega$ & $\alpha$ & $\beta$ & $\gamma$ & $\delta$ & $\mu$ \\ [0.5ex] 
 \hline
 2 & 1.3699 & 0.1432 & 0.0284 & 0.0081 & 0.0025 & 0.1888\\ 
 4 & 0.9363 & 0.2807 & 0.0329 & 0.0166 & 0.0044 & 0.0828\\
 6 & 0.3543 & 0.7064 & 0.1005 & 0.0363 & 0.0909 & 0.0909\\
 \hline\\[-8mm]
\end{tabular}
\end{center}
    \tablecomments{Weight factors $\omega$, optimized coefficients $\alpha$, $\beta$, $\gamma$, and $\delta$, and resulting smoothing factors $\mu$ for SPAI smoothers of different orders.}
    \label{table:coeffs}
\end{table}

To measure the convergence of the MG methods we want to measure the spectral radius of the MG iteration operator $\varrho(E_0)$. Since the spectral radii of the iteration matrices for the error and for the residual are the same we can get an estimate of $\varrho(E_0)$ by calculating the mean convergence rate of the error after $m$ V-cycles
\begin{equation} \label{eq:rho}
\varrho = \left(\frac{||r_m||_2}{||r_0||_2}\right)^{\frac{1}{m}},
\end{equation}
where $r_m$ is the final residual and $r_0$ the initial one. 

The convergence rate of the methods is of practical interest since it helps to gauge their effectiveness and more importantly it gives a good estimate on how accurately the coarse level problem needs to be solved to still maintain the same convergence rate as is done in \citet{trottenberg2001multigrid}. The level of accuracy of the coarsest solve can be analysed more rigorously as in \citet{vacek2024effect}, the analysis of which is in line with the fact that reducing the relative residual by a factor greatly more than $\varrho$ is overkill. Based on our measurements, solving the coarse problem to relative accuracy of $\varrho$ indeed still maintains the same asymptotic convergence rates. Importantly we found out that the ratio of increased computation compared to the increase of the convergence rate favoured having $\nu_{1,2}=3$, which can be seen from Table \ref{table:convergence_factors}. Additionally, the table showcases that by using compact stencils the performance is largely independent of the discretization order.

Similar to recent work of \citet{vacek2025mixed}, \citet{tamstorf2021discretization} and \citet{oo2020accelerating}, we take advantage of mixed-precision in our implementation of MG\@. As data movement is the performance bottleneck, all data on the coarse levels are stored in single precision, while all computations are always done in double precision. We also experimented in using half-precision but more work needs to be done to find out in which circumstances is it safe to use it.

\begin{center}
\begin{table}
    \caption{Multigrid Timing and Convergence Factors}
\hspace*{-.8cm}
    \begin{tabular}{c @{\hspace{6mm}} c c c 
    @{\hspace{8mm}}
    c c c 
    @{\hspace{8mm}}
    c c c} 
 \hline
 & \multicolumn{3}{c }{$\nu_{1,2} = 1$} 
  & \multicolumn{3}{c }{$\nu_{1,2} = 2$}
   & \multicolumn{3}{c }{$\nu_{1,2} = 3$}
  \\[1mm]
 \hline
 Order & $s$ & $\varrho$ & $s\varrho$ & $s$ & $\varrho$ & $s\varrho$ & $s$ & $\varrho$ & $s\varrho$ \\ [0.5ex] 
  \hline
 2 & 3.48e-2 & 0.102 & 3.54e-3 & 5.29e-2 &  0.028 & 1.48e-3 & 7.27e-2 & 0.013 & 9.45e-4 \\ 
 4 & 3.49e-2 & 0.065 & 2.27e-3 & 5.33e-2 & 0.017 & 9.06e-4 & 7.37e-2 & 0.011 & 8.11e-4\\
  6 & 3.53e-2 & 0.068 & 2.40e-3 & 5.46e-2 & 0.018 & 9.82e-4 & 7.47e-2 & 0.012 & 8.96e-4
\\

  \hline
\end{tabular}
    \tablecomments{Approximate convergence factors and timings of the V-cycles with seven coarse levels on a grid of size $511^3$ for different smoothing step numbers $\nu_1=\nu_2$ and discretization orders. Here, $s$ is the time in seconds a single V-cycle takes, and $\varrho$ is the mean error convergence rate given by Eq.~\eqref{eq:rho}. Lower values of the product $s\varrho$ show greater efficiency of the smoothers,
    giving evidence that higher values of $\nu_{1,2}$ lead to more efficient cycles overall.
    }
    \label{table:convergence_factors}
\end{table}
 \end{center}

\subsection{Boundary conditions} \label{sec:methods:BC}
As we have implemented them, these iterative solvers need Dirichlet boundary conditions (BCs) for the potential $\Phi$. For some testing purposes, we can use analytically known potentials for the BCs.
For a physically meaningful setup, one would ideally formulate a boundary condition that is equivalent to having a vacuum outside the computational domain with $\Phi$ vanishing as $1/r$ for $r \to \infty$. This could be achieved through an integral equation on the boundary (equivalent to a direct Newton sum), though that implies significant computational costs.
Hence, for production (and complete benchmarking) we choose here an approximate method based on multipole expansion for the boundary conditions of $\Phi$.
\citet{moon2019} have efficiently used a form of James's algorithm for this purpose. We instead choose a method that allows more flexible grid choices, and adapts more directly to the stencil-based structure of Astaroth.
This is the method of 3D multipole expansion, which we had previously implemented \citep{krasnopolsky2012} for nonuniform spherical coordinates with a user-adjustable number of terms $(\ell,m)$ in the expansion.
For this work, we reimplemented this for Astaroth, and extended the implementation to include Cartesian coordinates. \citet{tomida2023} implement this same method for Cartesian coordinates using  $\ell_{\mathrm{max}}=4$,
while D.\ Clarke's ZEUS-3D v3.6 code reaches
$\ell_{\mathrm{max}}=5$.

The multipolar expansion can be written for an outer point $\mathbf{x}$ as
\begin{equation}\label{eq:outer_expansion}
    \Phi(\mathbf{x}) = -G \sum_{\ell,m} \frac{q_{\ell,m}}{r^{\ell+1}} \sqrt{\frac{4\pi}{2\ell+1}} Y_{\ell,m}(\mathbf{x}) 
\end{equation}
where $Y_{\ell,m}$ are the real spherical harmonics, $q_{\ell,m}$ are the multipoles, $r=|\mathbf{x}|$, and the sum spans over pairs of integers with $0\leq\ell\leq\ell_{\mathrm{max}}$ and $|m|\leq\ell$.
The functional form, normalization, and indexing of the real spherical harmonics follow the conventions of \citet{tomida2023}, in which the sine spherical harmonics take indices $m < 0$ and cosine spherical harmonics take indices $m\geq0$.
The outer multipoles are written as volume integrals over points  $\mathbf{x}'$ within the mass distribution as
\begin{equation}\label{eq:outer_multipole}
    q_{\ell,m}=\int_V dV'\,\sqrt{\frac{4\pi}{2\ell+1}}\,\rho(\mathbf{x}') (r')^\ell Y_{\ell,m}(\mathbf{x}')\ ,
\end{equation}
where $r'=|\mathbf{x}'|$, and numerically the integrals are expressed as sums over cells.
We use this formula for all boundaries in Cartesian coordinates, and for the outer boundary $r=\rout$ in spherical coordinates. Truncation of this formula at finite $\ell_{\mathrm{max}}$ is accurate provided that the mass distribution does not approach the numerical domain boundaries. As we have been doing since \citet{krasnopolsky2012}, we set the density to zero in a region surrounding the mass distribution, which agrees with the recommendation in \citet{tomida2023}.
Equations \eqref{eq:outer_expansion} and \eqref{eq:outer_multipole} complete the Cartesian boundary conditions.
For spherical grids with a non-vanishing inner boundary $r=\rin>0$, we must also include the inner multipolar expansion
\begin{equation}\label{eq:inner_expansion}
    \Phi(\mathbf{x}) = -G \sum_{\ell,m} \bar{q}_{\ell,m} r^{\ell} \sqrt{\frac{4\pi}{2\ell+1}} Y_{\ell,m}(\mathbf{x})\ ,
\end{equation}
where
\begin{equation}\label{eq:inner_multipole}
    \bar{q}_{\ell,m}=\int_V dV'\,\sqrt{\frac{4\pi}{2\ell+1}}\,\frac{\rho(\mathbf{x}')}{(r')^{\ell+1}} Y_{\ell,m}(\mathbf{x}')\ ,
\end{equation}
with the same caveats about accuracy and use of zero-padding when necessary.
These equations can be used to express the outer and inner radial boundary conditions of a spherical grid $\rin<r<\rout$.
The angular boundary conditions in spherical coordinates are given by the usual coordinate symmetries.

\section{Results} \label{sec:results}
In this section, we show tests and benchmarks of the methods newly implemented in Astaroth.
We measure accuracy, convergence, and performance of the self-gravity solvers in \S\ref{sec:static}, using a combination of static spherical density distributions defined in \S\ref{sec:test_setup}.
In \S\ref{sec:dynamic} we solve in spherical coordinates a time-dependent problem of astrophysical interest having a known ODE solution, namely the collapse of the singular isothermal sphere (SIS) of \citet{shu1977}.

\subsection{Density distribution and analytical solution} \label{sec:test_setup}
We define a test problem combining two spherical density distributions.
Each sphere $s_\mathrm{p}$ of radius $a_\mathrm{p}$ is centered at a position $\mathbf{x}_\mathrm{p}=(x_\mathrm{p},y_\mathrm{p},z_\mathrm{p})$, and has a density parameter $\rho_\mathrm{p}$.
The density at a point $\mathbf{x}$ is
\begin{equation}
    \rho_S(\mathbf{x}\ |\ \mathbf{x}_\mathrm{p},a_\mathrm{p},\rho_\mathrm{p})=\left\{
    \begin{array}{ll}
         \rho_\mathrm{p} (1-w^2)^n  & \mbox{ if } d<a_\mathrm{p} \\
         0  & \mbox{ otherwise}
    \end{array}
    \right.\ ,
\end{equation}
where $d=|\mathbf{x}-\mathbf{x}_\mathrm{p}|=\left[(x-x_\mathrm{p}^2)+(y-y_\mathrm{p}^2)+(z-z_\mathrm{p}^2)\right]^{1/2}$, and $w=d/a_\mathrm{p}$. We use $n=2$ as in \citet{wang2020} and \citet{krasnopolsky2021}.
The corresponding potential is
\begin{equation}
    \Phi_S(\mathbf{x}\ |\ \mathbf{x}_\mathrm{p},a_\mathrm{p},\rho_{p})=\left\{
    \begin{array}{ll}
         4\pi G \rho_\mathrm{p} a_\mathrm{p}^2 (-1/6 + w^2/6 -w^4/10 +w^6/42)  & \mbox{ if } d<a_\mathrm{p} \\
         -G M_\mathrm{p}/d  & \mbox{ otherwise}
    \end{array}
    \right.\ ,
\end{equation}
where $M_\mathrm{p}=(32\pi/105)\rho_\mathrm{p} a_\mathrm{p}^3$ is the mass of $s_\mathrm{p}$.

For the tests in Cartesian coordinates, we use a cubic domain $-1\leq x,y,z\leq1$ with various resolutions, between  $63^3$ and $512^3$ cells.
For the self-gravity tests in Cartesian coordinates of \S\ref{sec:static}, we set up the two spheres, $S_1$ and $S_2$, with centers at ${\mathbf x}_{1}=(0.0,0.0,0.4)$ and ${\mathbf x}_{2}=(0.0,0.0,-0.2)$, equal radii $a_{1,2}=0.08$, and total masses $M_{1,2}=\{1,2\}$, implying density parameters  $\rho_{1,2}=M_{1,2}/(32\pi a_{1,2}^3 / 105)$.
Both spheres are located within the inner $\pm1/2$ of the Cartesian domain, 
reducing the errors of the outer multipole expansion by keeping a sufficiently thick layer of zones of zero density.
We take note of the recommendation of \citet{tomida2023} to keep the center of mass not far from the origin of the multipolar expansion for improved accuracy, which is a realistic mode of usage.

\begin{table}[hb]
     \caption{Execution speeds (CG and SOR)}
\begin{center}
    \begin{tabular}{crccrrrrrr}
    \hline\\[-3.5mm]
             Method & \parbox{3em}{\centering{}Grid\\$N$} \hspace*{-1.5mm} & Machine & \parbox{4.5em}{\centering{}Residual\\reduction}\hspace*{-2mm} & \parbox{3em}{\centering{}Steps\\$N_\mathrm{I}$}\hspace*{-1mm} & \parbox{4em}{\centering{}Multipole\\time\\$t_\mathrm{mp}$\\(ms)}\hspace*{-3mm} & \parbox{4em}{\centering{}Iter\\time\\$t_\mathrm{iter}$\\(ms)}\hspace*{-3mm} & \parbox{4em}{\centering{}Step\\time\\$t_\mathrm{s}$\\(ms)}\hspace*{-3mm} & \parbox{4.5em}{\centering{}Step speed\\$s_\mathrm{s}$ (Gzups)}\hspace*{-3mm} & \parbox{4.5em}{\centering{}Total speed\\$s_\mathrm{t}$\\(Gzups)} \\[4mm]
         \hline
         CG2 &  64 & nano5 & $<10^{-4}$ & 125 & 1.019 & 27.966 & 0.224 & 1.170 & 0.009 \\
            &     &         & $<10^{-6}$ & 171 & 1.019 & 38.636 & 0.226 & 1.160 & 0.006 \\
         CG2 & 128 & nano5 & $<10^{-4}$ & 239 & 1.421 & 75.157 & 0.314 & 6.679 & 0.027 \\
            &     &         & $<10^{-6}$ & 330 & 1.421 & 103.30 & 0.313 & 6.700 & 0.020\\
         CG2 & 256 & nano5 & $<10^{-4}$ & 406 & 3.745 & 353.03 & 0.726 & 23.129 & 0.047\\
            &     &         & $<10^{-6}$ & 624 & 3.745 & 542.34 & 0.869 & 19.306 & 0.030 \\
         CG2 & 512 & nano5 & $<10^{-4}$ & 763 & 21.990  & 4091.2 & 5.362 & 25.031 & 0.032 \\
            &     &         & $<10^{-6}$ & 1214& 21.990  & 6510.2 & 5.363 & 25.027 & 0.020 \\        
            
         SOR  & 64 & nano5 & $<10^{-4}$ & 130 & 1.011 & 32.453 & 0.250 & 1.049 & 0.008 \\
              &    &         & $<10^{-6}$ & 187 & 1.011 & 46.687 & 0.250 & 1.049 & 0.005 \\
         SOR  & 128 & nano5 & $<10^{-4}$ & 246 & 1.191 & 73.752 & 0.300 & 6.991 & 0.028 \\
            &     &         & $<10^{-6}$ & 353 & 1.191 & 105.88 & 0.300 & 6.991 & 0.020 \\
         SOR  & 256 & nano5 & $<10^{-4}$ & 469 & 3.469 & 332.22 & 0.708 & 23.697 &  0.050\\
              &     &         & $<10^{-6}$ & 677 & 3.469 & 480.16 & 0.709 & 23.663 & 0.035 \\
         SOR  & 512 & nano5 & $<10^{-4}$ & 903 & 20.071 & 3478.4 & 3.852 & 34.844 &  0.038\\
             &     &         & $<10^{-6}$ & 1312 & 20.071 & 5054.0 & 3.852 & 34.844 & 0.026 \\

         CG6 & 64 & nano5 & $<10^{-4}$ & 151 & 1.554 & 33.346 & 0.221  & 1.187 & 0.008 \\
         &  &  & $<10^{-6}$ & 207 & 1.554 & 45.735 & 0.221  & 1.186 & 0.006 \\
         CG6 & 128 & nano5 & $<10^{-4}$ & 291 & 2.098 & 99.026 & 0.340 & 6.163 & 0.021 \\
         &  &  & $<10^{-6}$ & 401 & 2.098 & 136.07 & 0.339 & 6.180 & 0.015 \\
         CG6 & 256 & nano5 & $<10^{-4}$ & 489 & 4.528 & 521.81  & 1.067  & 15.722 & 0.032 \\
         &  &  & $<10^{-6}$ & 761 & 4.528 & 812.17 & 1.067 & 15.720 &  0.021\\
         CG6 & 512 & nano5 & $<10^{-4}$ & 920 & 24.357 & 6523.6 & 7.091 & 18.928 & 0.021 \\
         &  &  & $<10^{-6}$ & 1477 & 24.357 & 10472 & 7.090 & 18.930 & 0.013 \\

         CG2 &  64 & DGX & $<10^{-4}$ & 125 & 2.1  & 25.4  & 0.203   & 1.29 & 0.010 \\
            &     &         & $<10^{-6}$ & 171 & 2.1  & 36.1 & 0.211 & 1.24 & 0.007 \\
         CG2 & 128 & DGX & $<10^{-4}$ & 239 & 11.4 & 253  & 1.06 & 1.98 & 0.008 \\
            &     &         & $<10^{-6}$ & 330 & 11.4 & 352 & 1.07 & 1.97  & 0.006 \\
         CG2 & 256 & DGX & $<10^{-4}$ & 406 & 74.4 & 3608 & 8.89 & 1.89 &  0.005\\ 
            &     &         & $<10^{-6}$ & 624 & 74.4  & 5550 & 8.89  & 1.89  & 0.003 \\
         CG2 & 512 & DGX & $<10^{-4}$ & 763 & 552   & 44233  & 58.0  & 2.32  & 0.003 \\
            &     &         & $<10^{-6}$ & 1214& 552  & 70558  & 58.1 & 2.31  & 0.002 \\

         SOR &  64 & DGX & $<10^{-4}$ & 130 & 1.88 & 34.3  & 0.264  & 0.993  & 0.007 \\
            &     &         & $<10^{-6}$ & 187 & 1.88  & 49.6 & 0.265  & 0.989  & 0.0051\\
         SOR & 128 & DGX & $<10^{-4}$ & 246 & 10.2 & 283  & 1.15  & 1.82  & 0.007\\
            &     &         & $<10^{-6}$ & 353 & 10.2 & 407  & 1.15 & 1.82  &  0.005\\
         SOR & 256 & DGX & $<10^{-4}$ & 469 & 77.3 & 4390 & 9.36  & 1.79 &  0.004\\
            &     &         & $<10^{-6}$ & 677 & 77.3 & 6304  & 9.31 & 1.80 &  0.003\\

         CG6 & 64 & DGX & $<10^{-4}$ & 151 & 5.78 & 46.0 & 0.305  & 0.860 &  0.005\\
         &  &  & $<10^{-6}$ & 207 & 5.78 & 59.4 & 0.287  & 0.914 &  0.004\\
         CG6 & 128 & DGX & $<10^{-4}$ & 291 & 14.9 & 562 & 1.93 & 1.09 &  0.004\\
         &  &  & $<10^{-6}$ & 401 & 14.9 & 775 & 1.93 & 1.09 & 0.003 \\
         CG6 & 256 & DGX & $<10^{-4}$ & 489 & 85.2 & 6106  & 12.5  & 1.34 & 0.003 \\
         &  &  & $<10^{-6}$ & 761 & 85.2 & 9504 & 12.5 & 1.34 &  0.002\\
         CG6 & 512 & DGX & $<10^{-4}$ & 920 & 594 & 72779 & 79.1 & 1.70 & 0.002 \\
         &  &  & $<10^{-6}$ & 1477 & 594 & 117362 & 79.5 & 1.69 & 0.001 \\
         \hline
     \end{tabular}
     \label{tab:timings1}
\end{center}
    \tablecomments{Timings (in ms) and code speeds in Gzups (billion zone updates per second) at four resolutions of isotropic grids, on the H200 GPUs available on the NCHC cluster nano5, on an in-house DGX Spark system, and on one MI250X GPU of LUMI\@.
    The methods shown in this table are CG2, CG6 (CG 2nd and 6th order), and SOR (2nd order, with $\omega_\mathrm{opt,a}$). Total speed $s_\mathrm{t} = N^3 / (t_\mathrm{mp} + t_\mathrm{iter})$. Step speed is computed based on one iteration step. See text in \S\ref{sec:static} for details of iteration timing.
    }

\end{table}

\begin{table}[hb]
\begin{center}
     \caption{Execution speeds (MG)}
    \begin{tabular}{crccc@{\hspace{6mm}}rrrrr}
    \hline\\[-3.5mm]
         $\nu_{1,2}$ & \parbox{3em}{\centering{}Grid\\$N$} \hspace*{-1.5mm} & Machine & \parbox{4.5em}{\centering{}Residual\\reduction}\hspace*{-2mm} & \parbox{3em}{\centering{}Steps\\$N_\mathrm{I}$}\hspace*{-1mm} & \parbox{4em}{\centering{}Multipole\\time\\$t_\mathrm{mp}$\\(ms)}\hspace*{-3mm} & \parbox{4em}{\centering{}Iter\\time\\$t_\mathrm{iter}$\\(ms)}\hspace*{-3mm} & \parbox{4em}{\centering{}Step\\time\\$t_\mathrm{s}$\\(ms)}\hspace*{-3mm} & \parbox{4.5em}{\centering{}Step speed\\$s_\mathrm{s}$ (Gzups)}\hspace*{-3mm} & \parbox{4.5em}{\centering{}Total speed\\$s_\mathrm{t}$\\(Gzups)} \\[4mm]
         \hline
          1 & 63 & DGX & $< 10^{-6}$ & 7 & 11.8 & 23.87 & 3.41 & 0.0720 & 0.007 \\
          2 & 63 & DGX & $< 10^{-6}$ & 4 & 11.8 & 16.12 & 4.03 & 0.0620 & 0.009 \\ 
          3 & 63 & DGX & $< 10^{-6}$ & 3 & 11.8 & 13.92 & 4.64 & 0.0538 & 0.010\\

          1 & 127 & DGX & $< 10^{-6}$ & 7 & 17.7 & 45.49 & 6.49 & 0.315 &  0.032\\
          2 & 127 & DGX & $< 10^{-6}$ & 4 & 17.7 & 39.52 & 9.88 & 0.205 & 0.036\\ 
          3 & 127 & DGX & $< 10^{-6}$ & 3 & 17.7 & 39.6 & 13.2 & 0.151 &  0.036\\ 

          1 & 255 & DGX & $< 10^{-6}$ & 7 & 93.5 & 234.5 & 33.5 & 0.483 & 0.051\\
          2 & 255 & DGX & $< 10^{-6}$ & 4 & 93.5 & 243.6 & 60.9 & 0.272 & 0.049\\
          3 & 255 & DGX & $< 10^{-6}$ & 3 & 93.5 & 204 & 81.0 & 0.483 &  0.056\\

          1 & 511 & DGX & $< 10^{-6}$ & 6 & 653.4 & 1344 & 224 & 0.603 & 0.067\\ 
          2 & 511 & DGX & $< 10^{-6}$ & 5 & 653.4 & 1915 & 383 & 0.348 & 0.052 \\ 
          3 & 511 & DGX & $< 10^{-6}$ & 4 & 653.4 & 2184 & 546 & 0.244 & 0.047 \\

          1 & 63 & LUMI & $< 10^{-6}$ & 7 & 4.92 & 43.19 & 6.17 & 0.040 & 0.005\\     
          2 & 63 & LUMI & $< 10^{-6}$ & 4 & 4.92 & 28.84 & 7.21 & 0.034 & 0.007\\ 
          3 & 63 & LUMI & $< 10^{-6}$ & 3 & 4.92 & 21.9 & 7.30 & 0.034 & 0.009 \\ 

          1 & 127 & LUMI & $< 10^{-6}$ & 7 & 5.66 & 69.8 & 6.98 & 0.293 & 0.027\\ 
          2 & 127 & LUMI & $< 10^{-6}$ & 4 & 5.66 & 32.36 & 8.09 & 0.253 & 0.054\\ 
          3 & 127 & LUMI & $< 10^{-6}$ & 3 & 5.66 & 24.96 & 8.32 & 0.246 & 0.067 \\ 

          1 & 255 & LUMI & $< 10^{-6}$ & 7 & 9.63 & 74.2 & 10.6 & 1.56 & 0.198 \\ 
          2 & 255 & LUMI & $< 10^{-6}$ & 4 & 9.63 & 54.8 & 13.7 & 1.21 & 0.257\\ 
          3 & 255 & LUMI & $< 10^{-6}$ & 3 & 9.63 & 50.1 & 16.7 & 0.99 & 0.278\\ 

          1 & 511 & LUMI & $< 10^{-6}$ & 6 & 41.2 & 225 & 37.5 & 3.55 &  0.501\\ 
          2 & 511 & LUMI & $< 10^{-6}$ & 5 & 41.2 & 229 & 59.8 & 2.23 &  0.494\\ 
          $3$ & 511 & LUMI & $< 10^{-6}$ & 4 & 41.2 & 328 & 82.0 & 1.62 & 0.361\\

          3 & 63 & nano5 & $< 10^{-4}$ & 3 & 0.783 & 16.4 & 5.46 & 0.0457 & 0.015 \\ 
           &  &  & $< 10^{-6}$ & 4 & 0.783 & 20.6 & 5.14 & 0.0486 & 0.012 \\ 
          
          3 & 127 & nano5 & $< 10^{-4}$ & 3 & 1.29 & 18.76 & 6.25 & 0.328 & 0.102 \\ 
           &  &  & $< 10^{-6}$ & 4 & 1.29 & 24.56 & 6.14 & 0.334 & 0.079 \\ 
          
          3 & 255 & nano5 & $< 10^{-4}$ & 3 & 3.69 & 27.9 & 9.30 & 1.78 &  0.525\\ 
           &  &  & $< 10^{-6}$ & 4 & 3.69 & 37.0 & 9.24 & 1.79 &  0.408\\ 
          
          3 & 511 & nano5 & $< 10^{-4}$ & 3 & 22.1 & 103.7 & 34.6 & 3.86 & 1.060 \\ 
           &  &  & $< 10^{-6}$ & 4 & 22.1 & 135.4 & 33.9 & 3.94 &  0.847\\

         \hline
     \end{tabular}
     \label{tab:timings2}
\end{center}
    \tablecomments{Continuation of Table \ref{tab:timings1}, showing test results for the 6th order
    MG method, for various machines and values of $\nu_{1,2}$. Here, ``steps'' refers to the number of V-cycles for MG\@. See text in \S\ref{sec:static} and notes of Table \ref{tab:timings1} for details of iteration timing.
    }
    
\end{table}

\begin{figure}
    \centering
    \includegraphics[width=0.24\linewidth]{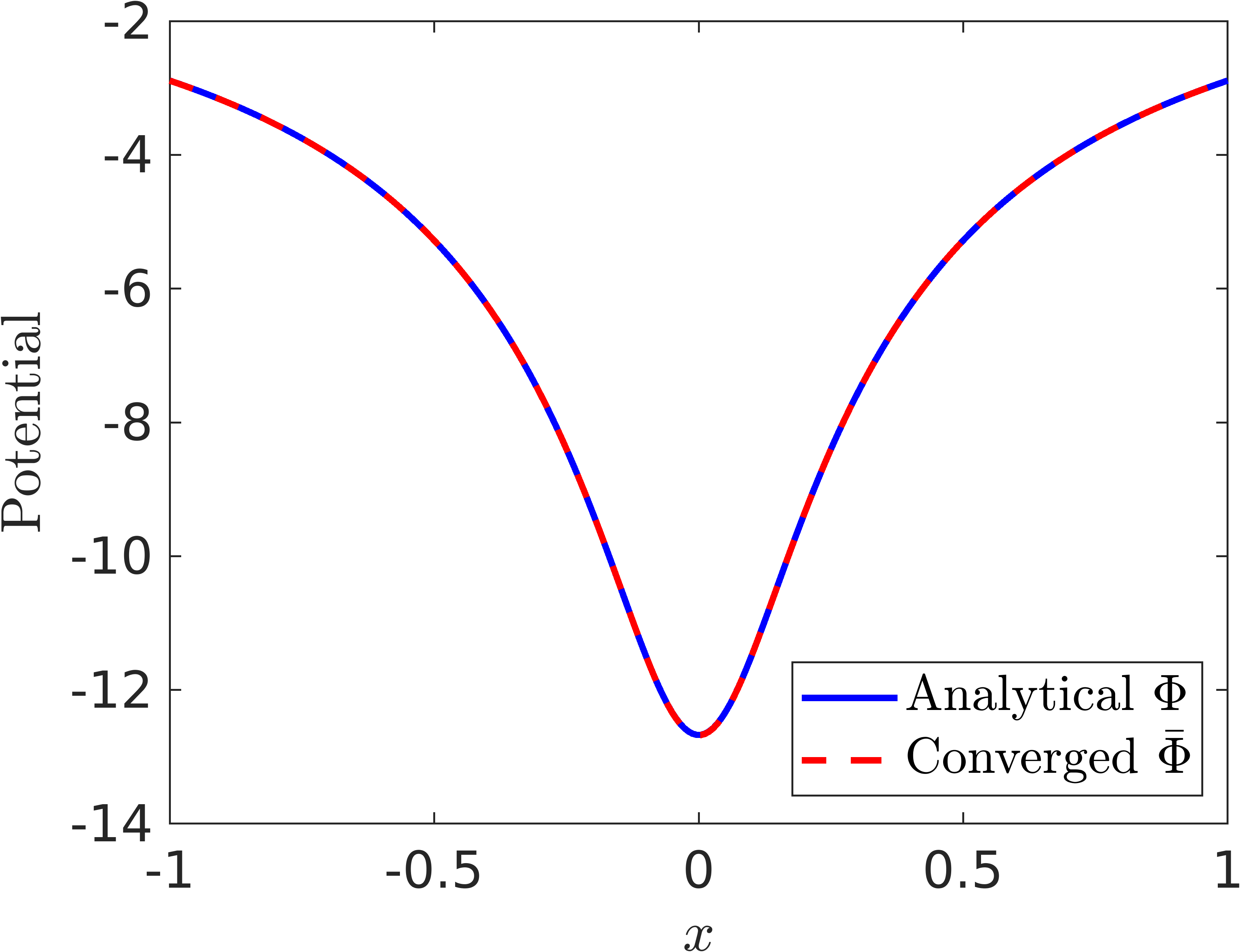}
    \includegraphics[width=0.24\linewidth]{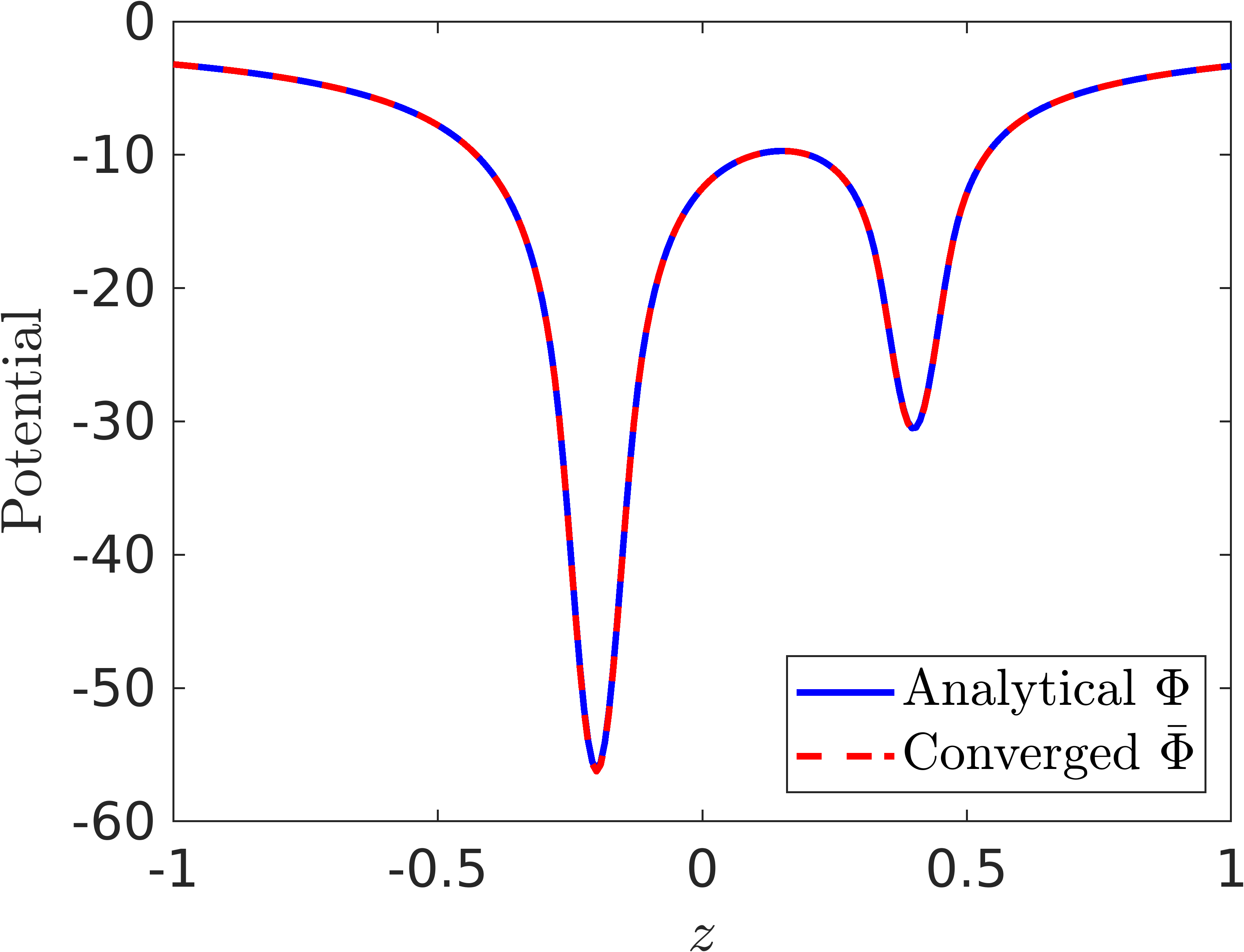}
    \includegraphics[width=0.24\linewidth]{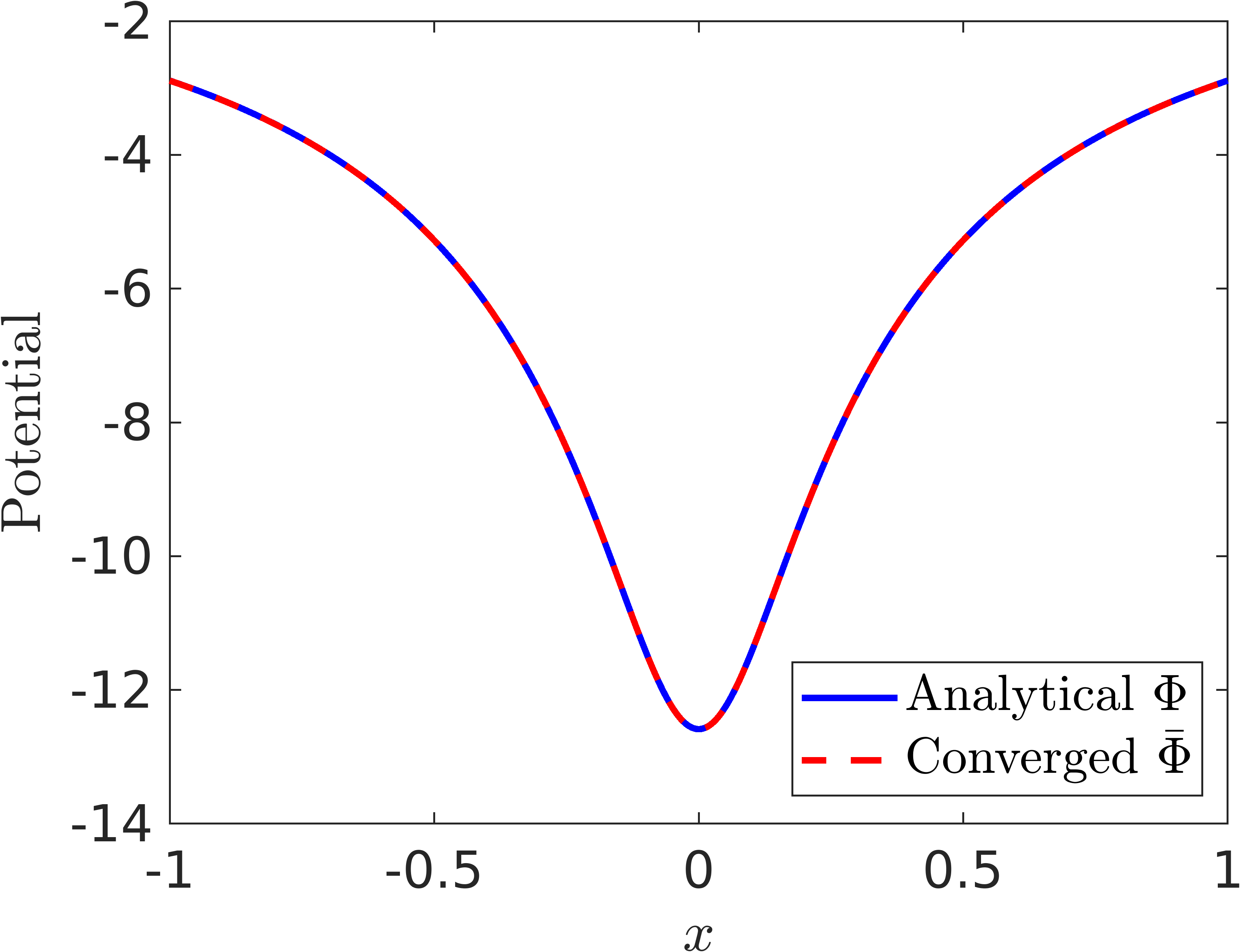}
    \includegraphics[width=0.24\linewidth]{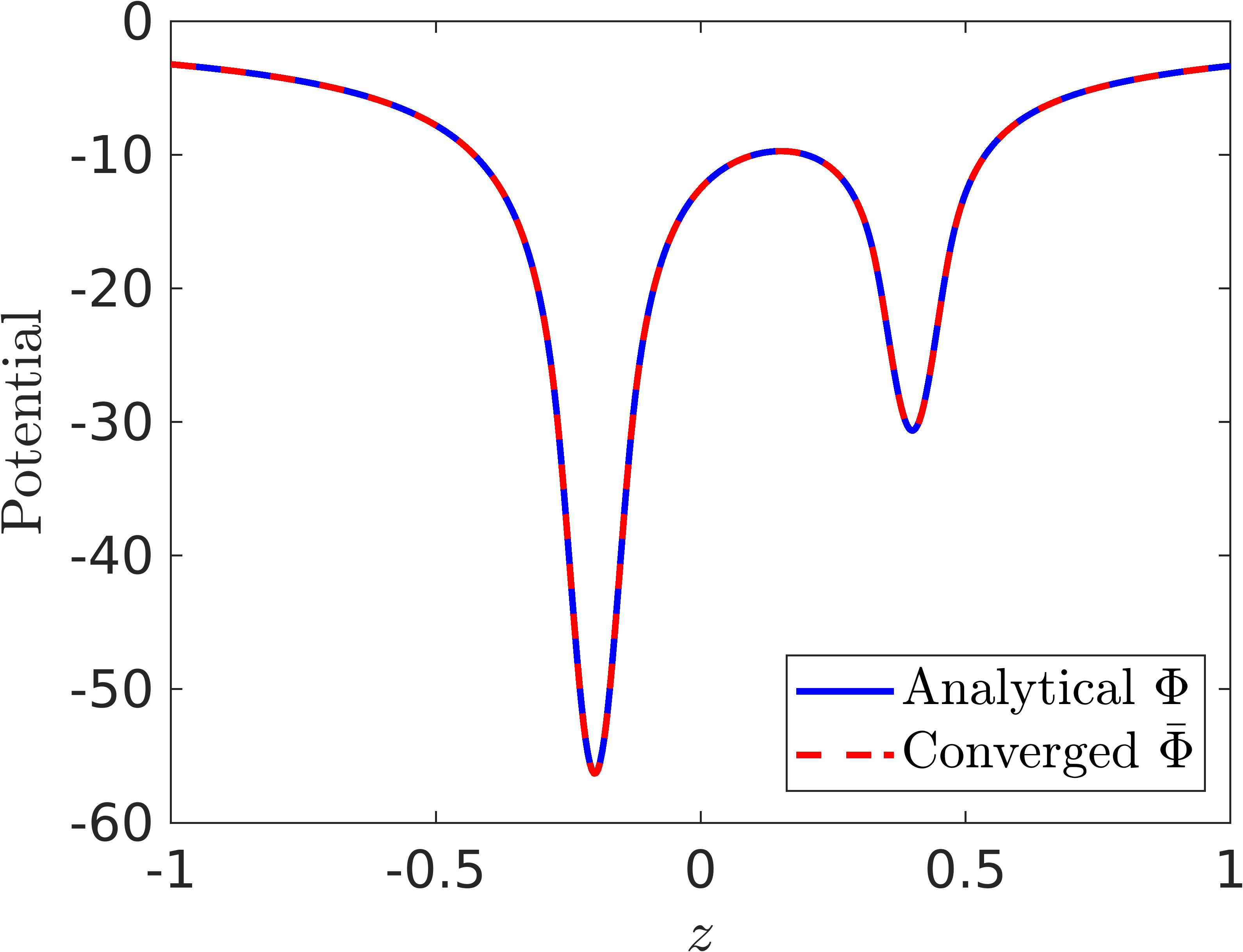} \\
    \includegraphics[width=0.24\linewidth]{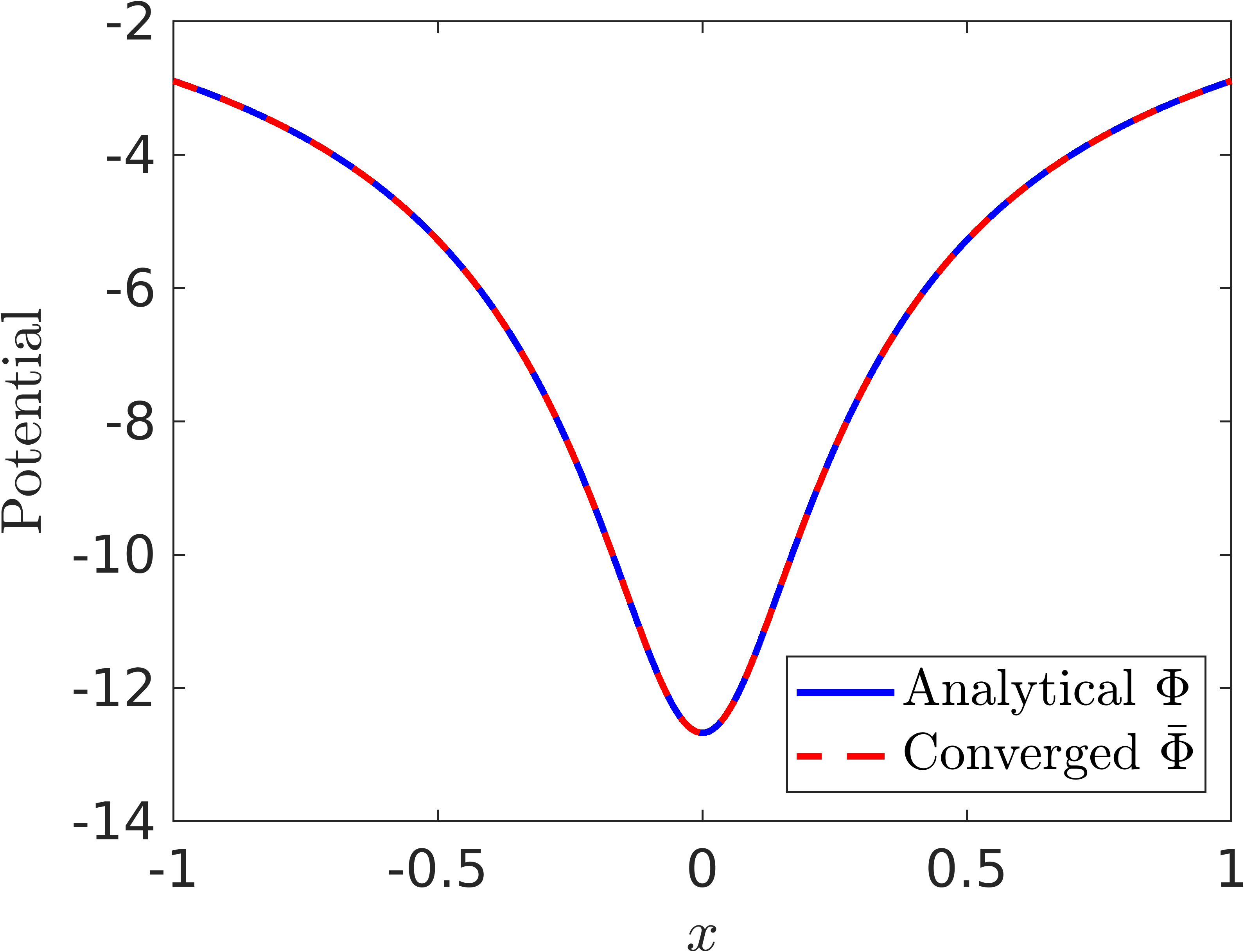}
    \includegraphics[width=0.24\linewidth]{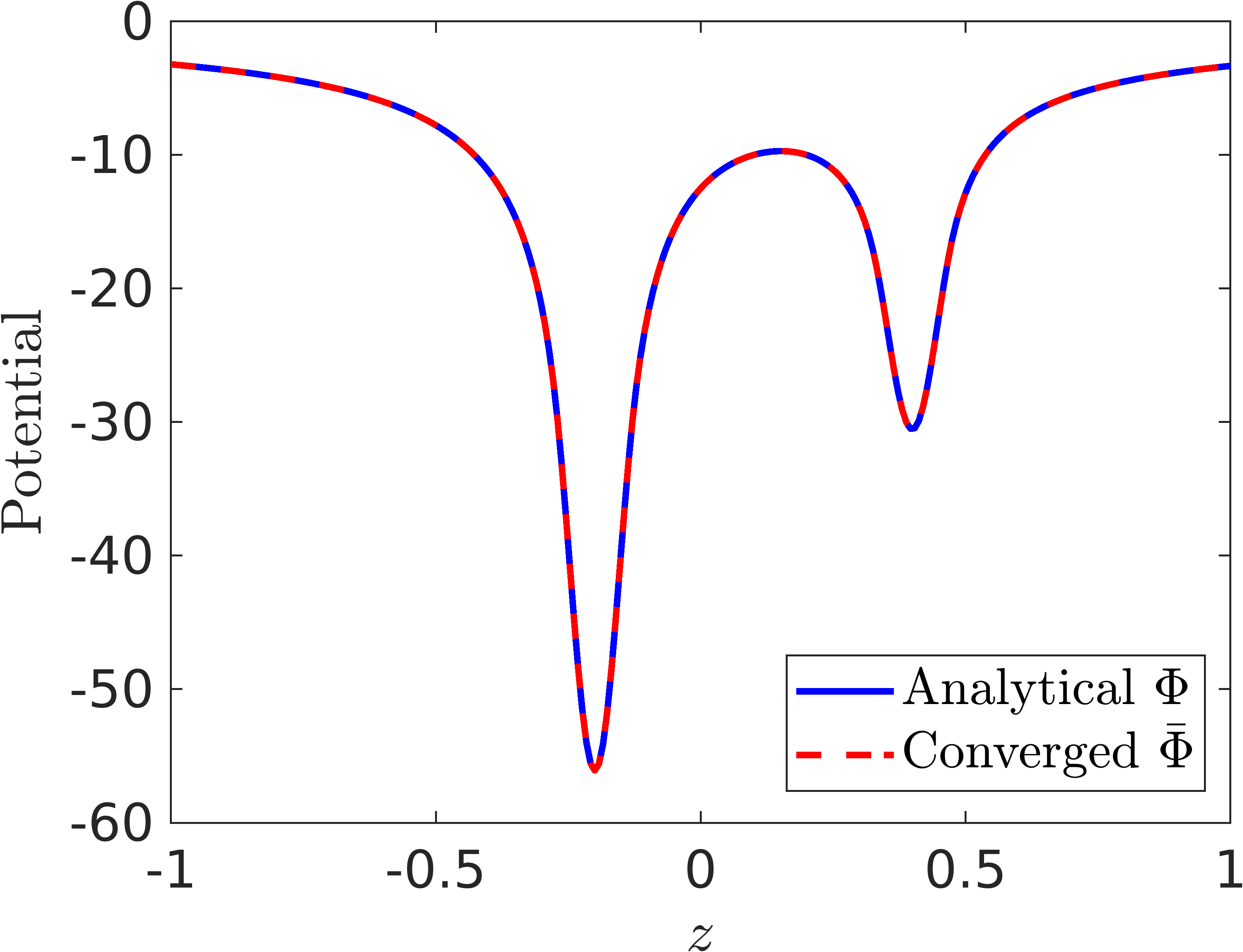}
    \includegraphics[width=0.24\linewidth]{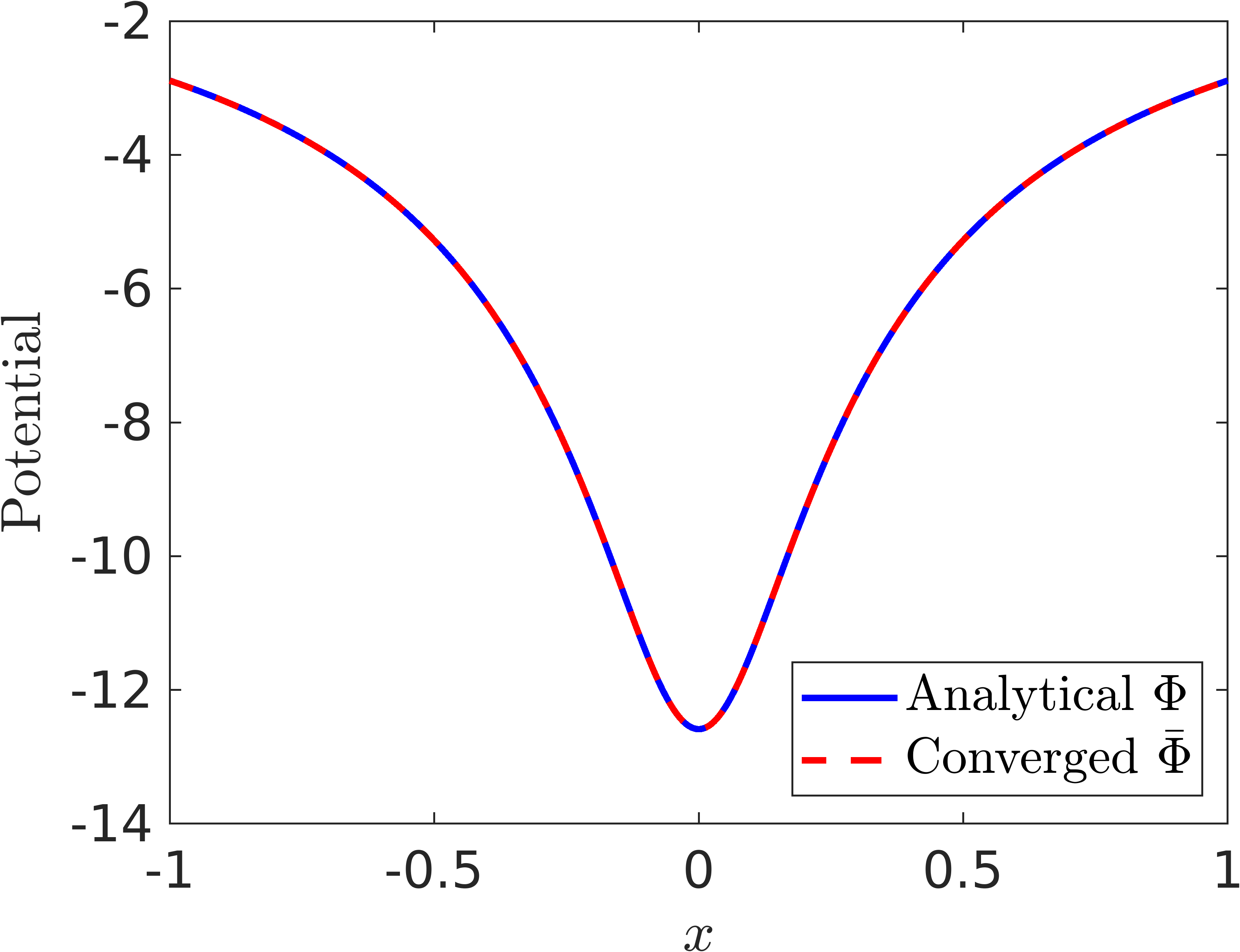}
    \includegraphics[width=0.24\linewidth]{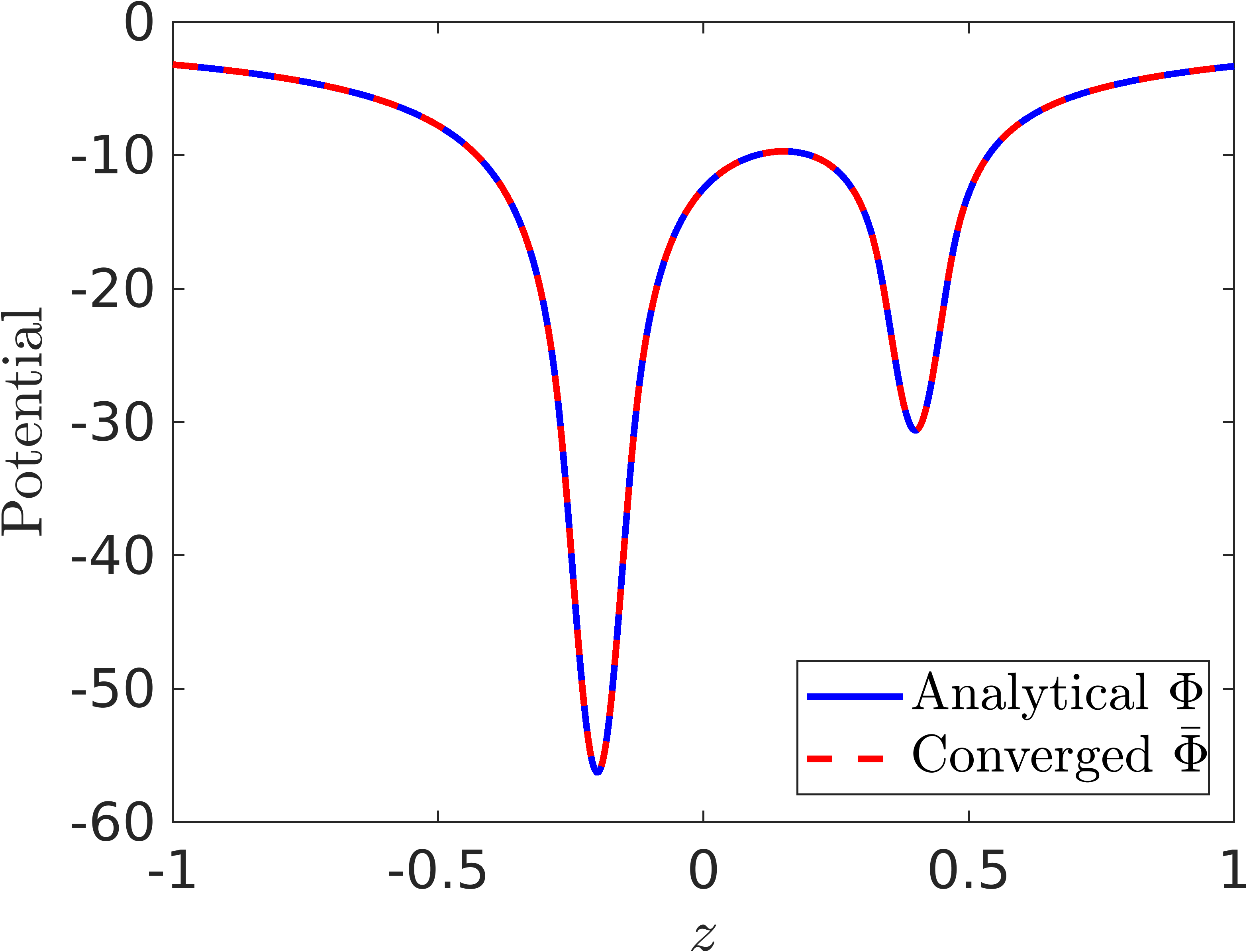} \\
    \includegraphics[width=0.24\linewidth]{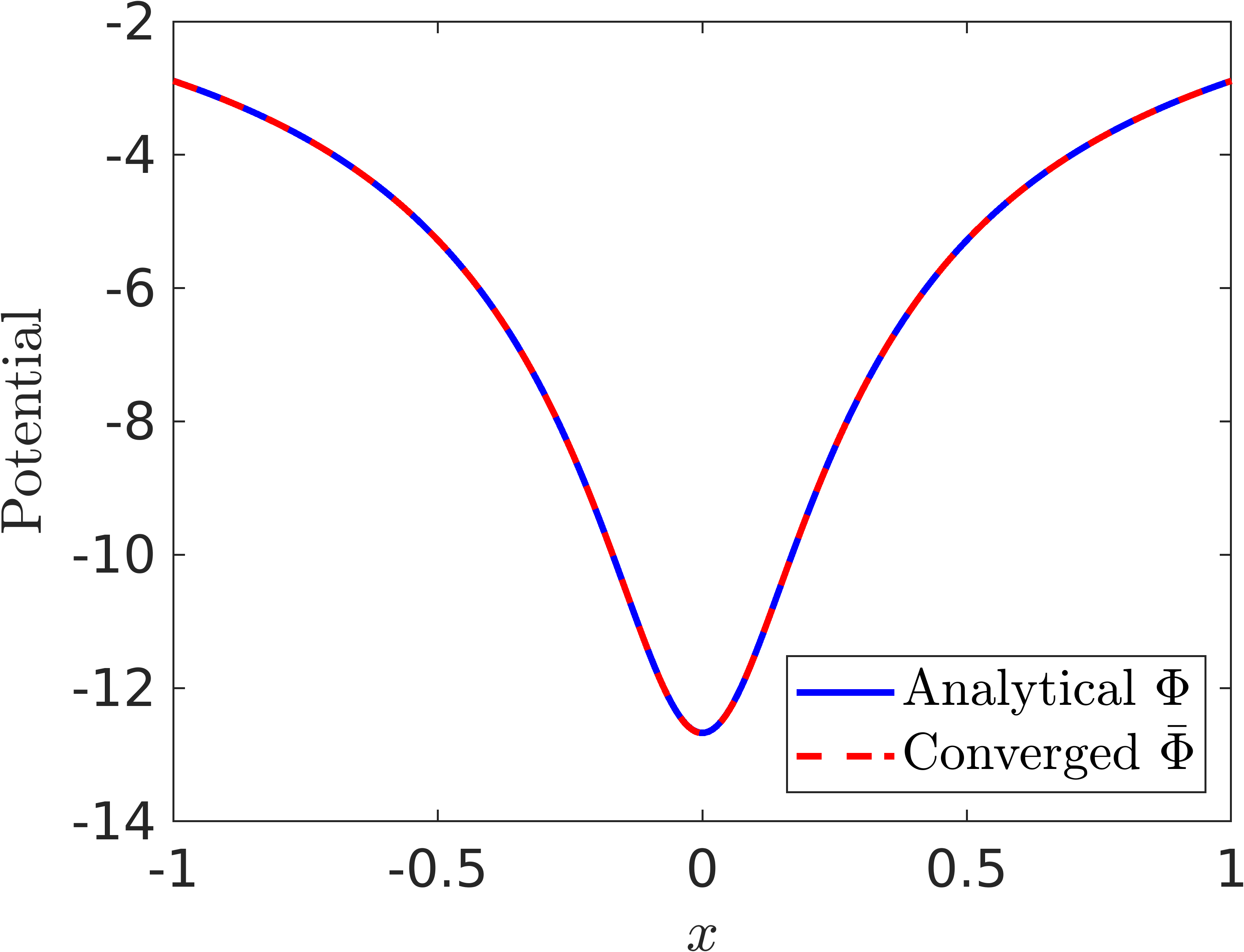}
    \includegraphics[width=0.24\linewidth]{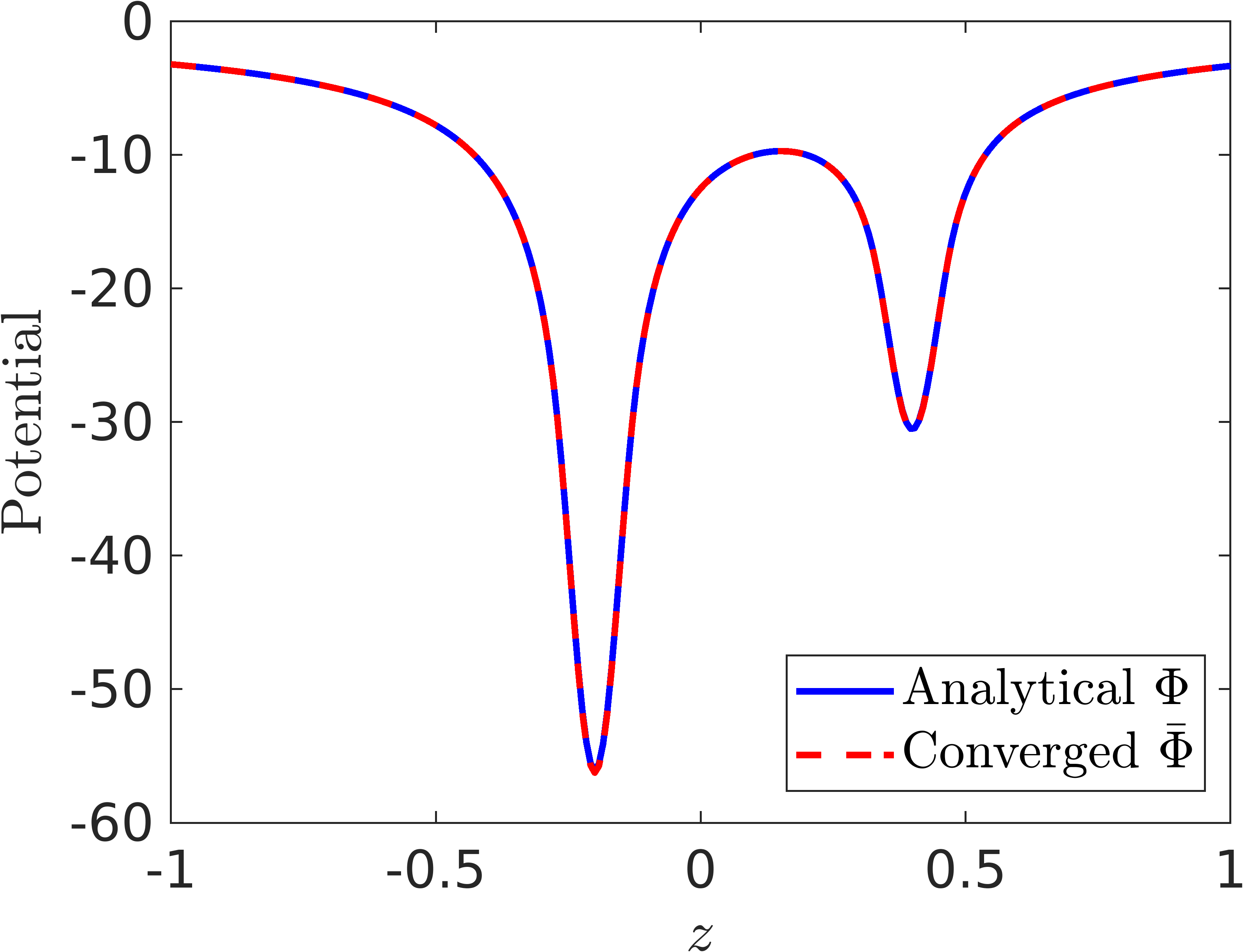}
    \includegraphics[width=0.24\linewidth]{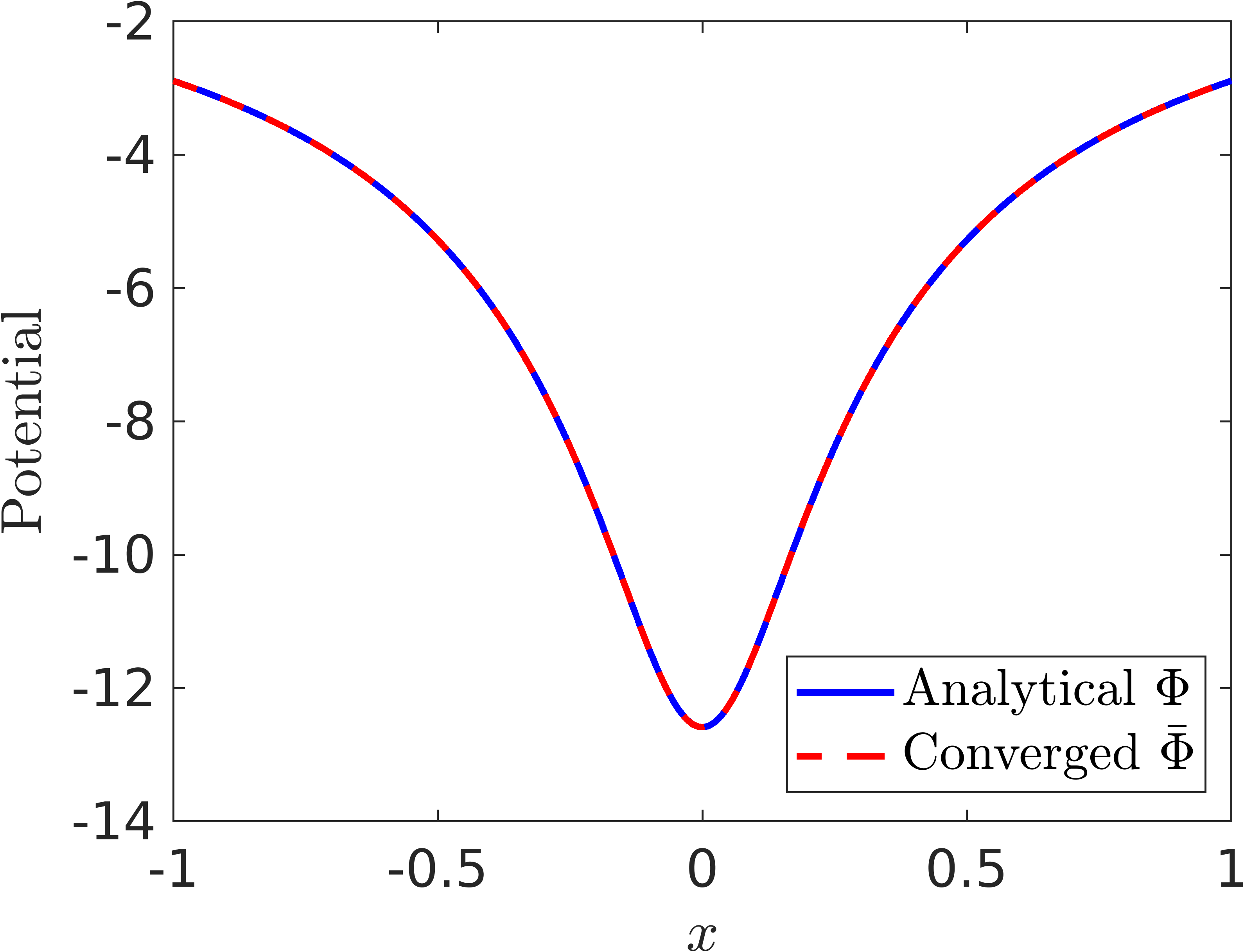}
    \includegraphics[width=0.24\linewidth]{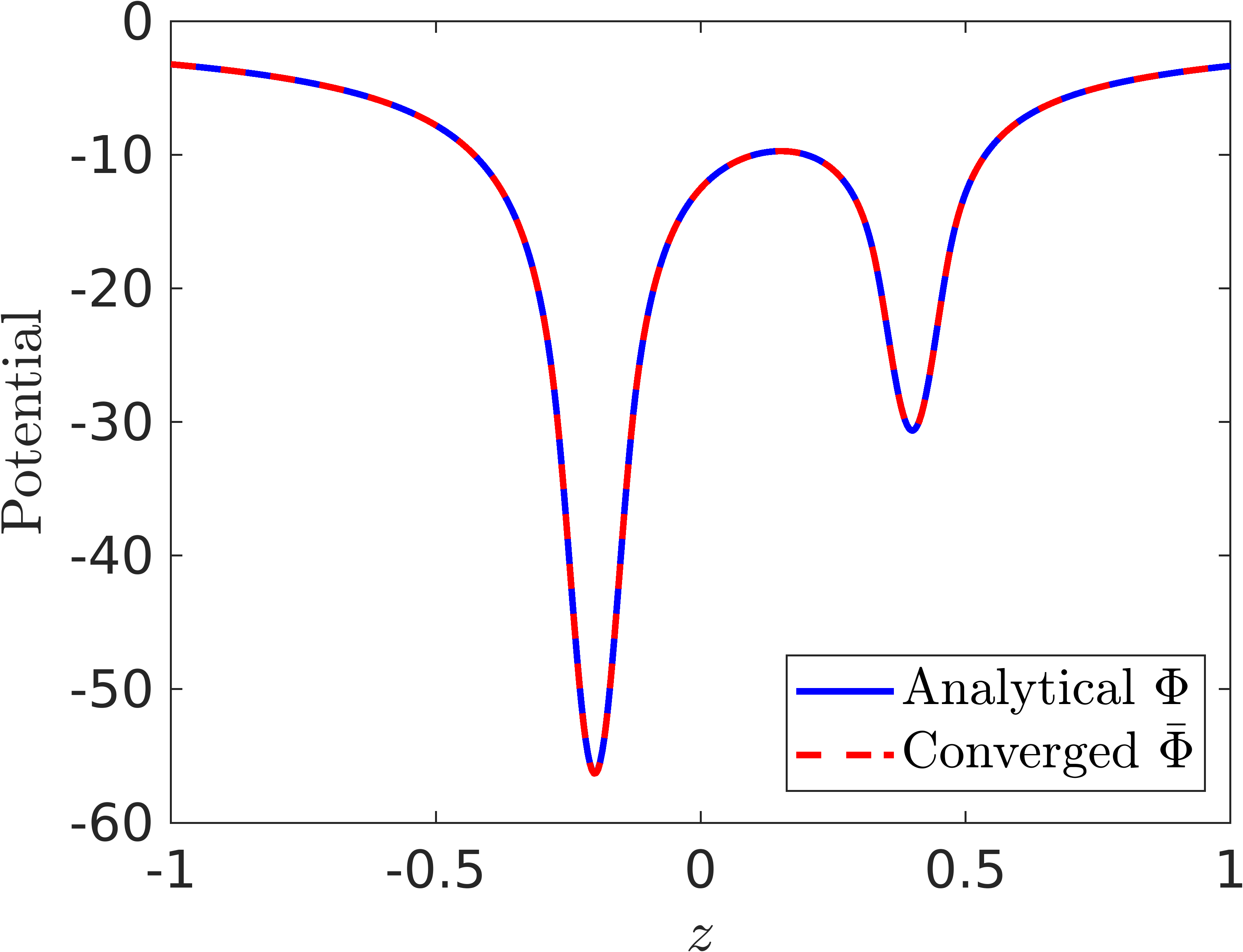} \\
    \includegraphics[width=0.24\linewidth]{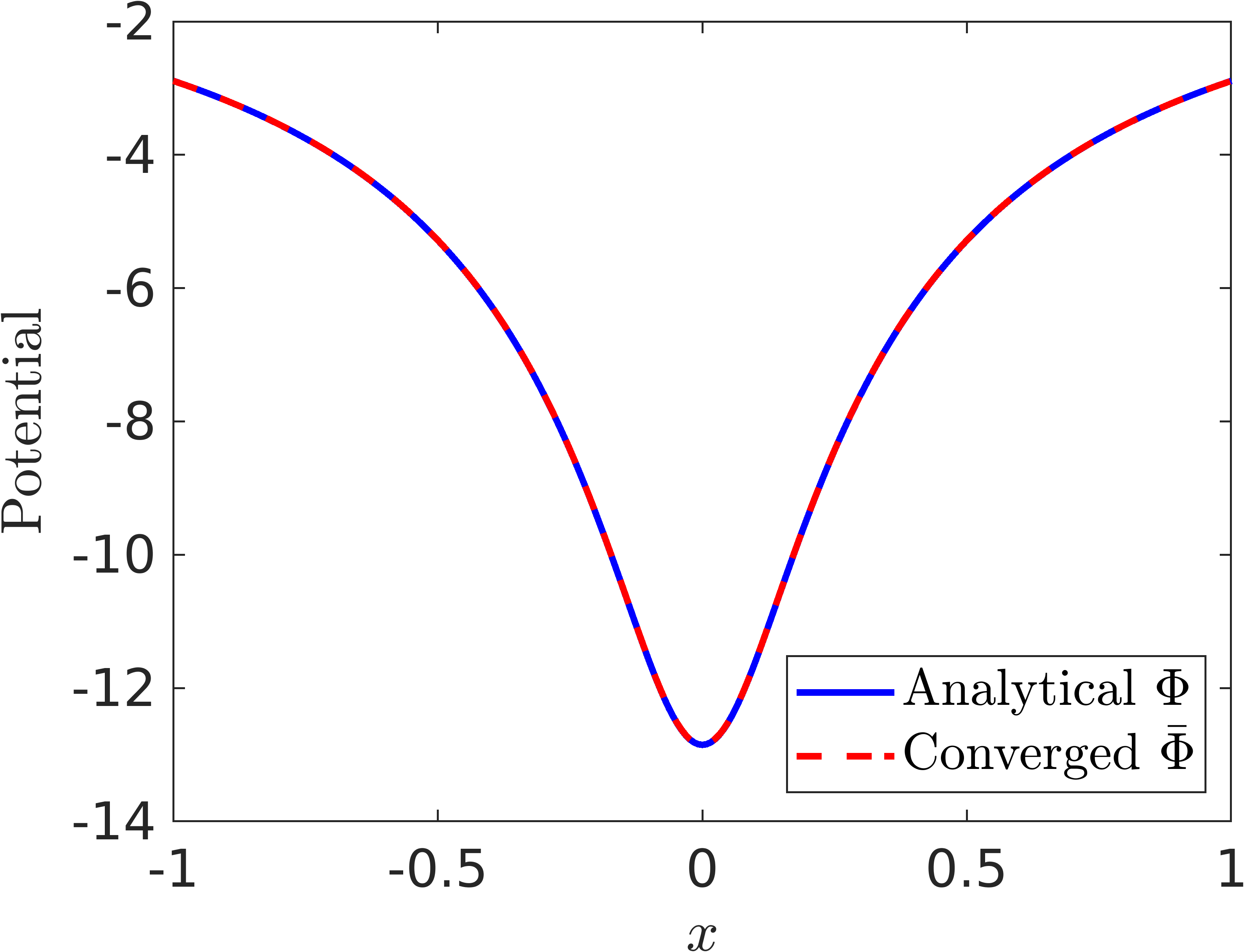}
    \includegraphics[width=0.24\linewidth]{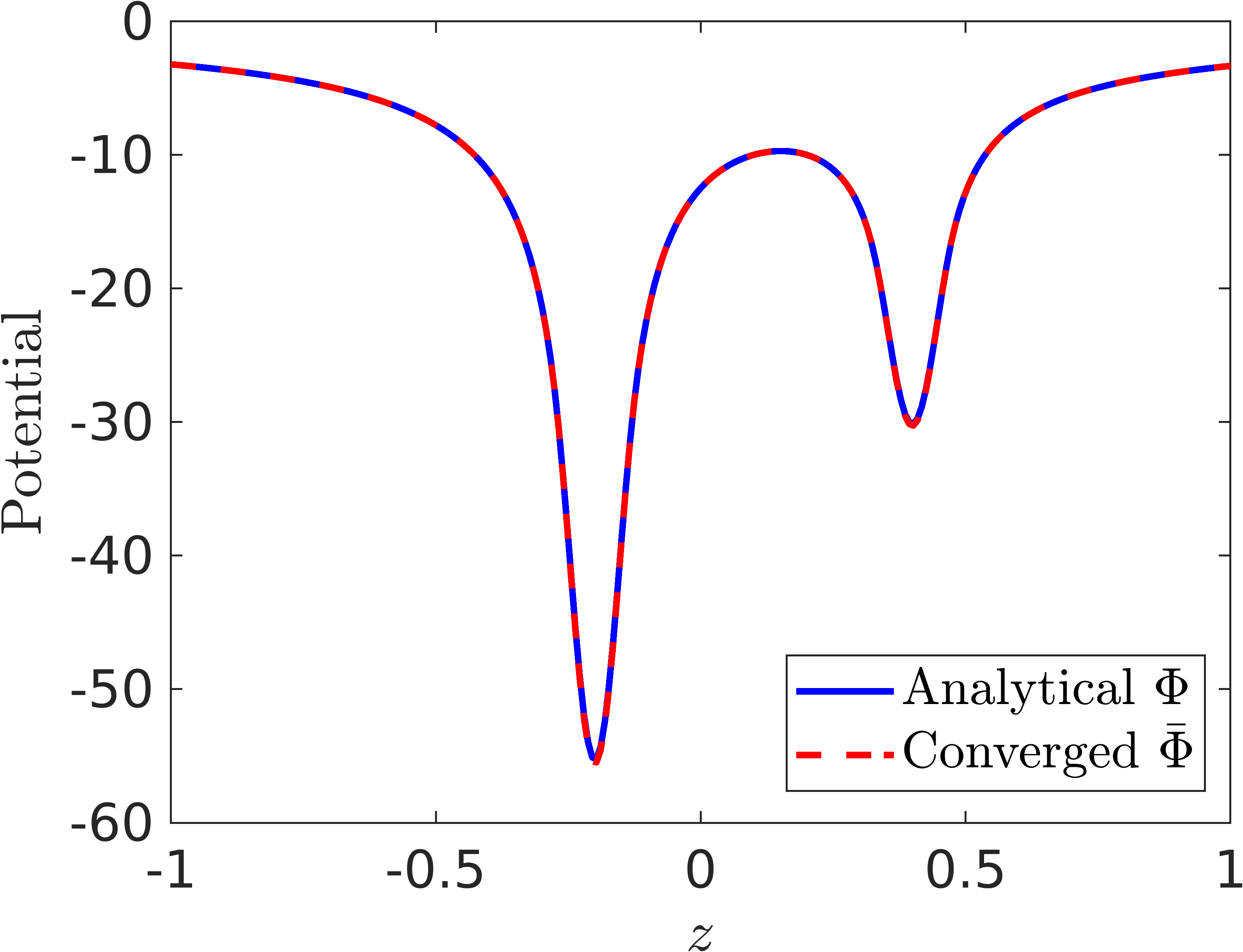}
    \includegraphics[width=0.24\linewidth]{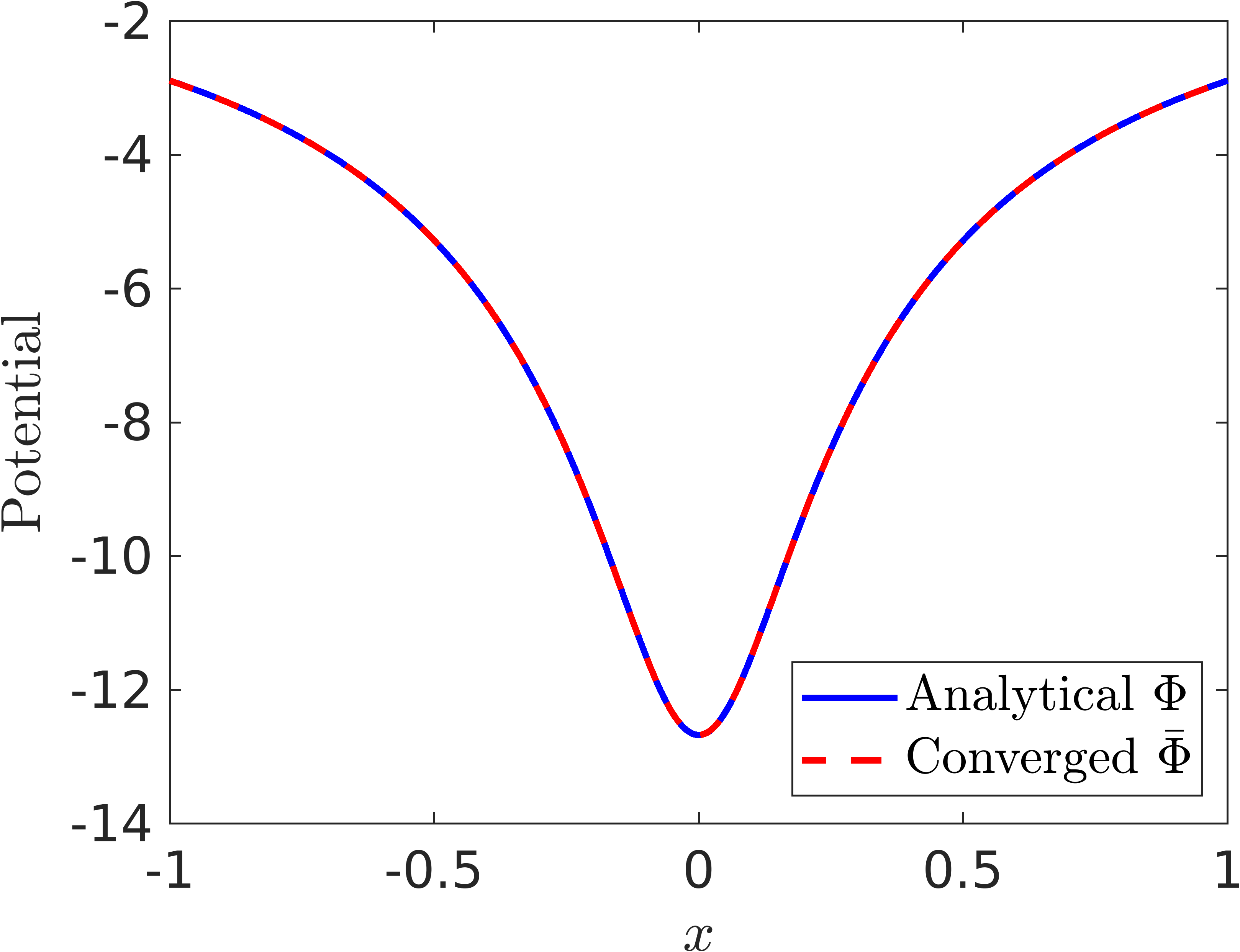}
    \includegraphics[width=0.24\linewidth]{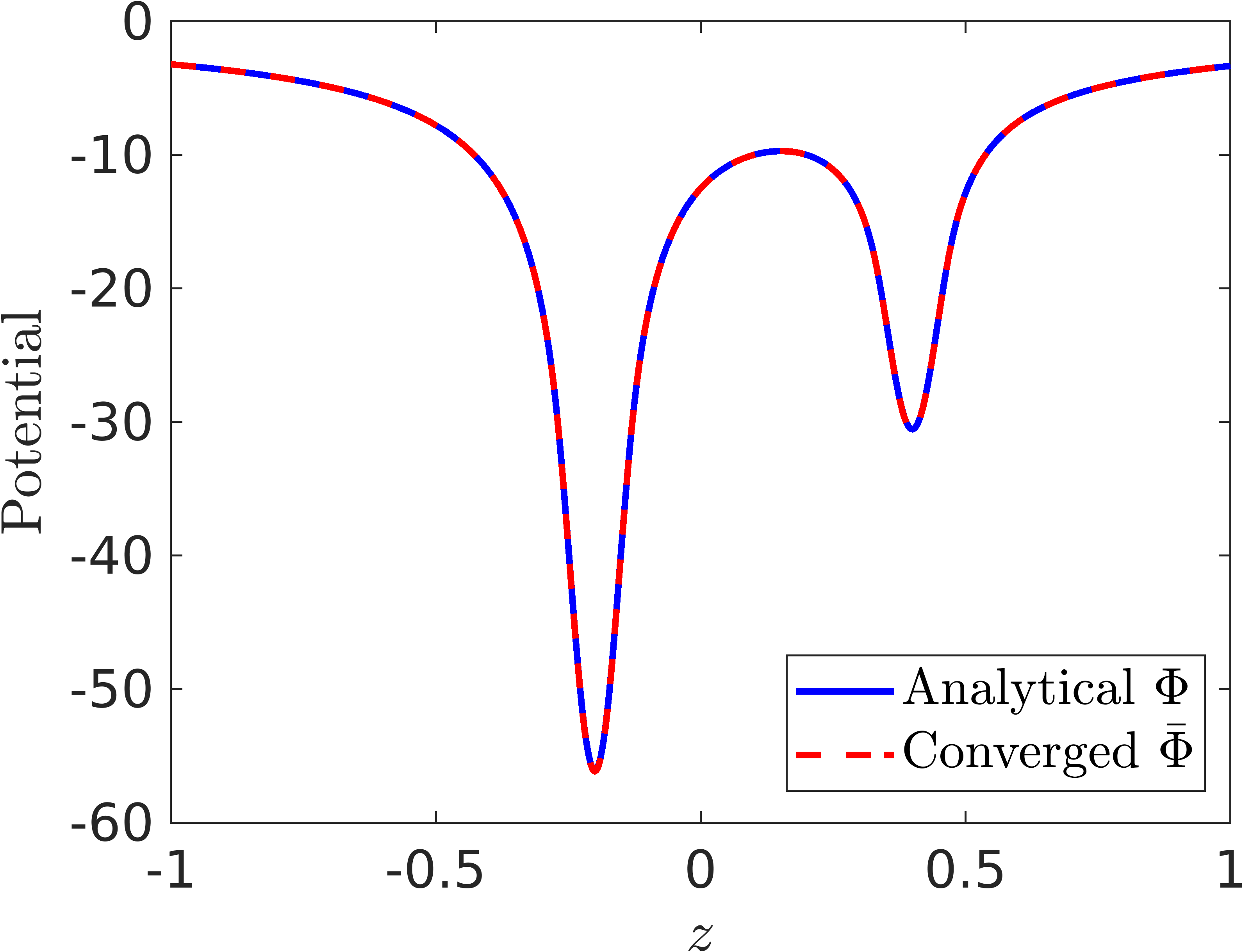} \\
    \caption{
    Computed potential $\bar{\Phi}$ and analytical potential $\Phi$ traced along the $x$ and $z$ axes for the problem stated in \S\ref{sec:static}.
    From top to bottom rows, results for CG 2nd order, CG 6th order, SOR 2nd order, and MG 6th order\@.
    Column 1: $x$-axis, resolution $256^3$. Column 2: $z$-axis, ($256^3$). Columns 3: $x$-axis ($512^3$). Column 4: $z$-axis ($512^3$). On the bottom row (MG results), the grid sizes of Columns 1--4 are $255^3$, $255^3$, $511^3$, and $511^3$, respectively.
    }
    \label{fig:test1}
\end{figure}
\begin{figure}
    \centering
    \includegraphics[width=0.24\linewidth]{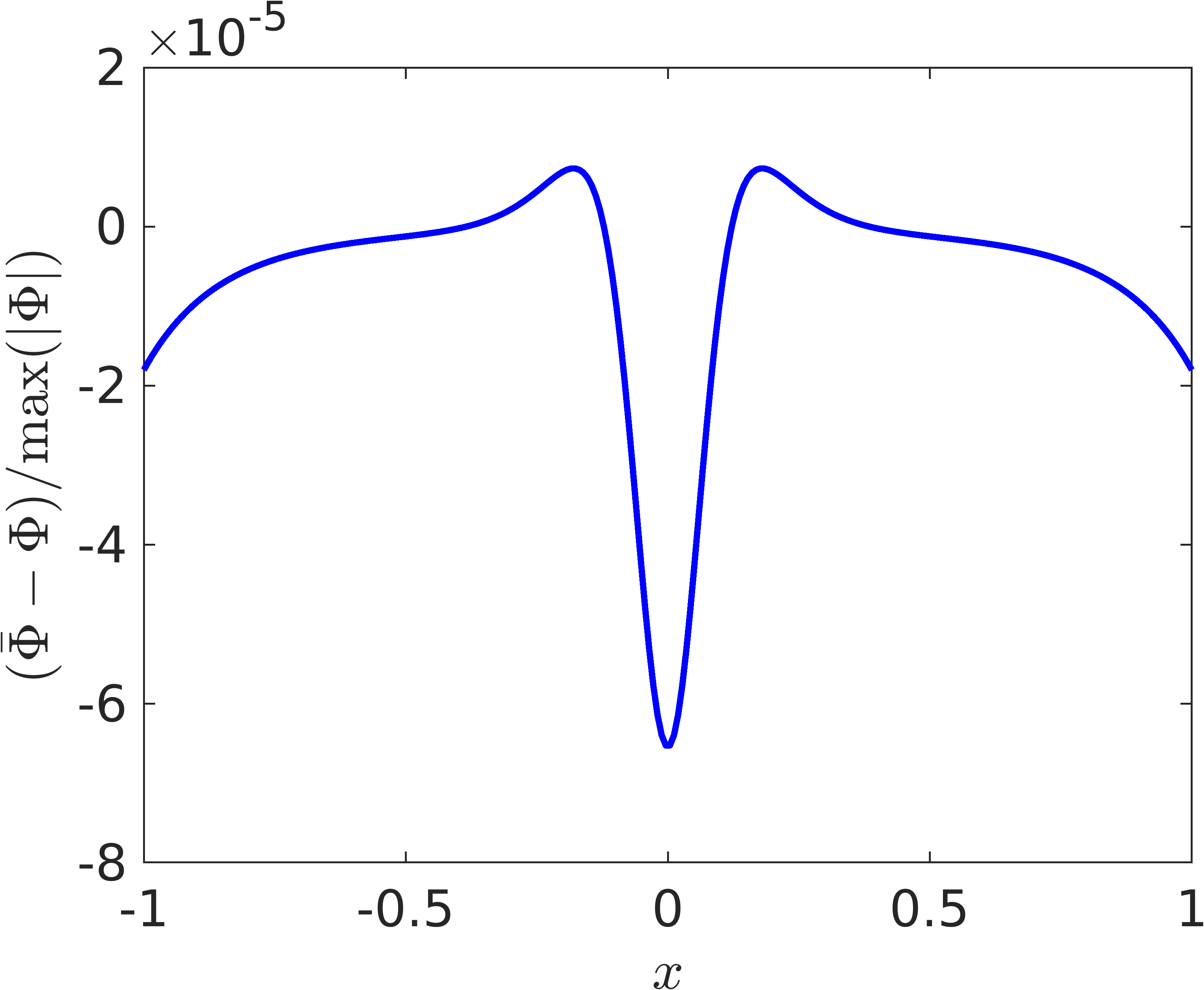}
    \includegraphics[width=0.24\linewidth]{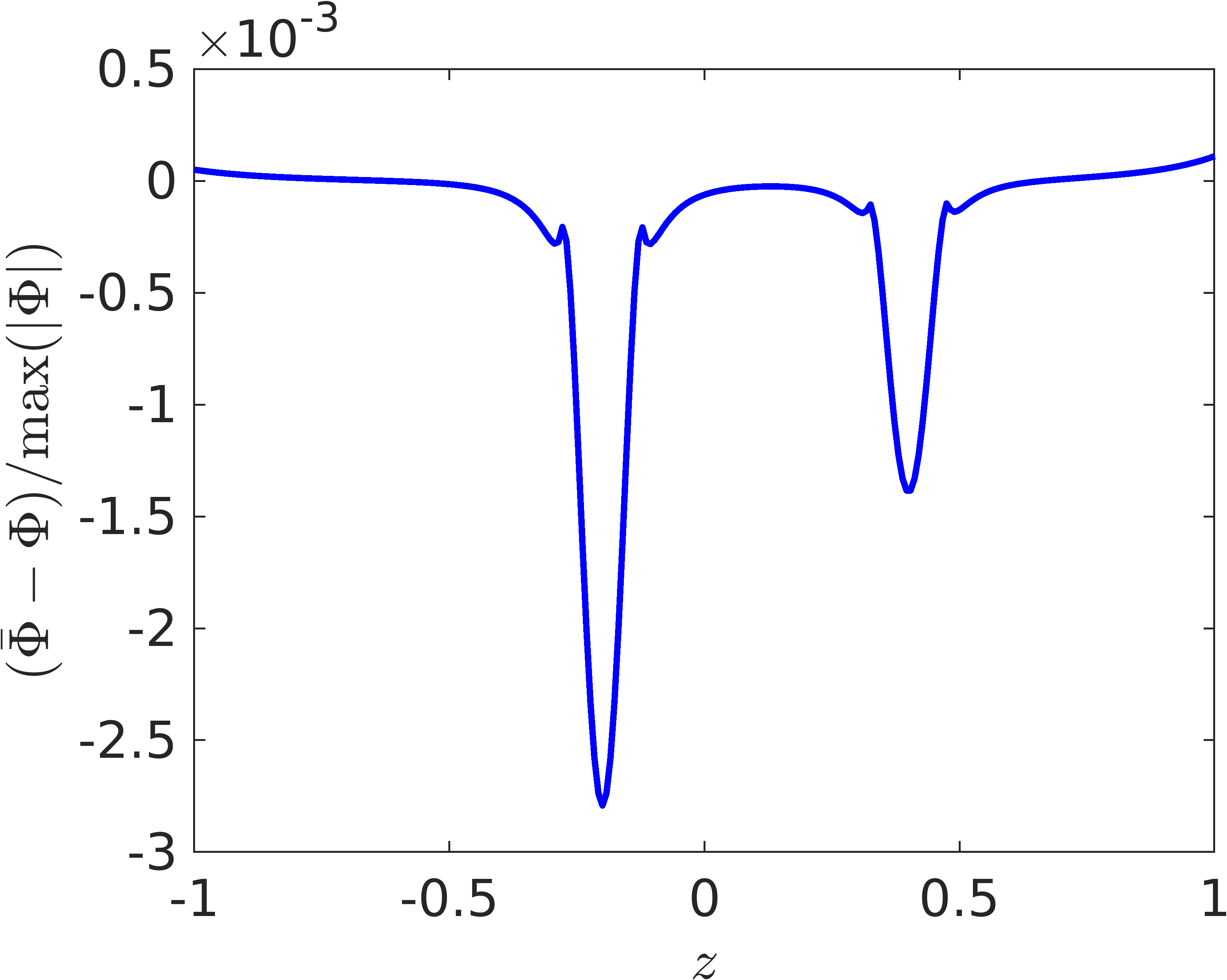}
    \includegraphics[width=0.24\linewidth]{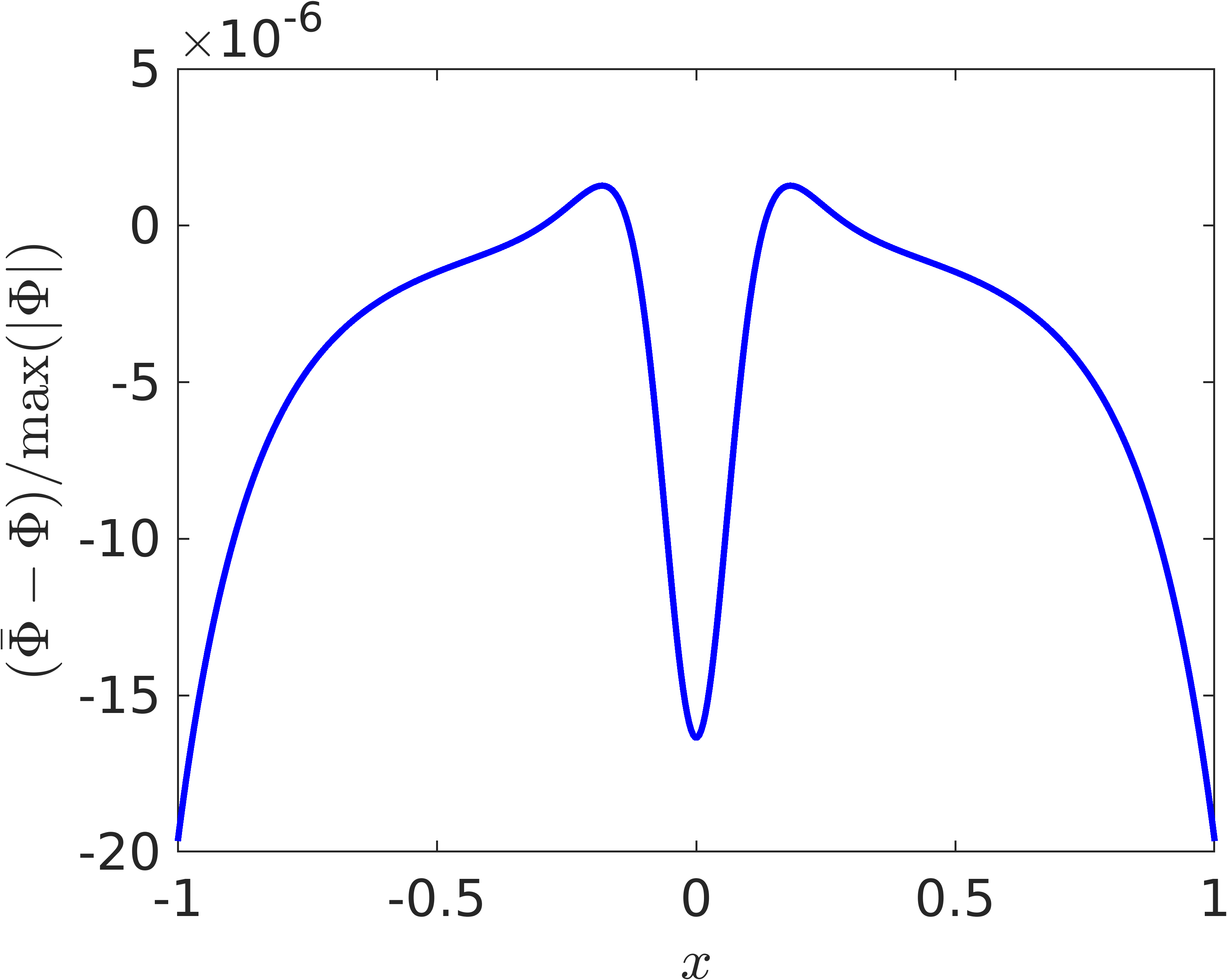}
    \includegraphics[width=0.24\linewidth]{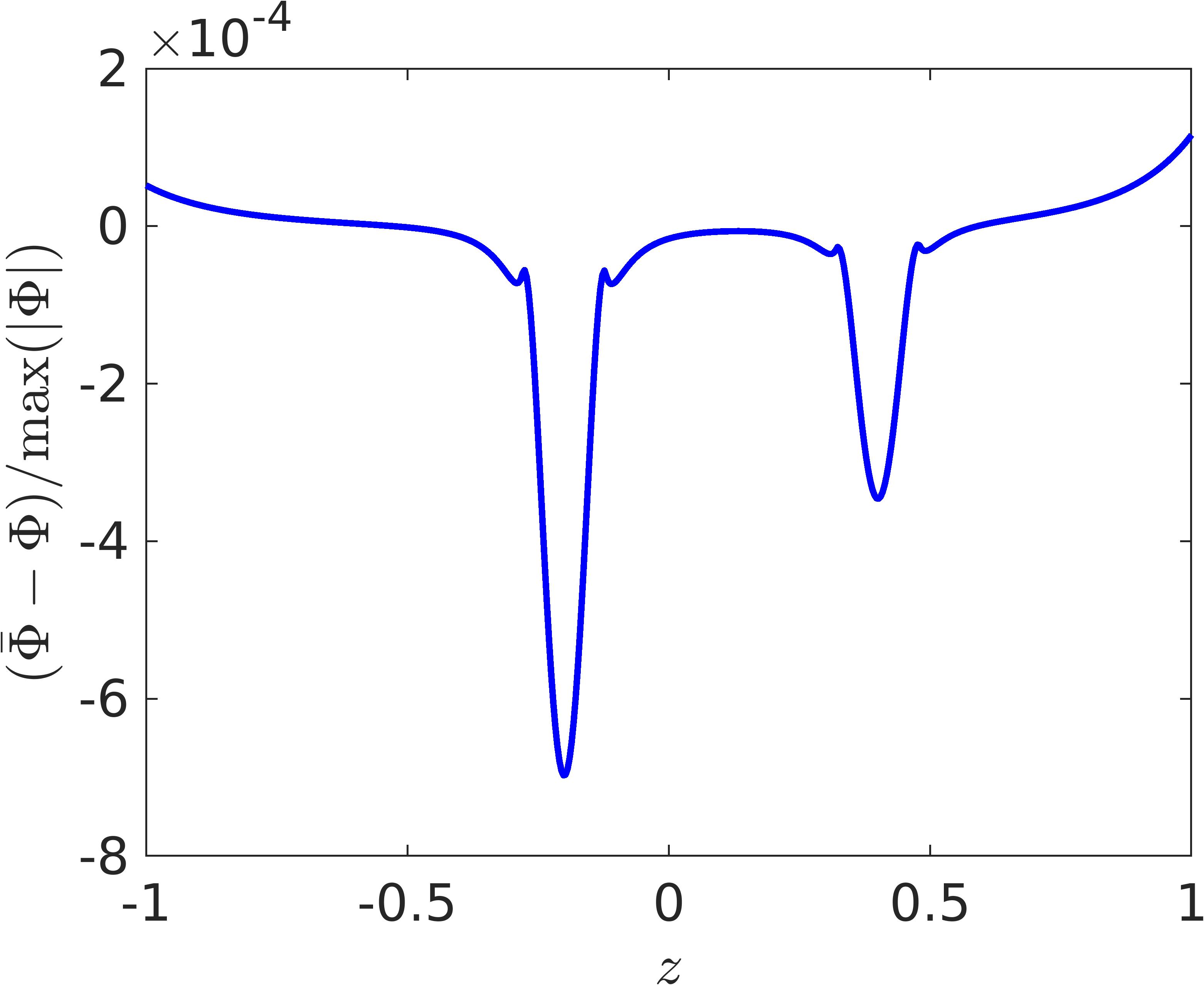}\\
    \includegraphics[width=0.24\linewidth]{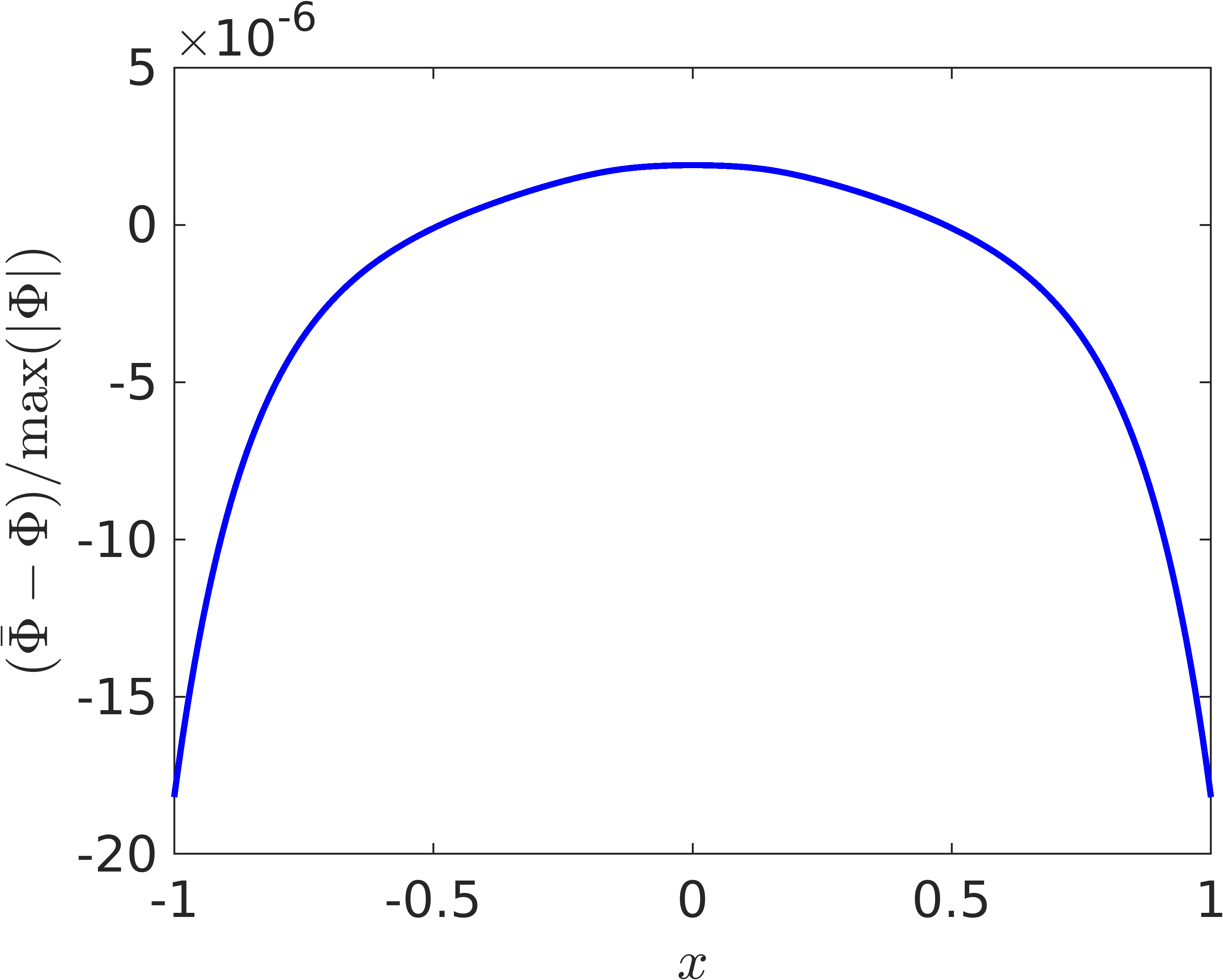}
    \includegraphics[width=0.24\linewidth]{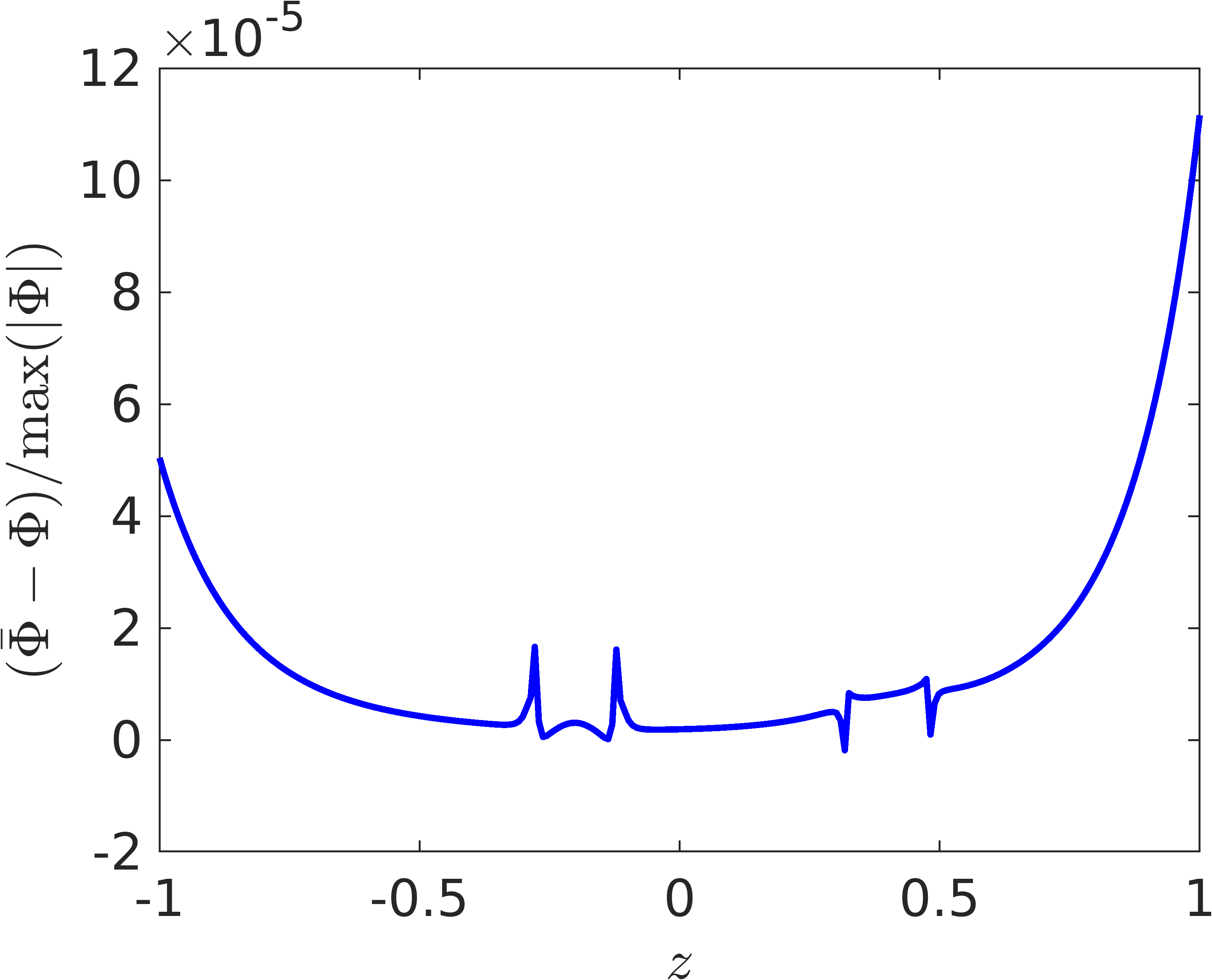}
    \includegraphics[width=0.24\linewidth]{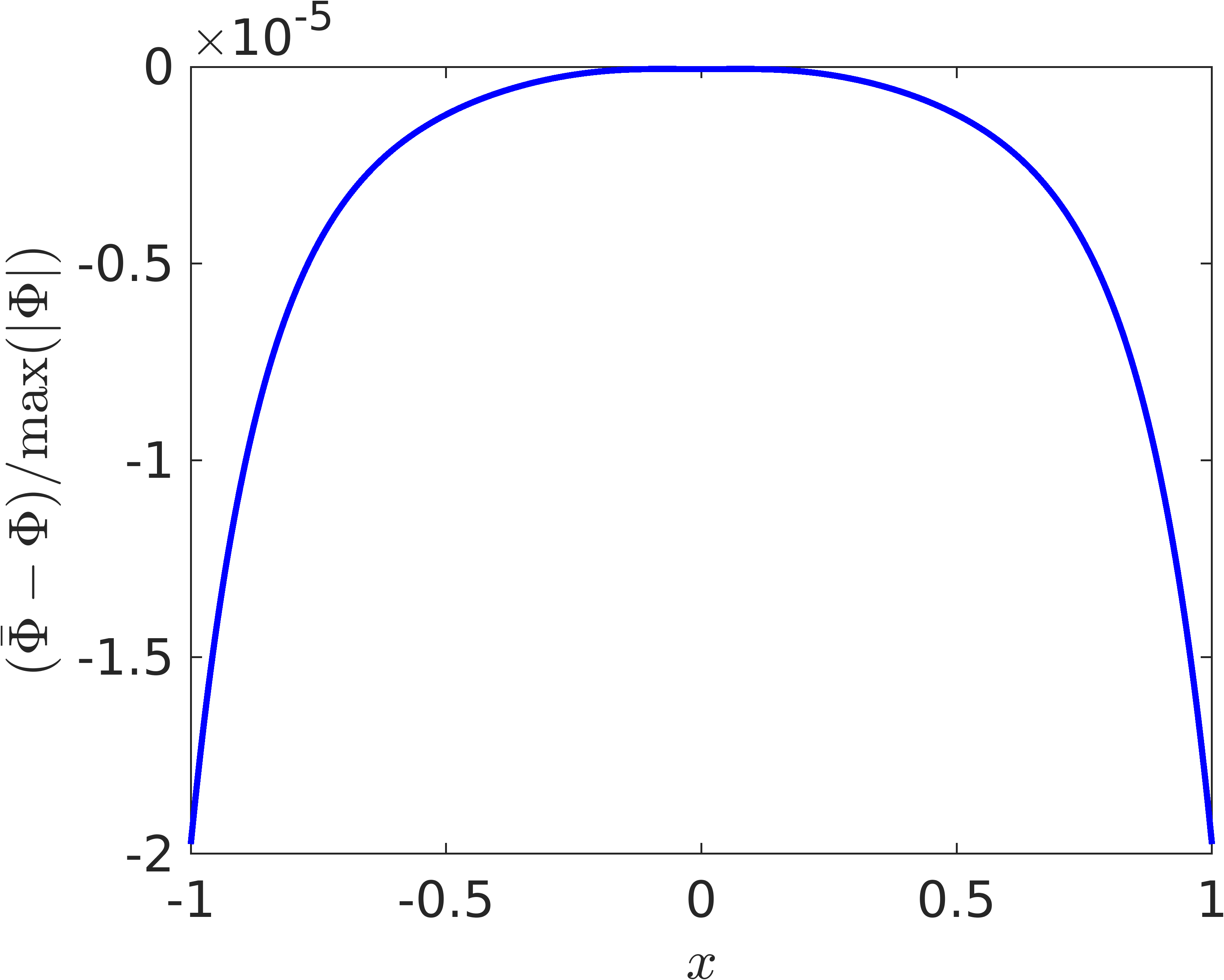}
    \includegraphics[width=0.24\linewidth]{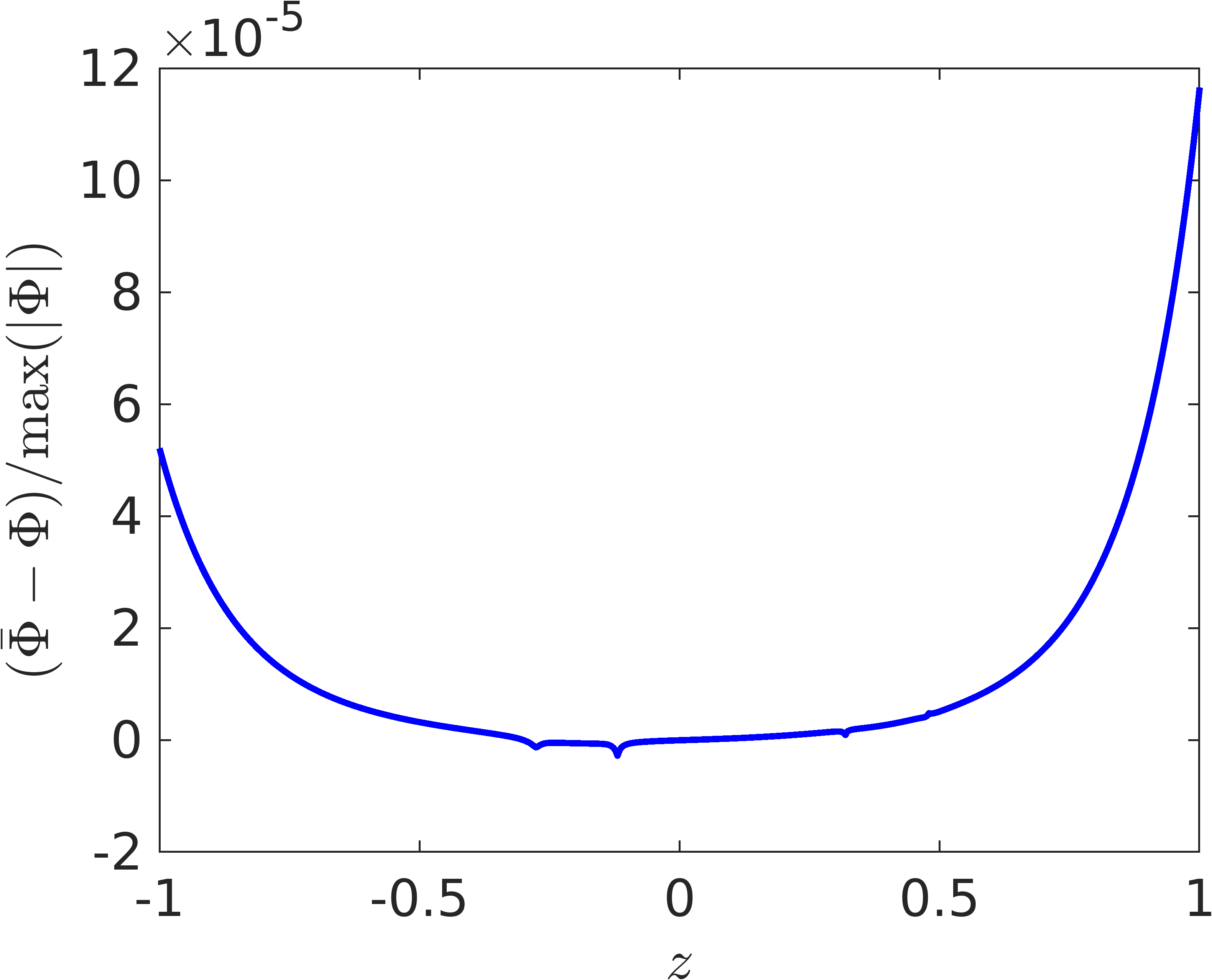}\\
    \includegraphics[width=0.24\linewidth]{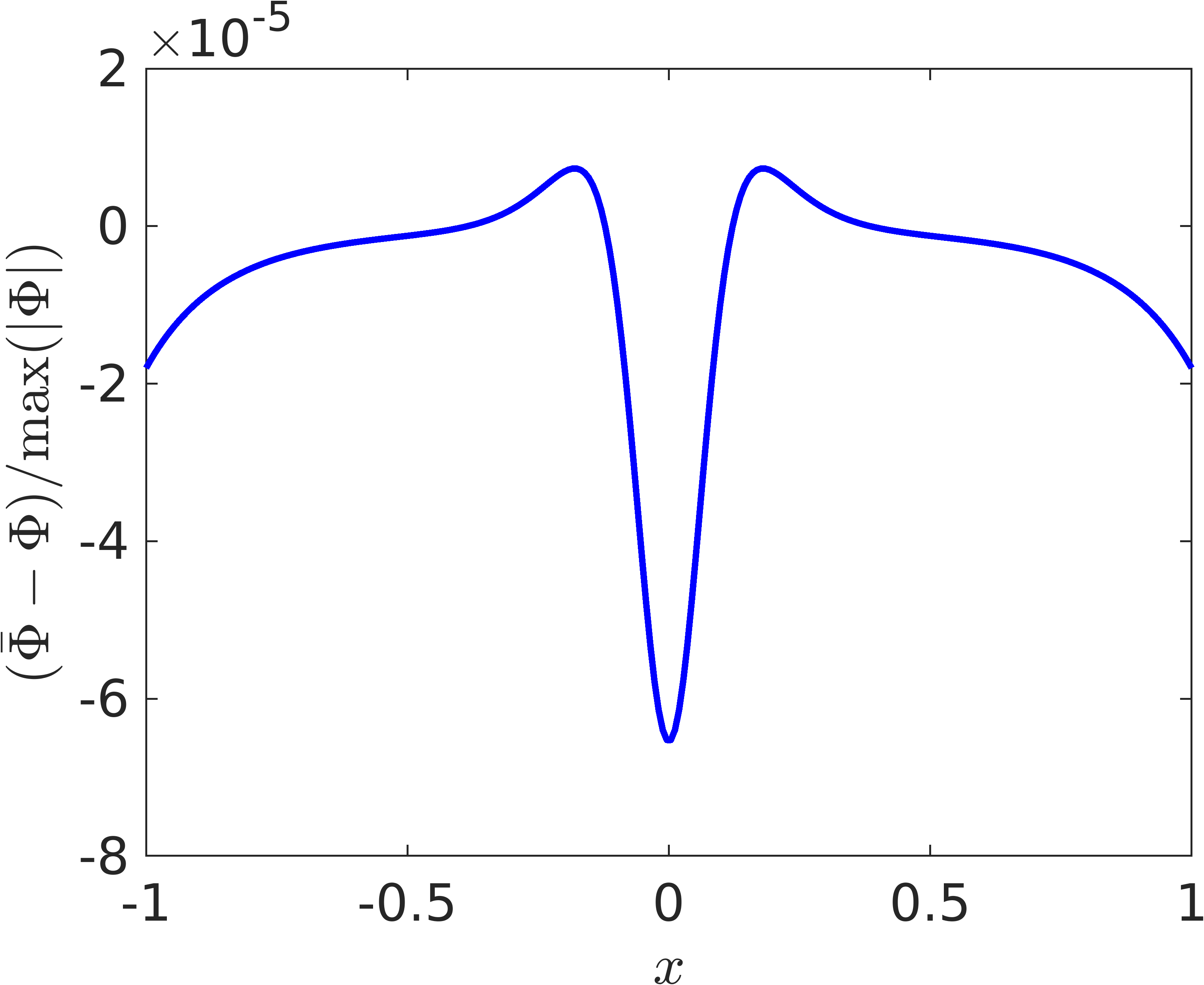}
    \includegraphics[width=0.24\linewidth]{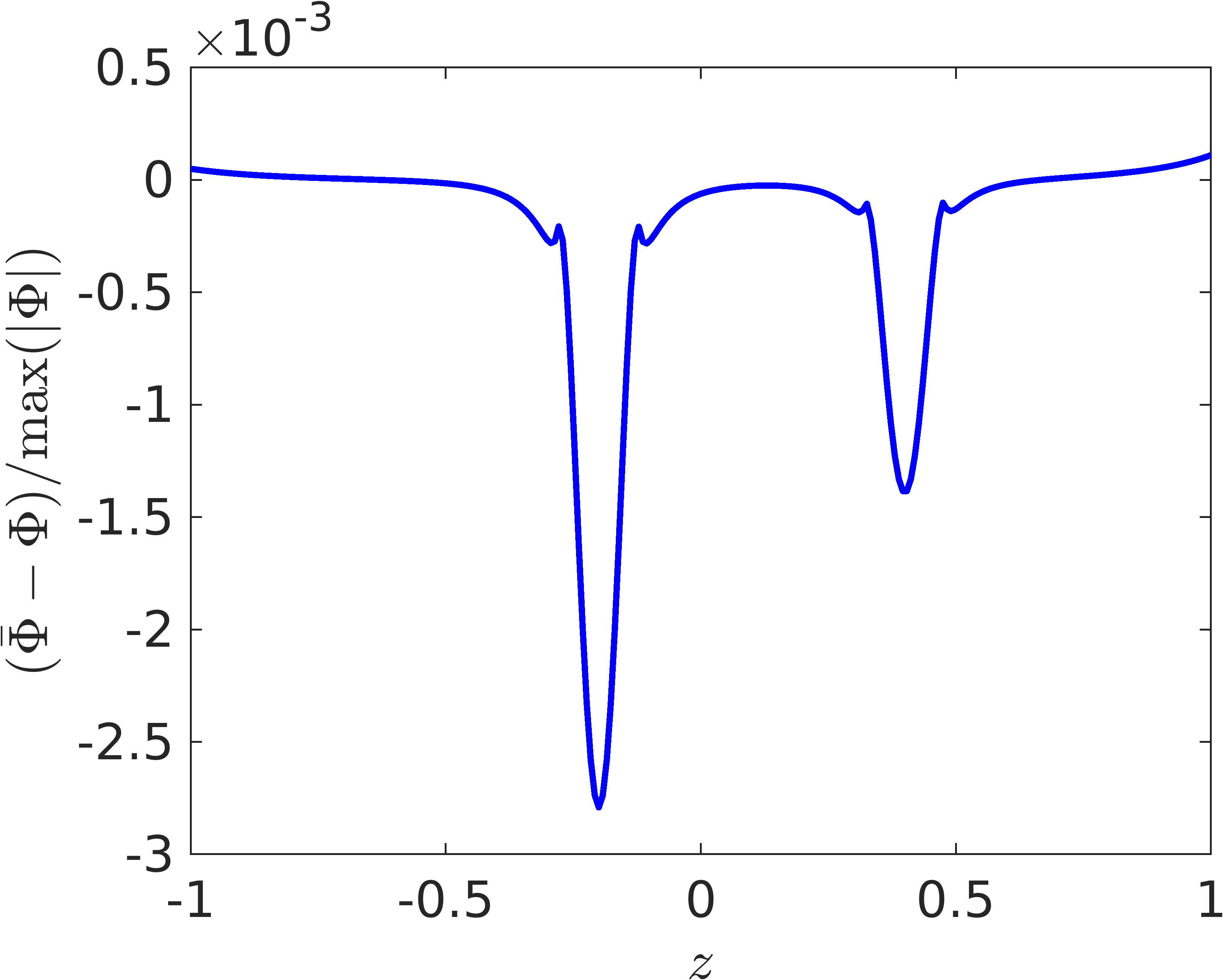}
    \includegraphics[width=0.24\linewidth]{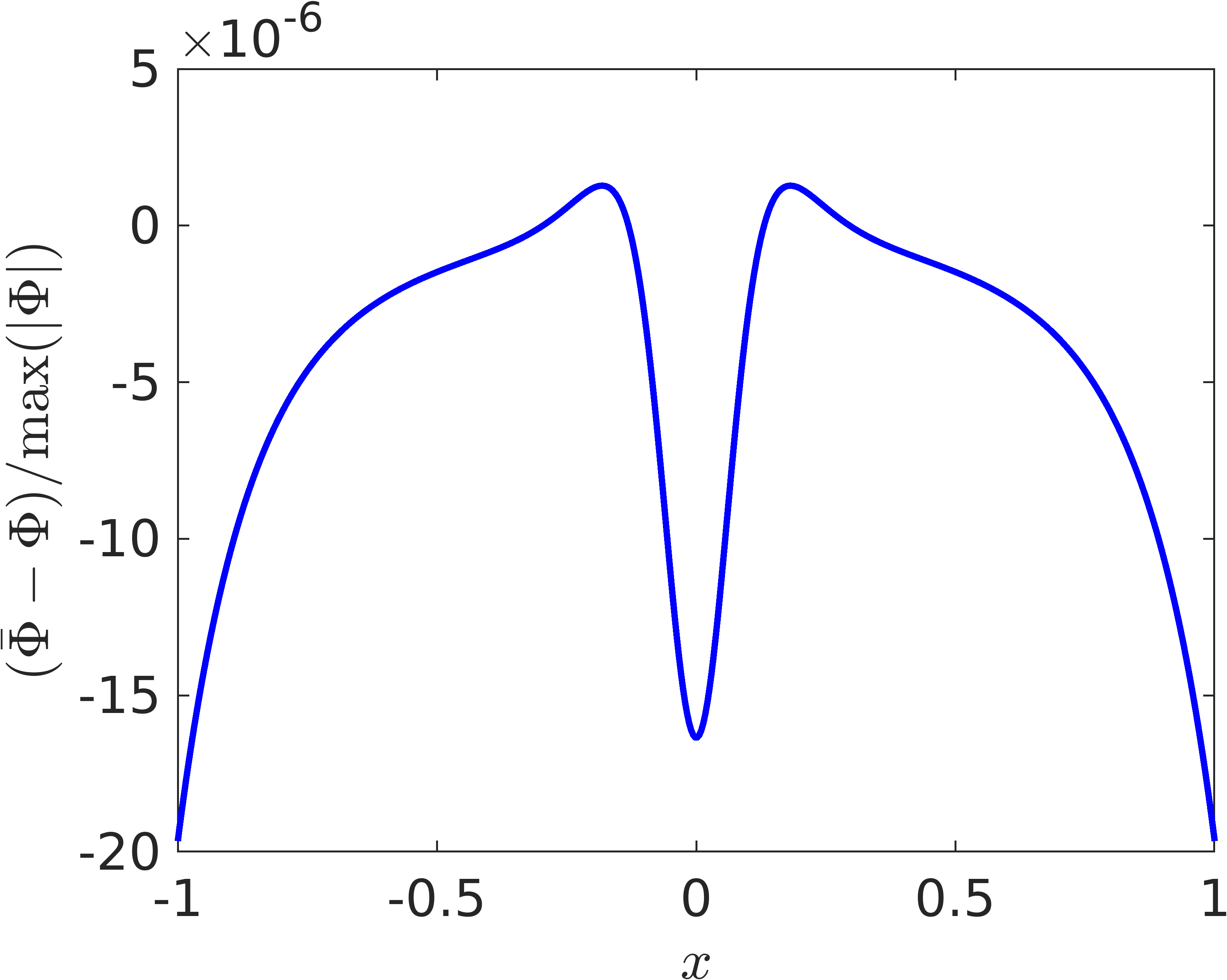}
    \includegraphics[width=0.24\linewidth]{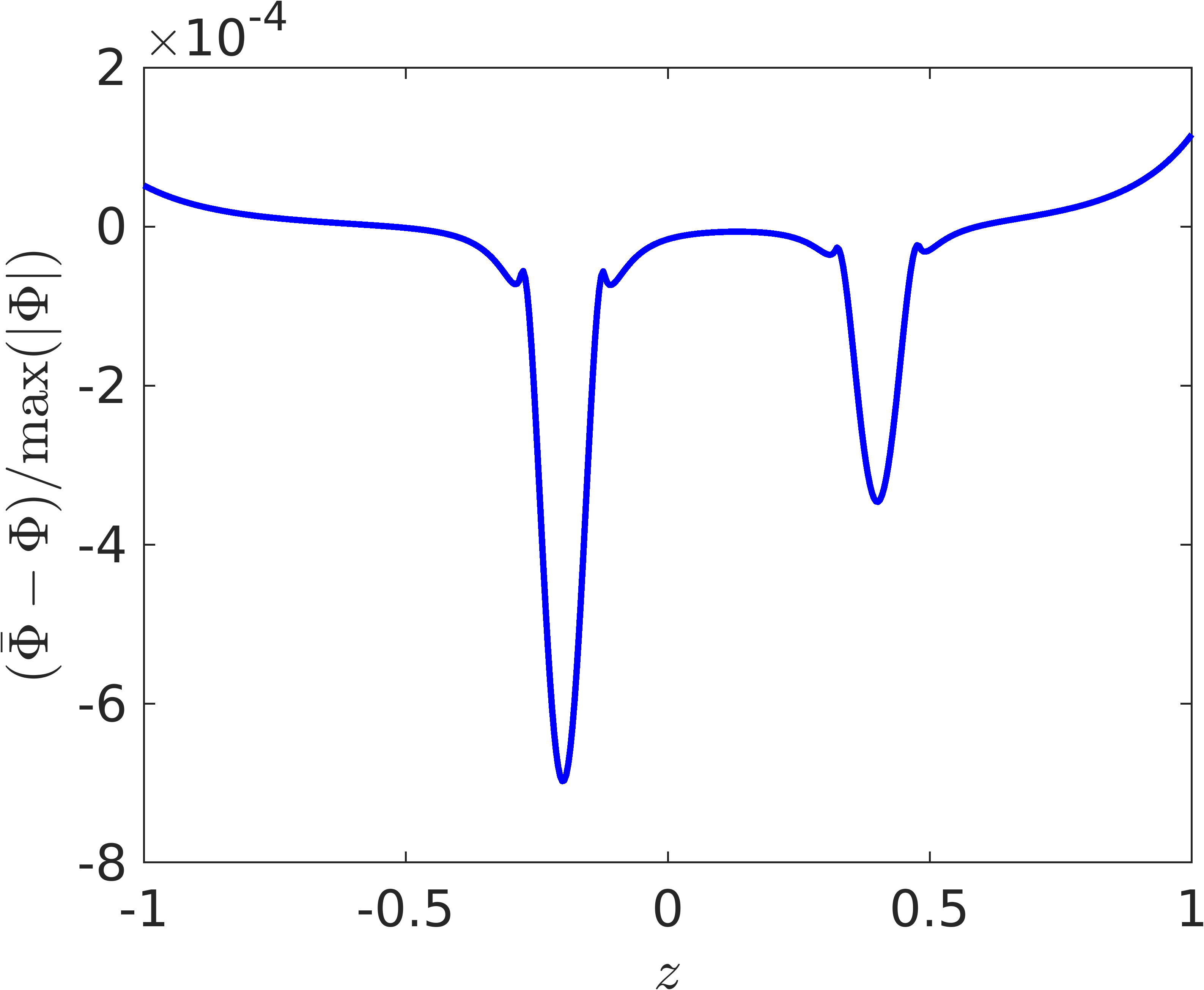}\\
    \includegraphics[width=0.24\linewidth]{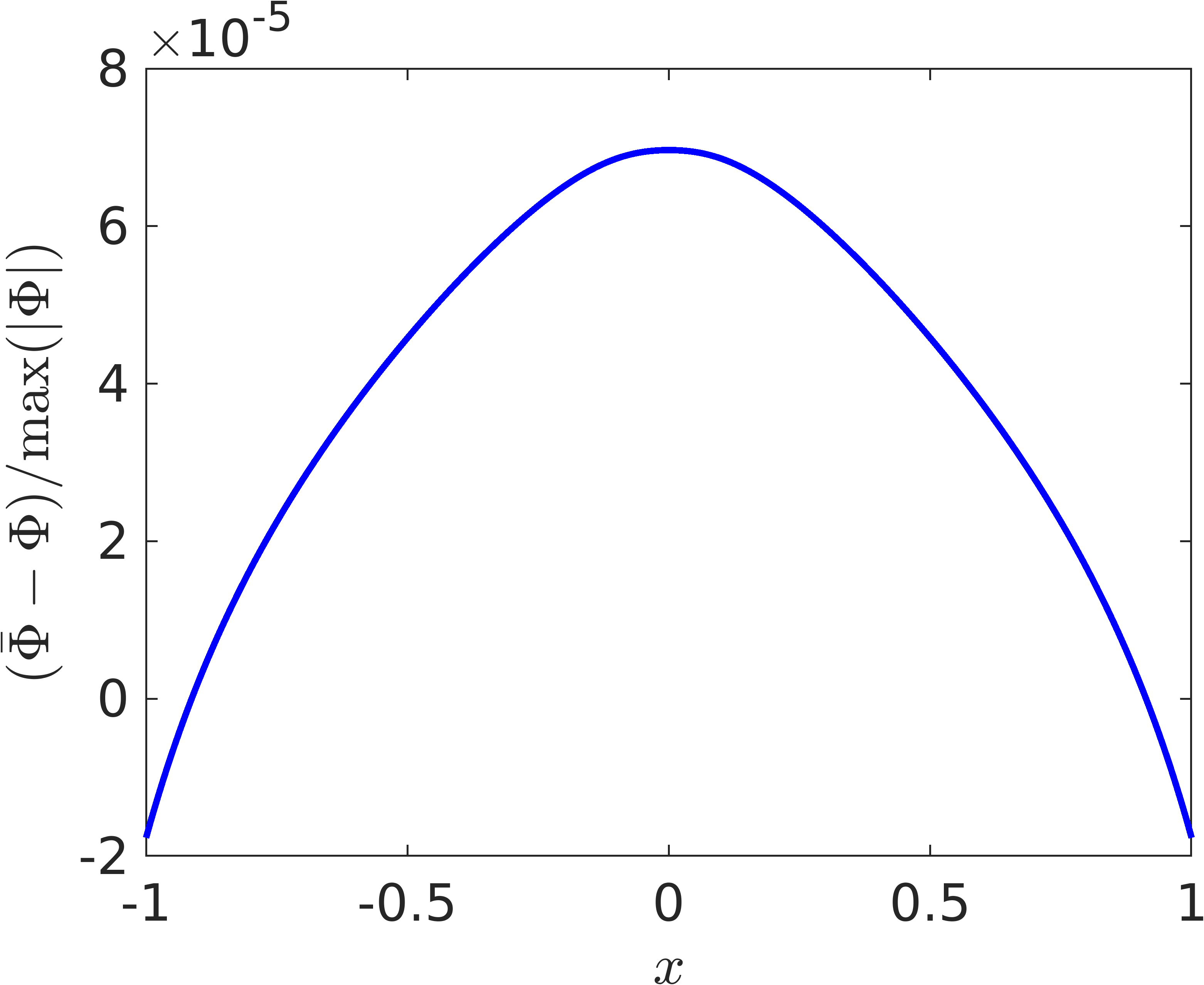}
    \includegraphics[width=0.24\linewidth]{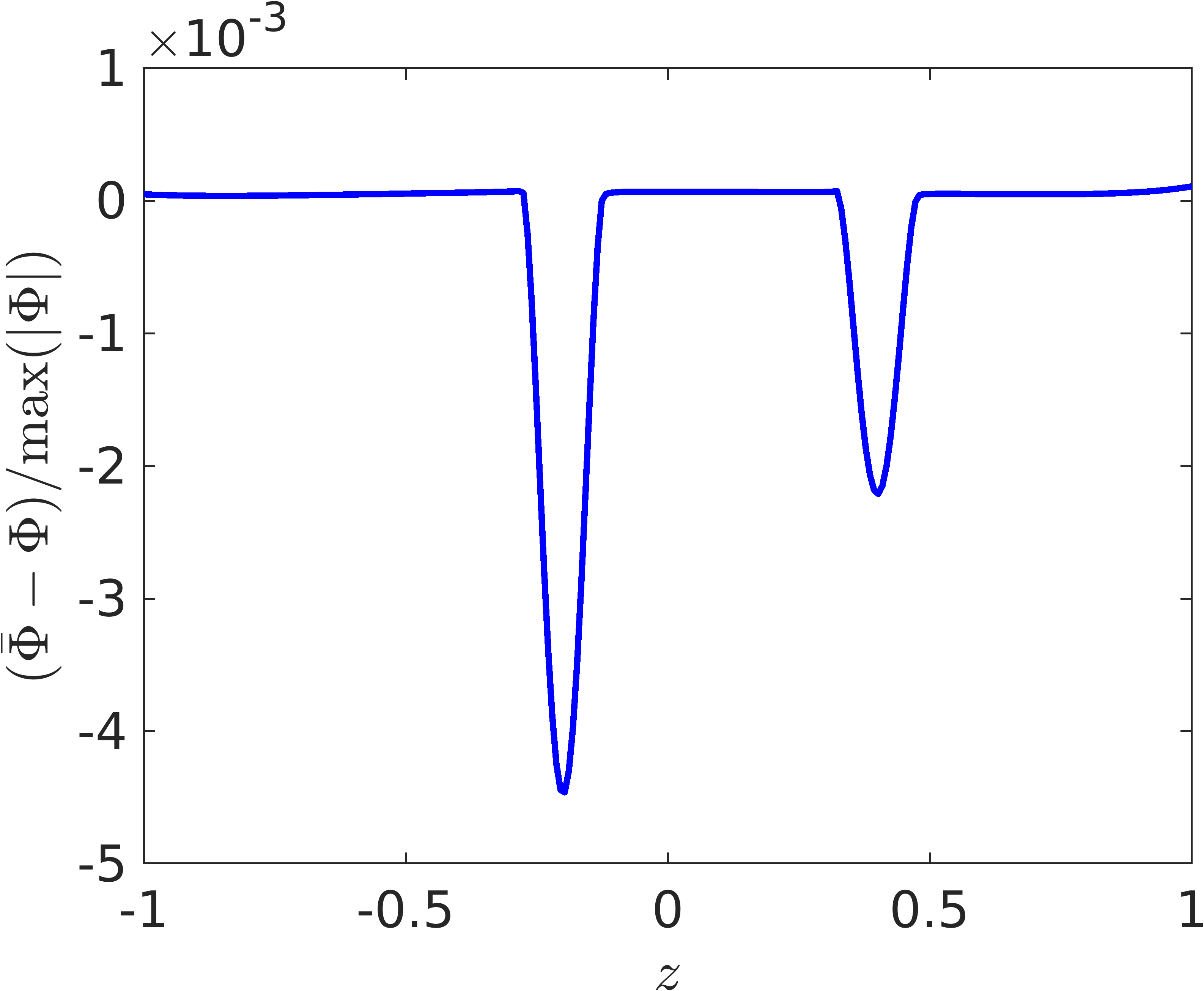}
    \includegraphics[width=0.24\linewidth]{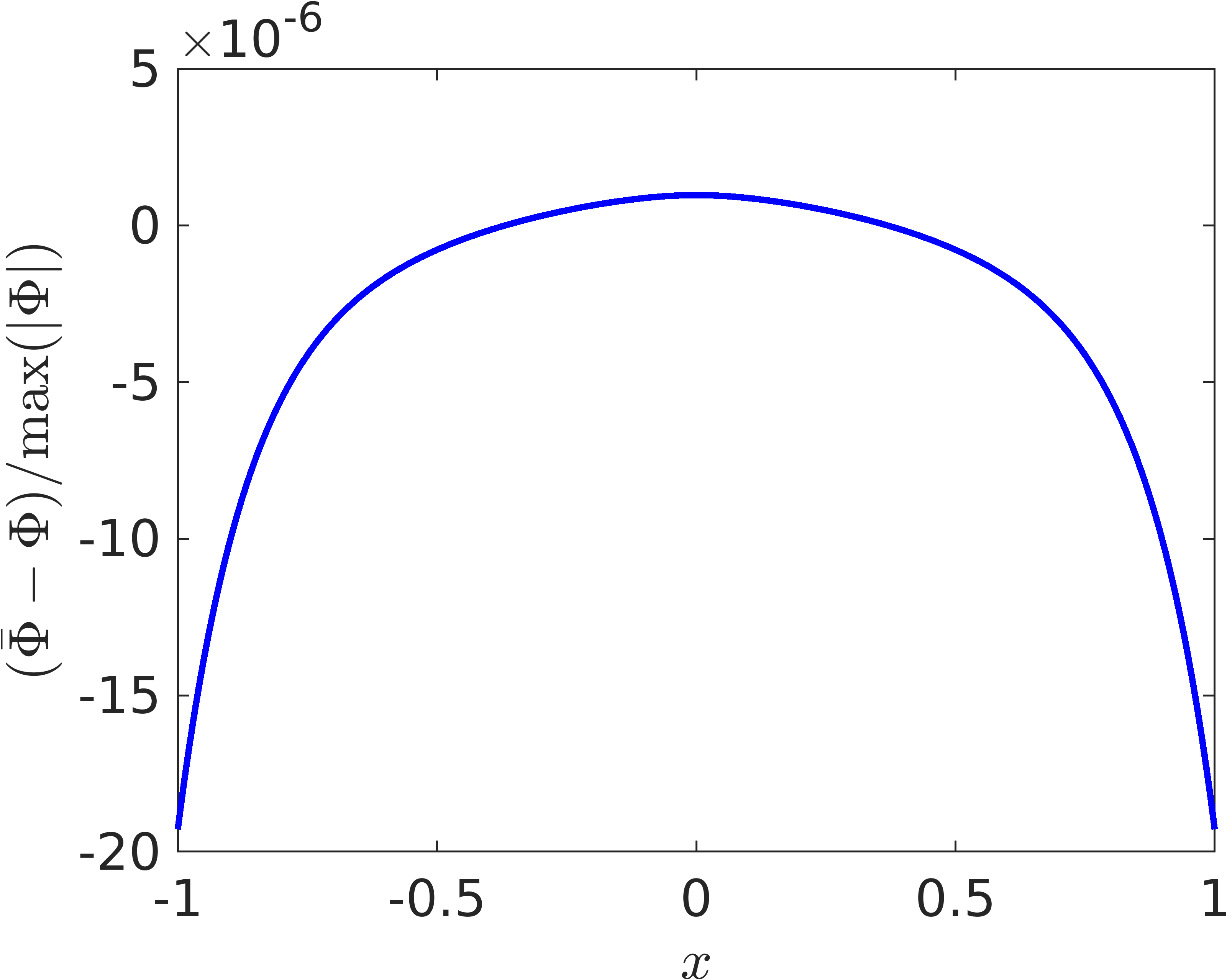}
    \includegraphics[width=0.24\linewidth]{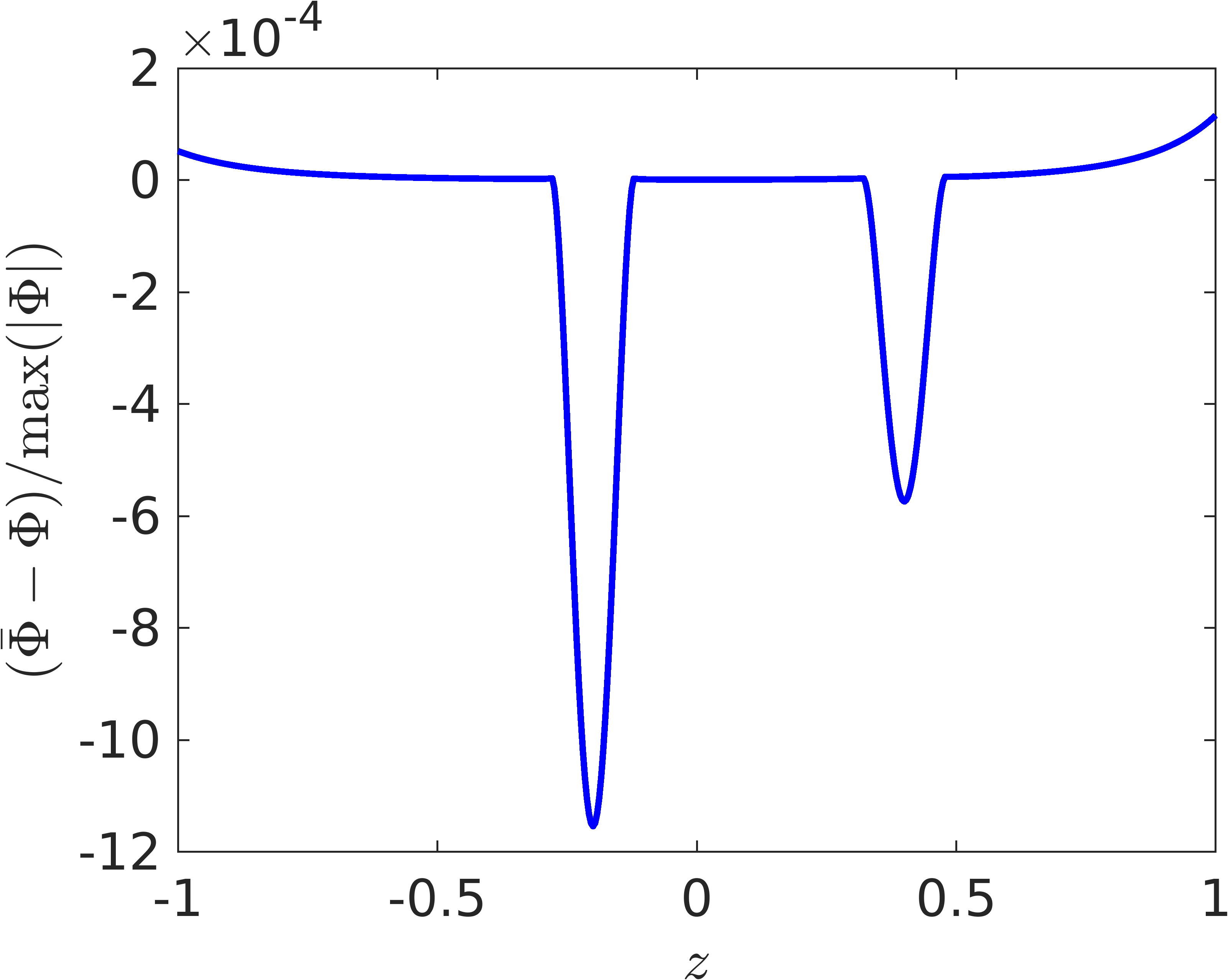}
    \caption{
    Relative errors of the solutions shown in Figure \ref{fig:test1}, presented in the same order.
    The relative errors are traced along the $x$ and $z$ axes for the problem stated in \S\ref{sec:static}.
    From top to bottom rows, results for CG 2nd order, CG 6th order, SOR 2nd order, and MG 6th order\@.
    Column 1: $x$-axis, resolution $256^3$. Column 2: $z$-axis, ($256^3$). Columns 3: $x$-axis ($512^3$). Column 4: $z$-axis ($512^3$). On the bottom row (MG results), the grid sizes of Columns 1--4 are $255^3$, $255^3$, $511^3$, and $511^3$, respectively.
    Note, that the $x$ axis avoids the masses, while the $z$ axis passes through the mass centers.
    }
    \label{fig:test_errors}
\end{figure}

\subsection{Cartesian tests} \label{sec:static}
We test the code performance of the algorithms by finding solutions $\Phi$ for the combination of two spheres in Cartesian coordinates.
Tables \ref{tab:timings1} and \ref{tab:timings2} show the timings, number of steps, and speeds for selected methods and resolutions, as well as for various computing platforms with different GPU hardware.
Each run has a phase in which the multipole expansion of $\Phi$ in the ghost zones is calculated in time $t_\mathrm{mp}$, followed by a number of iteration steps or V-cycles $N_\mathrm{I}$ from an initial guess $\Phi_0=0$, taking a time $t_\mathrm{iter}$.  We report the value of $N_I$ required to reach a residual goal, the timings $t_\mathrm{mp}$ and $t_\mathrm{s}=t_\mathrm{iter}/N_\mathrm{I}$, the iteration step speed $s_\mathrm{s}=N^3/t_\mathrm{s}$, and the total speed $s_\mathrm{t} = N^3 / (t_\mathrm{mp} + t_\mathrm{iter})$, both in billions of zone updates per second (Gzups).

The assumption of a cold start for the potential $\Phi_0=0$ yields a poorly performing case for $N_I$, growing as $O(N)$ for the classical solvers (SOR, CG, BICGSTAB) but remaining grid-size independent for MG\@.
In practice, during HD/MHD simulations, the previous time step provides a $\Phi_0$ much closer to the solution, reducing the practical step count $N_\mathrm{p}$, for example, to $1$--$30$ using BICGSTAB in Sect.~\ref{sec:dynamic}, so the practical speed $s_\mathrm{p}\approx \min(s_\mathrm{s}/N_\mathrm{p},\,s_\mathrm{t})$ lies between $s_\mathrm{s}$ and $s_\mathrm{t}$.

We benchmark against the Astaroth MHD speed $s_\mathrm{MHD}\approx0.9$ Gzups per Runge-Kutta substep, measured on LUMI for a 6th-order 3D MHD model at $256^3$ on one GPU, with three Runge-Kutta substeps per MHD step.
For MG at comparable resolution, Table \ref{tab:timings2} gives $s_\mathrm{s}\approx1$ Gzups and $s_\mathrm{t}\approx0.28$ Gzups, so self-gravity would consume roughly one-half to three-quarters of the run time, an acceptable cost given its physical importance, and a conservative one, since simulations naturally provide a better $\Phi_0$ initial guess.
For the classical second-order solvers SOR and CG, Table \ref{tab:timings1} gives $s_\mathrm{s}\approx20$ Gzups and $s_\mathrm{t}\approx0.03$ Gzups; for $N_\mathrm{p} =1$, 30, and 50 this yields $s_\mathrm{p}\approx20$, 0.6, and 0.4 Gzups, and self-gravity fractions of $\sim0.04$, 0.6, and 0.7 of the run time.
Thus classical solvers offer negligible-to-tolerable cost depending on how good the initial guess is, while MG provides a tighter guarantee of low-to-tolerable cost regardless of $N_\mathrm{p}$.

Results and errors of various solvers are shown in Figures \ref{fig:test1} and \ref{fig:test_errors} as line graphs of $\Phi$ along the $x$- and $z$-axes. The $z$-axis, which passes through the centers of the two spheres, is the more demanding direction. Details of the respective run setups are described in the caption of Figure \ref{fig:test1}.
In general, regions with larger errors in Figure \ref{fig:test_errors} fall near the peak positions of the potential, where $\Phi$ has rapid variations, and also regions near the boundaries.

\begin{figure}
\centering
    \includegraphics[width=0.24\linewidth]{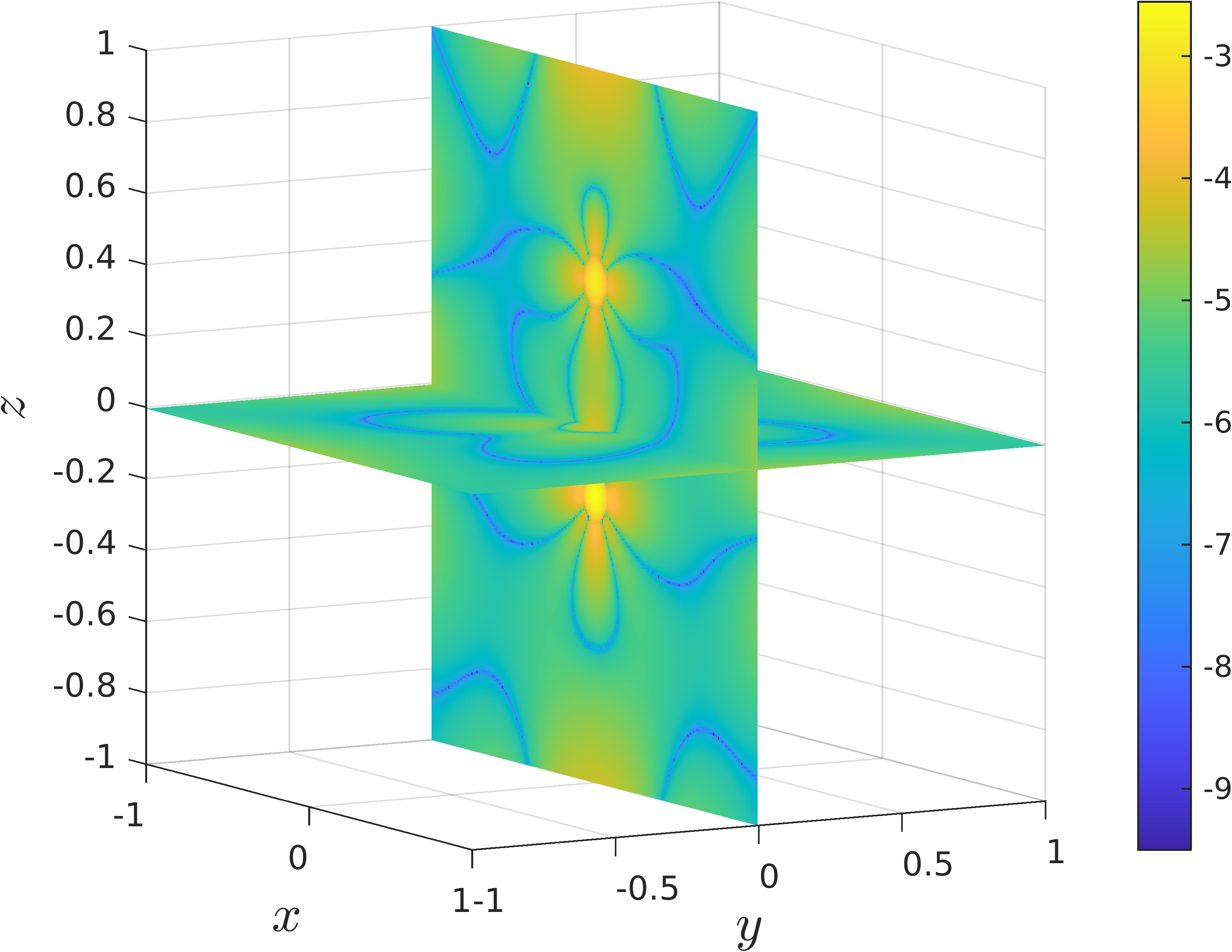}
    \includegraphics[width=0.24\linewidth]{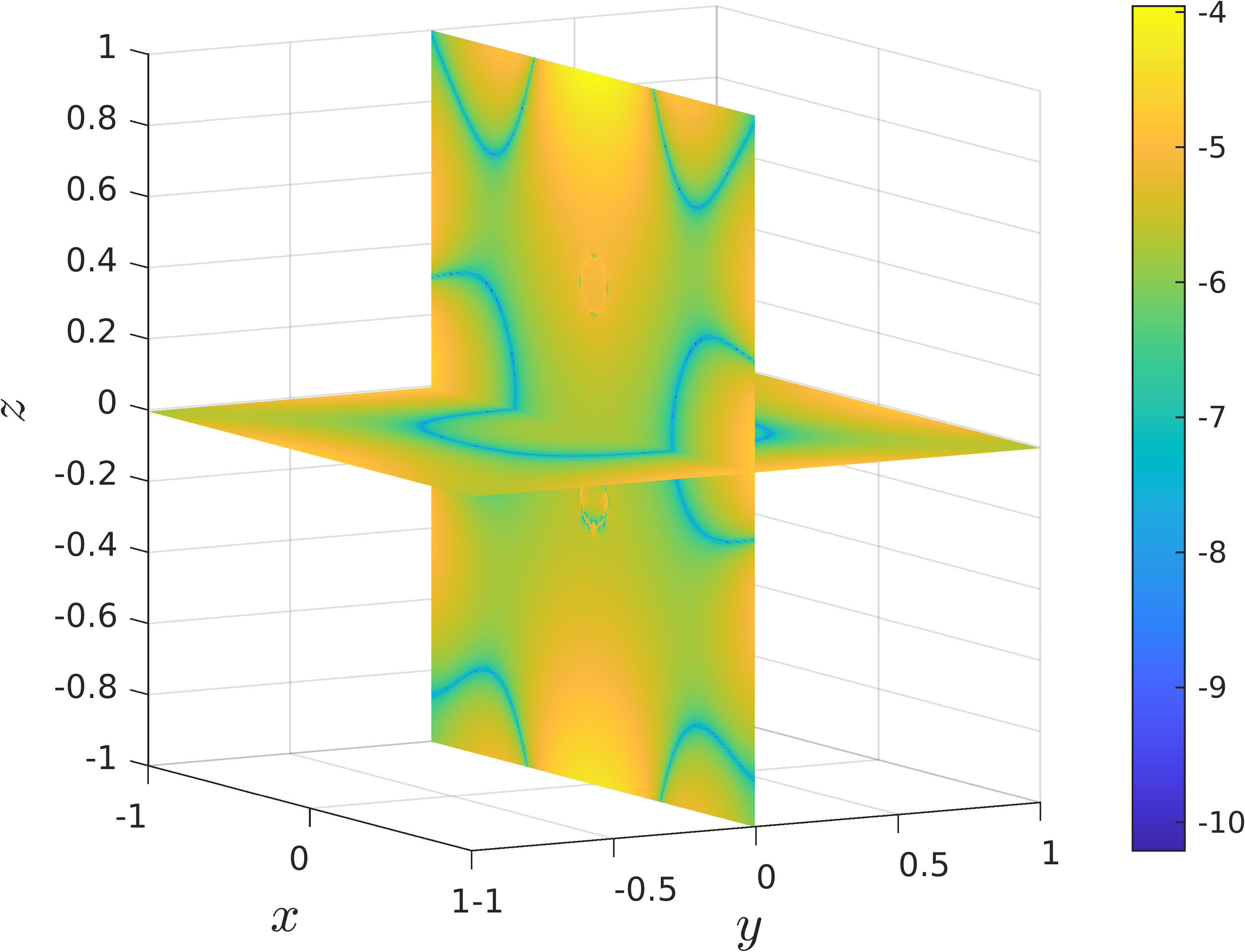}
    \includegraphics[width=0.24\linewidth]{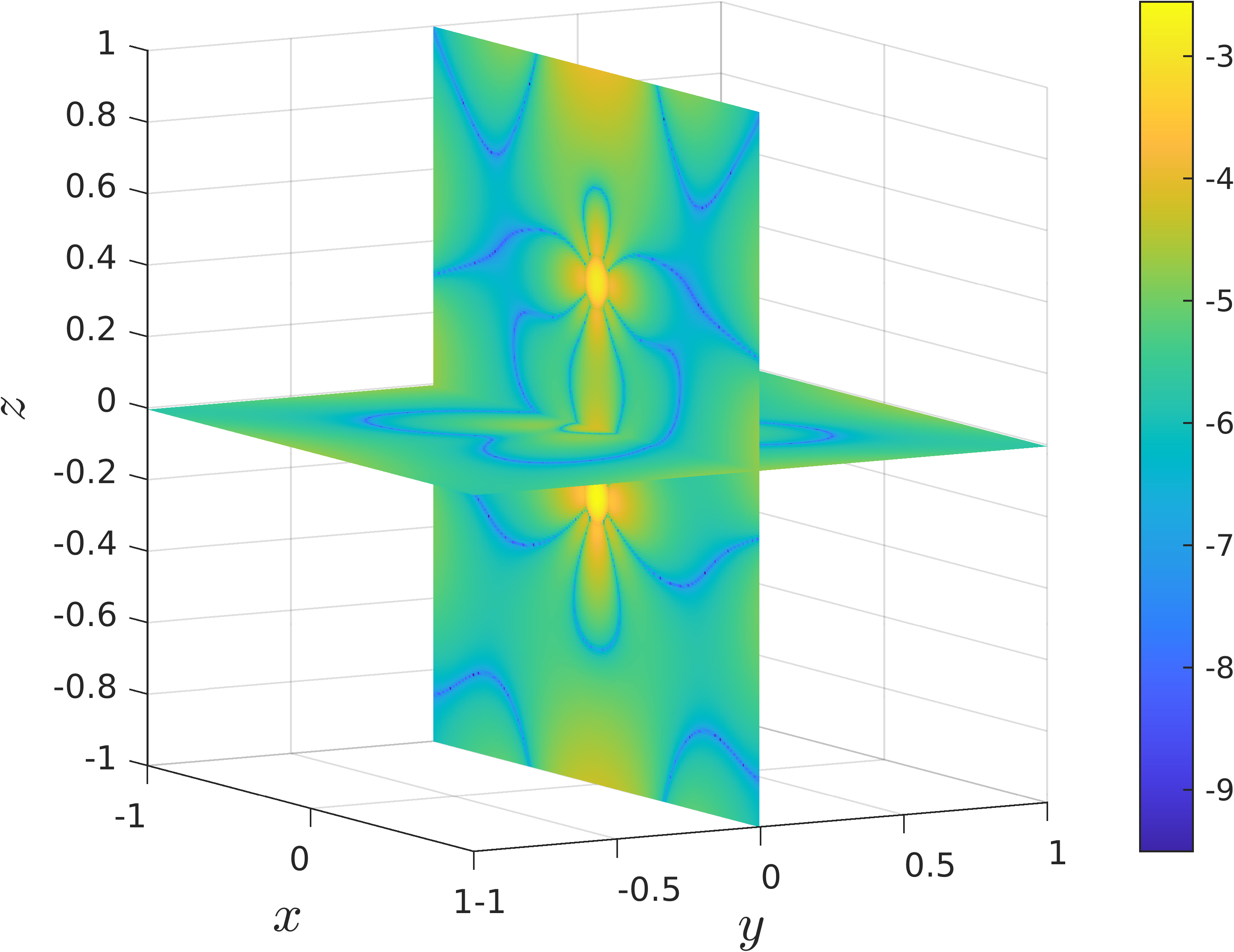}
    \includegraphics[width=0.24\linewidth]{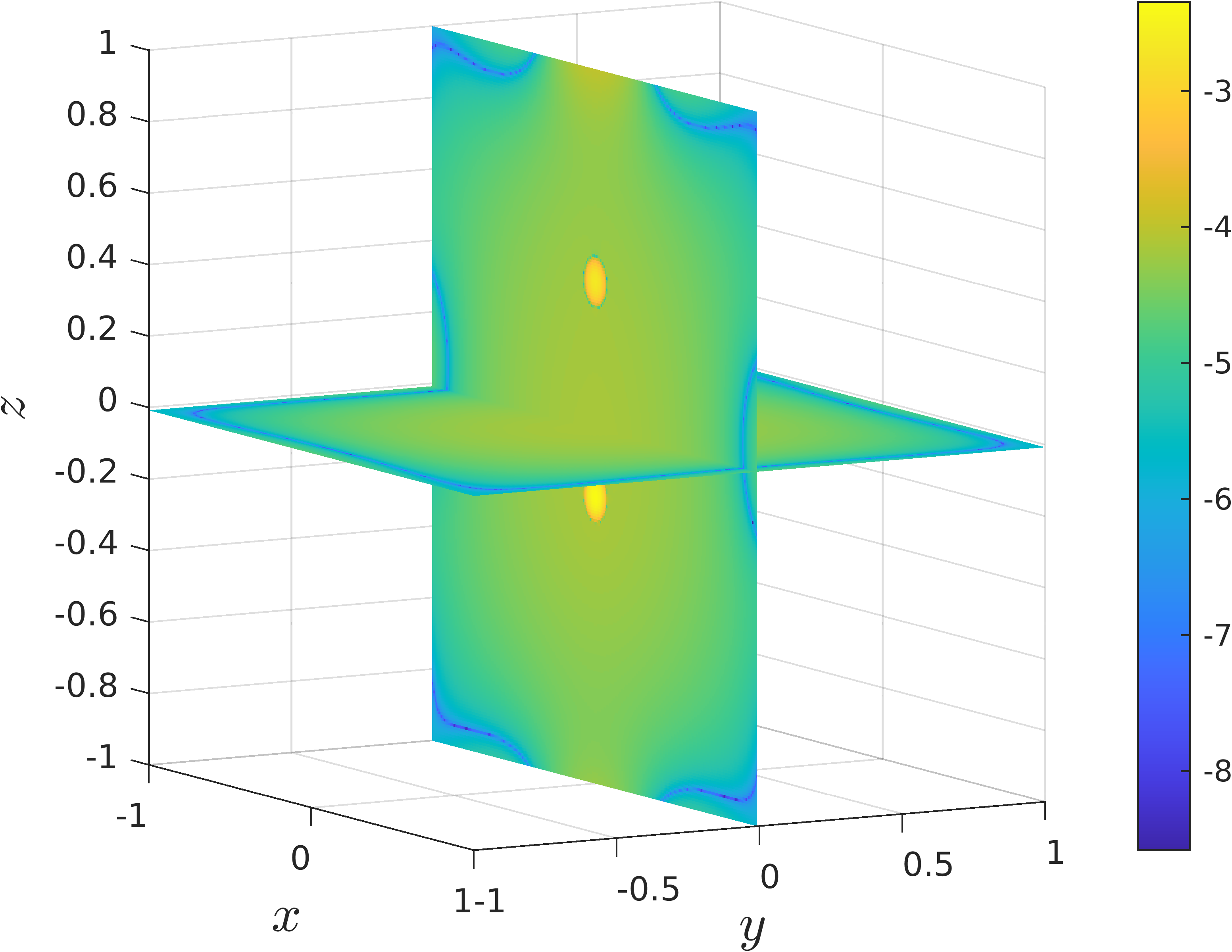}\\
    \includegraphics[width=0.24\linewidth]{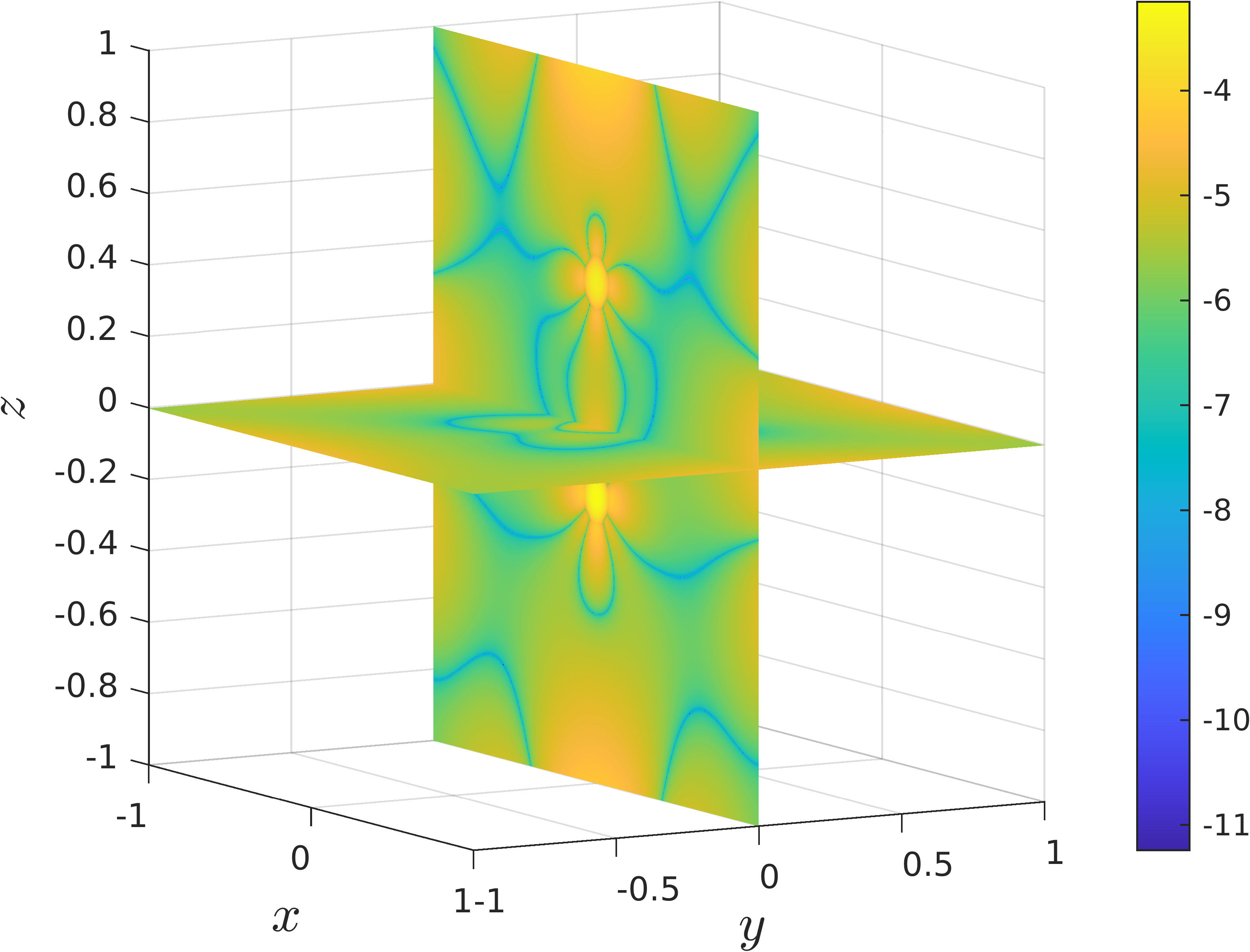}
    \includegraphics[width=0.24\linewidth]{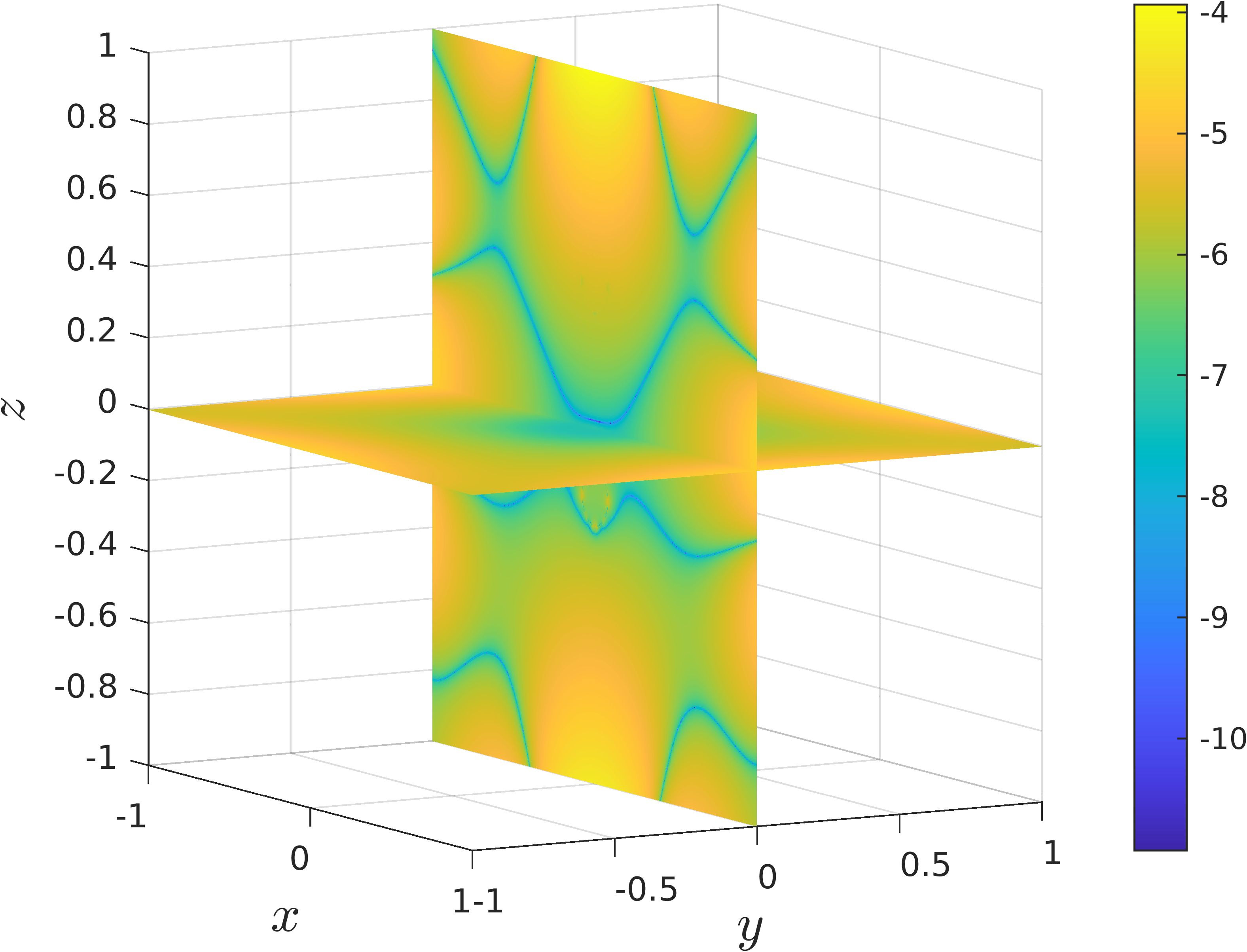}
    \includegraphics[width=0.24\linewidth]{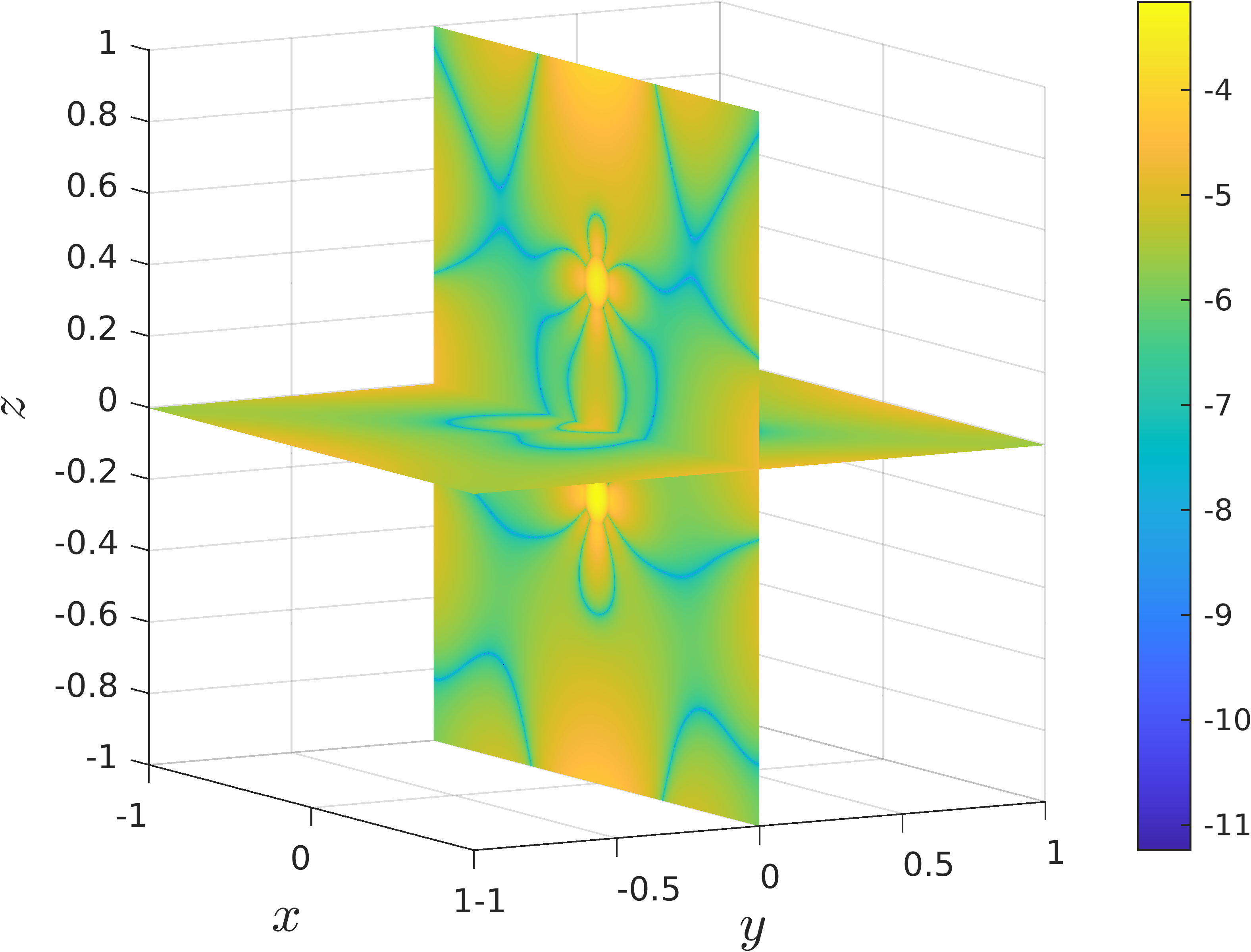}
    \includegraphics[width=0.24\linewidth]{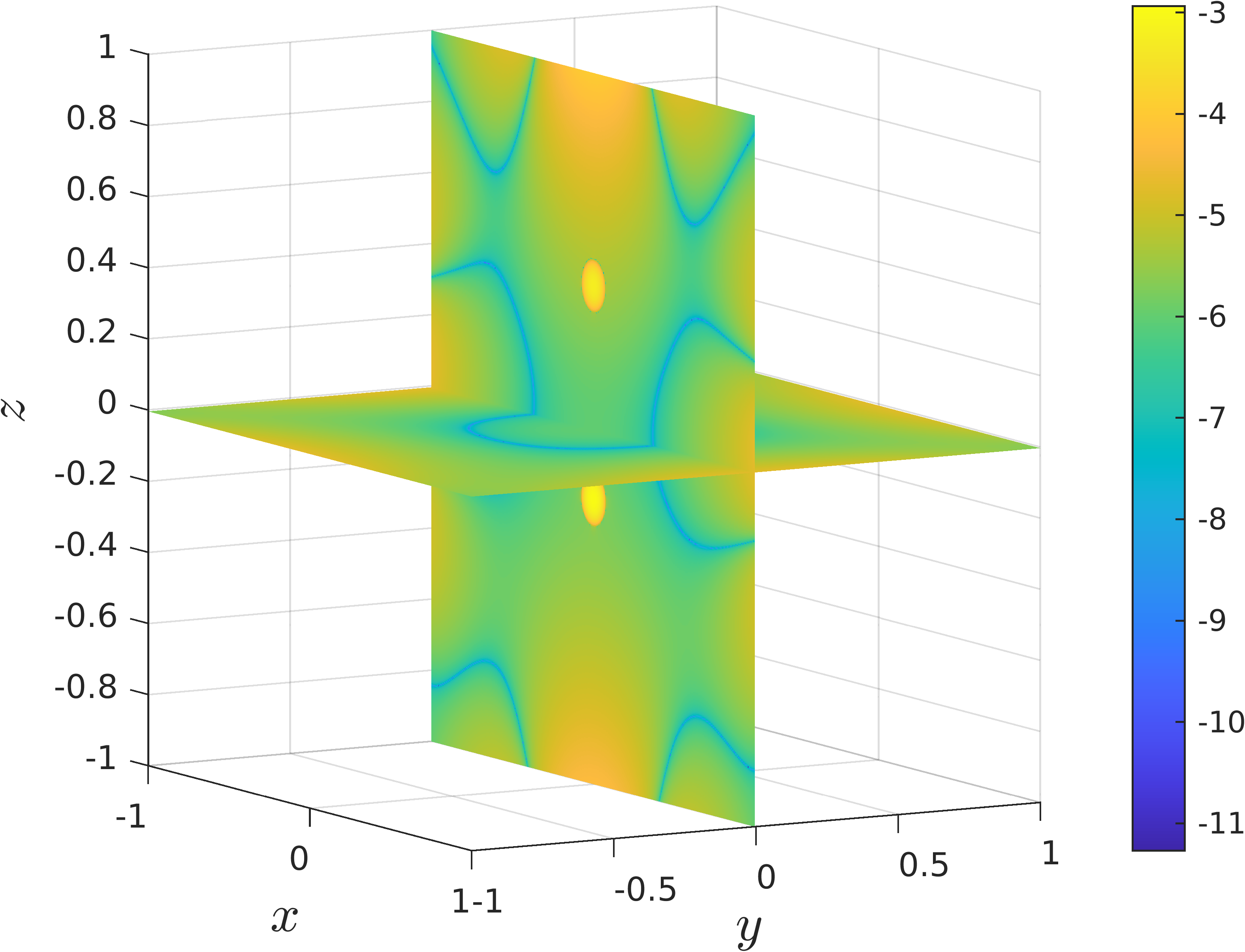}
    \caption{Relative error of the converged solutions, shown as colormaps of the logarithm of the error between the analytical solution $\Phi$ and the converged numerical solution $\bar{\Phi}$, computed as $\log_{10}(|(\bar{\Phi}-\Phi)/\max{(|\Phi|)}|)$ in \S\ref{sec:static}. From left to right: CG 2nd order, CG 6th order, SOR 2nd order, and MG 6th order.
    Top row: resolution $256^3$;  bottom row: $512^3$ ($255^3$ and $511^3$, respectively, for MG)\@.}
    \label{fig:errormap}
\end{figure}

We perform a convergence study of the iterative methods, using $\Phi_0=0$ for the initial guess. Following the methodology of \S4.2.1 of \citet{tomida2023}, we define two numerical estimates of convergence of a numerical iterate $\Phi_n$, measuring either the approach to the analytical solution $\Phi$ or to the fully converged numerical solution $\bar{\Phi}$ of a given numerical scheme. These estimates are the relative errors $\epsilon_n=(\Phi_n-\Phi)/\Phi$ and $\delta_n=(\Phi_n-\bar{\Phi})/\bar{\Phi}$, respectively.
Figure \ref{fig:test_errors} shows the relative error $(\bar{\Phi}-\Phi)/\textrm{max}(|\Phi|)$ as a function of position along the $x$- and $z$-axes, while Figure \ref{fig:errormap} shows, as a function of position along the planes $z=0$ and $y=0$, a logarithmic color map of the final relative error $(\bar{\Phi}-\Phi)/\textrm{max}(|\Phi|)$ of the converged solutions for the various methods.
The fully converged solution $\bar{\Phi}$ is experimentally computed as $10^4$ steps of CG and SOR iteration, and as 64 V-cycles (``steps'') of MG iteration.

\begin{figure}
\centering
    \includegraphics[width=0.24\linewidth]{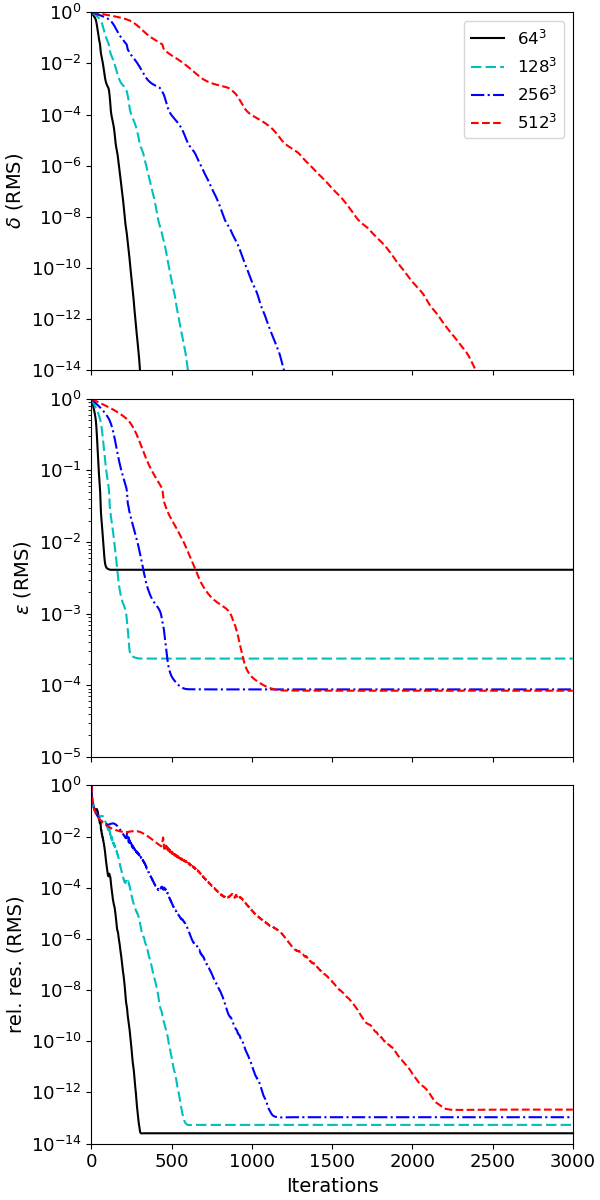}
    \includegraphics[width=0.24\linewidth]{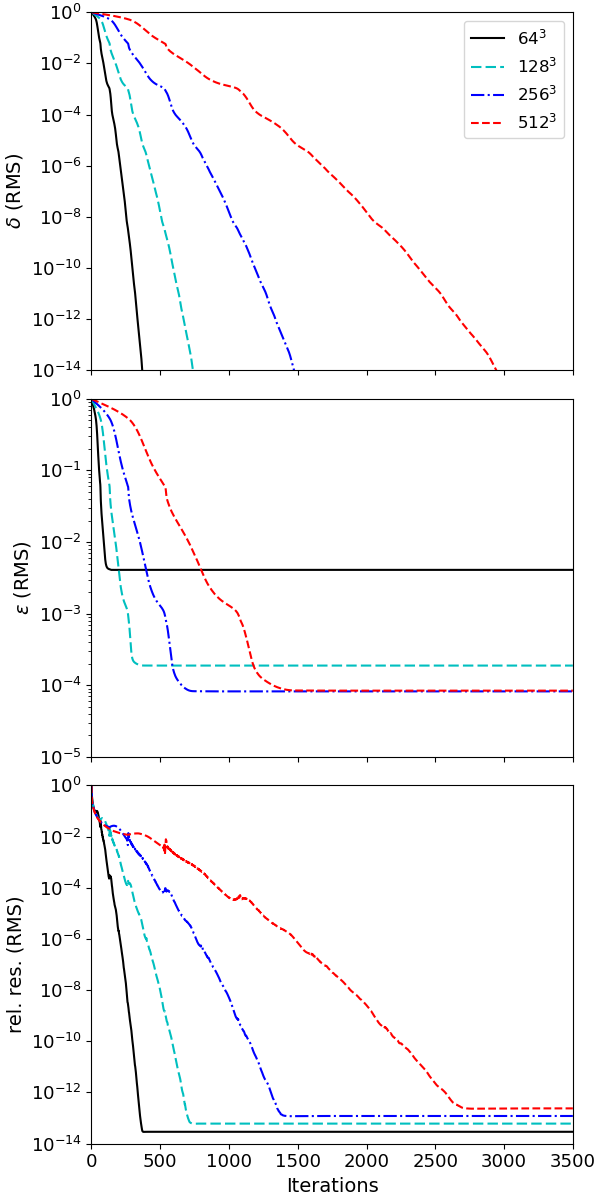}
    \includegraphics[width=0.24\linewidth]{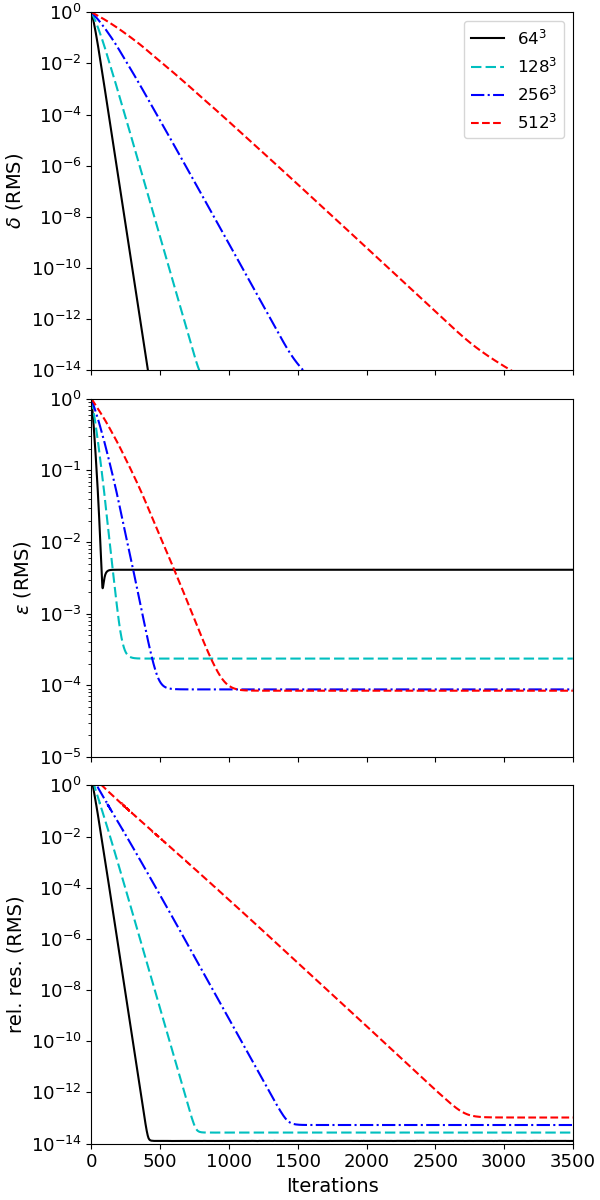}
    \includegraphics[width=0.24\linewidth]{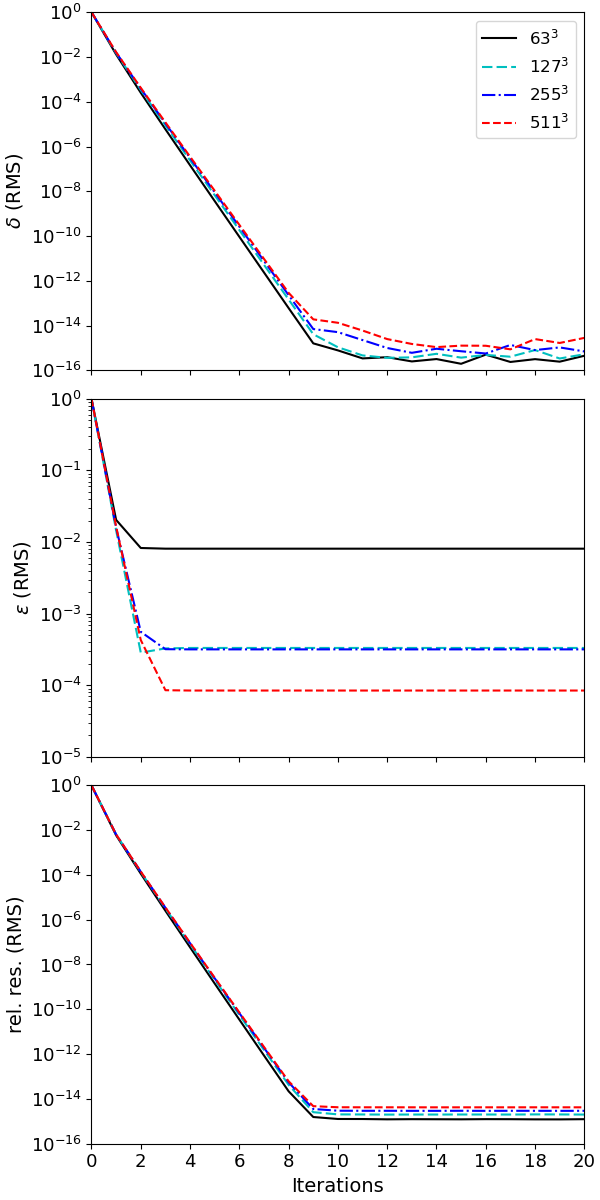}
    \caption{Convergence of the RMS relative errors and residuals, as the number of iterations increases.. 
    From top to bottom rows: RMS relative numerical error $\delta$, RMS relative analytical error $\epsilon$, and the RMS relative residuals. From left to right columns: CG 2nd order, CG 6th order, SOR 2nd order, and MG 6th order.
    Within each panel, the line colors indicate the grid resolution (64: red; 128: cyan; 256: blue; 512: black) for CG and SOR, and one zone less for MG\@. Details in \S\ref{sec:static}.}
    \label{fig:err1}
\end{figure}

The panels of Figure \ref{fig:err1} show the decrease of the relative RMS of the analytical error $\epsilon_n$, the numerical error $\delta_n$, and the residual as a function of the number $n$ of iterations. We notice that the relative analytical error for a given resolution does not decrease below a plateau value (depending on resolution, and controlled by a truncation error), but the numerical error and the residual can converge very strongly, to values controlled by the round-off error. This behavior is consistent with the results shown in Figure 7 of \citet{tomida2023}. 
We remark that during a simulation of a potential without a known analytical representation, only $\delta$ and the residual can be obtained.

\subsection{Dynamic example} \label{sec:dynamic}
\begin{figure}
\centering
    \includegraphics[height=0.3\linewidth]{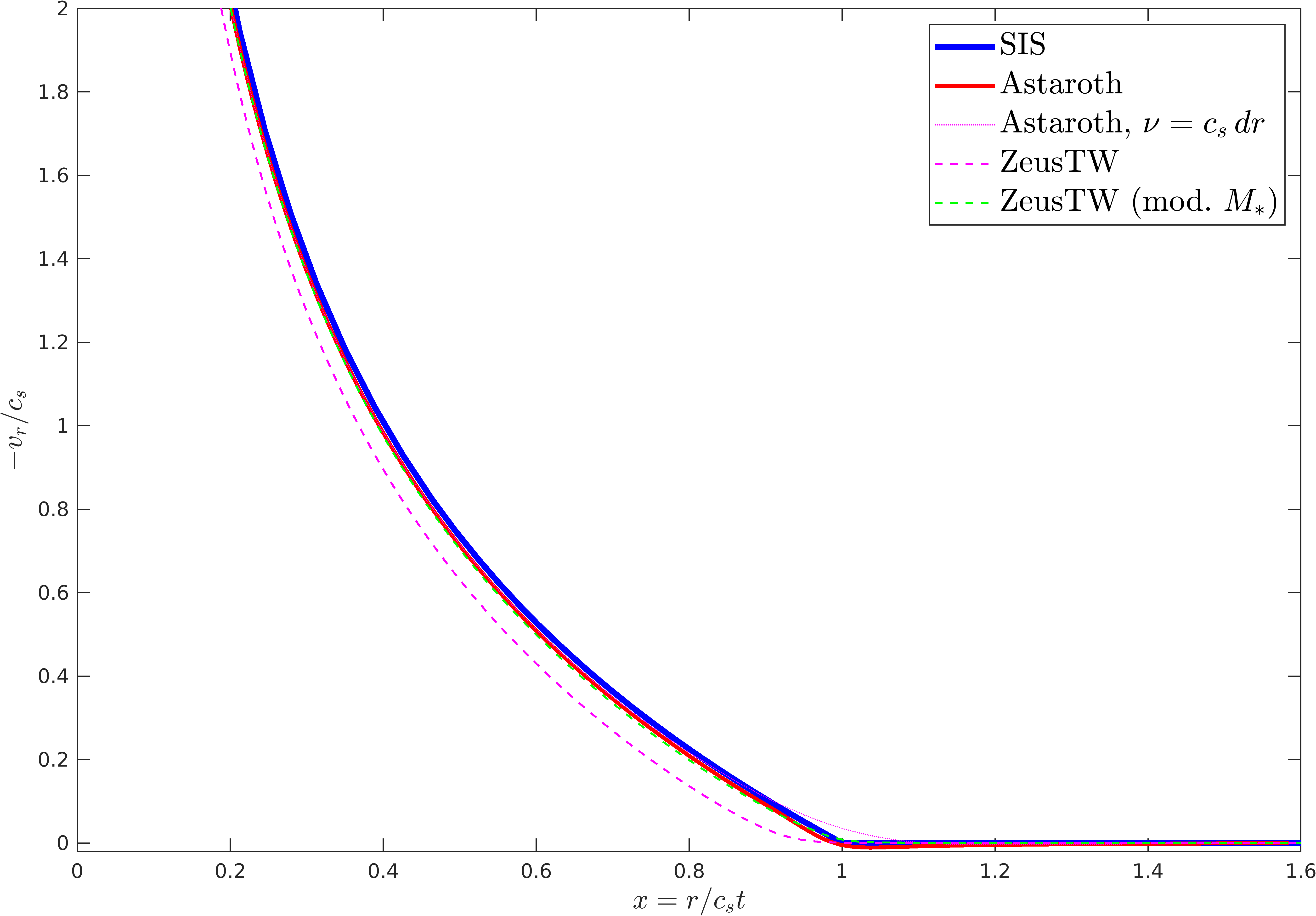}
    \includegraphics[height=0.3\linewidth]{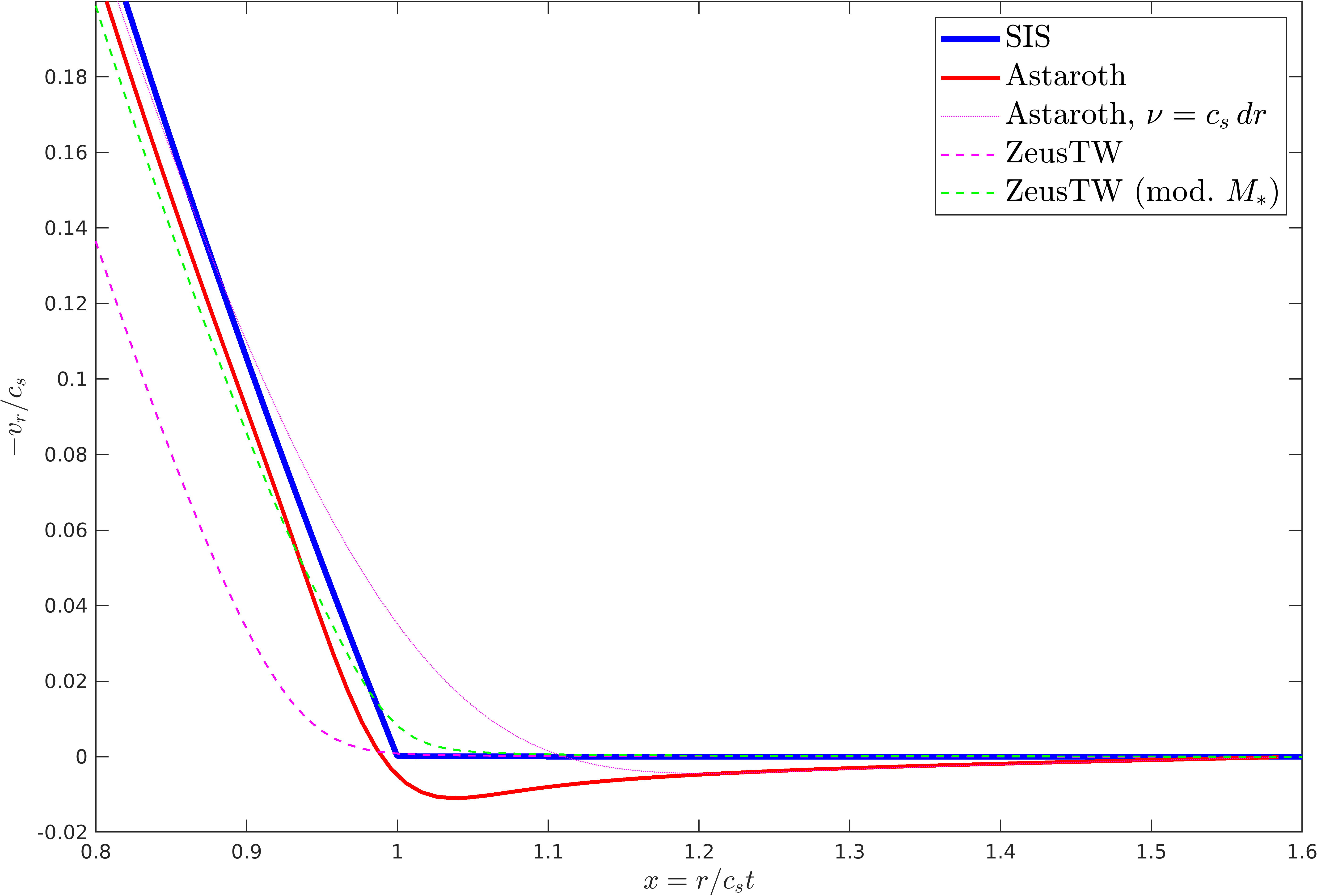}
    \caption{Collapse of an isothermal sphere. Dimensionless infall velocity $-v_r/c_s$ vs.\ dimensionless radial coordinate $x=r/(c_s t)$ at $t=1.5\times10^4\yr$. Left: full scale; right:  zoom into the region where the analytical SIS solution (blue solid line) features a discontinuous angle at $x=1$, numerically very demanding for both Astaroth (red solid line) and ZEUS (magenta dashes).  We have also run ZEUS using an initial mass $M_\ast$ and grid density locations aligned with Astaroth (green dashes). An Astaroth run with a small additional kinematic viscosity $\nu=c_s\,\Delta r$ (magenta dots) has a smaller deviation of velocity near the angle at $x=1$, but a slightly displaced position. See text in \S\ref{sec:dynamic} for further details.}
    \label{fig:sis}
\end{figure}

For our time-dependent test, we solve a problem of great astrophysical interest having a known ordinary differential equation (ODE) solution, namely the collapse of the singular isothermal sphere of \citet{shu1977}.
One crucial stage in the formation of a new young star is the collapse of a dense, essentially spherical pre-stellar core of radius $\rout$ within a molecular cloud. A pre-stellar core having efficient cooling and having a mass even slightly exceeding the  Bonnor--Ebert critical value naturally develops into a concentration of mass towards the center, with a density profile $\propto r^{-2}$ (Sect. I of \citealt{shu1977}).
This profile is represented, at the critical point of unstable equilibrium, by the profile of the singular isothermal sphere (SIS), which at a radius $r<\rout$ has a density
\begin{equation}\label{eq:sis}
\rho(r)=(c_s^2/2\pi G) r^{-2}\ ,
\end{equation}
where $c_s$ is the isothermal sound speed \citep[Eq.\ (2) of][]{shu1977}.
Despite the formal infinity of Equation (\ref{eq:sis}) at $r=0$, the enclosed mass $M(r)=(2c_s^2/G) \min(r,\rout)$ at any radius $r$ is finite.
In the test of this section, we perform the collapse of the SIS density profile given by Equation \eqref{eq:sis}.
Immediately after SIS collapse starts, a central point mass $M_\ast>0$ appears, representing the mass of a nascent young star.
Material falls from the inner parts of the collapsing SIS and feeds $M_\ast$, within an expanding rarefaction wave of radius $r=c_s t$, counting the initial time $t=0$ as the first moment of the formation of a non-zero $M_\ast$. An ODE system can be written to solve for the density and infall velocity as a function of radius and time, and the evolution is self-similar, depending on the dimensionless variable $x=r/(c_s t)$, until approaching the radius $\rout$ of the cloud.
The self-similar solution for the infall velocity is shown in the solid blue line of Figure \ref{fig:sis}.

In our numerical setting, a spherically symmetric density profile following Equation \eqref{eq:sis} is allowed to collapse under its own self-gravity.
Accurate numerical solutions of the collapse can be obtained by implementing the problem on a 3D spherical grid with $0<\rin<r<\rout$, and employing a central sink particle.\footnote{This sink particle, located outside the computational domain at $r=0<\rin$, is conceptually similar but numerically different from the sink-particle technique previously implemented into Astaroth for simulations of pseudodisk formation in Cartesian coordinates, which is located within the Cartesian domain \citep{vaisala2023}.}
In this work, we port to Astaroth the central sink particle as implemented in ZEUS-TW \citep{krasnopolsky2012}.  
(That implementation also includes a technique of zero-padding of the density.)
The central mass $M_\ast$ is set to be initially equal to the mass that is trapped within $0\leq r<\rin$, which is $(2c_s^2/G)\rin$ \citep[Eq.\ (2) of][]{shu1977} and then to grow with time as matter flows inwards through the numerical boundary $r=\rin$, by setting the numerical growth of the central mass to the rate $\dot{M}_\ast=\int\!\!\int\max(0,-v_r) \rho\,\rin^2\sin\theta\,d\theta\,d\phi$.
The analytical self-similar solution of the ODE, as computed in \citet{shu1977}, yields $M(t)=t m_0c_s^3/G$ with $m_0=0.975$.

The numerical initial condition deviates from self-similarity because it has explicit length scales $\rin$ and $\rout$. However, as the simulation progresses between these two radii, it approaches self-similarity, and the growth rate of the numerical sink particle evolves towards its self-similar solution. The numerical
central mass approaches $M(t)=t m_0c_s^3/G$ and its growth rate approaches $\dot{M}(t)=m_0' c_s^3/G$, with the numerical $m_0$ and $m_0'$ values in ranges near the analytical result.
Equation (10) of \citet{shu1977} shows that the self-similar solution for the dimensionless enclosed mass within a radius $r$ is $m(r)=4\pi r^2 (r/t-v_r)\rho (G/c_s^3)$, and the inner values of $m$ approximate the central mass  $m_\mathrm{in}=4\pi r^3(\rin/t-v_r)\rho G/c_s^3$.

When computing the numerical integrals of Equations (\ref{eq:outer_multipole})--(\ref{eq:inner_multipole}), a zero-padding technique is applied, intended to improve the accuracy of the multipole expansion by extending the grid for the potential both in the inner and outer directions to an extended range $r_\mathrm{in,pad}<r<r_\mathrm{out,pad}$. In the ranges $r_\mathrm{in,pad}<r<r_\mathrm{in}$ and $r_\mathrm{out}<r<r_\mathrm{out,pad}$ no hydrodynamics is computed, and the density is set to zero. The multipole expansions are computed at $r=r_\mathrm{in,pad}$ and $r=r_\mathrm{out,pad}$ based only on the density located within $\rin<r<\rout$. Avoidance of mass near the boundaries where the multipole expansion is computed increases the accuracy of the boundary condition for the potential.

The collapse of the SIS features a wave front propagating outwards from inside at the sound speed \citep{shu1977}. Hence, the simulations must be stopped at a time $t<\rout/c_s$, before the expansion wave leaves the computational domain and the numerical solutions lose accuracy.
However, the approach of the numerical solution to self-similarity takes time, so to reach those late times, a radially uniform grid would need to be large. We avoid that by implementing into Astaroth a logarithmic grid over $\rin<r<\rout$ similar to the one already present in ZEUS\@. We design this nonuniform grid by using the mapping function implemented in the Pencil Code, allowing a high degree of grid smoothness.
The extended portions of the grid, $r_\mathrm{in,pad}<r<\rin$ and $\rout<r<r_\mathrm{out,pad}$ do not require the same degree of accuracy as the rest of the grid, because they are used only to improve the BC accuracy for the potential. Therefore, as an extension of the technique, we use different spacing rules on the extended portions. When using a logarithmic grid the spacing in the extension at the inner radius becomes unnecessarily small, since the spacing at the interior radius is already sufficiently small for accuracy. Thus in the inner extension, we do not continue the logarithmic grid and instead continue the grid smoothly with a fourth-order polynomial, which allows us to separately control the spacing in the extension.
For this test, the polynomial for the extended inner grid is applied to 20 cells in $r_\mathrm{in,pad}<r<\rin$, with $r_\mathrm{in,pad}=\rin/2$. A similar technique can be used for the outer extended grid; however, for this particular test, having an initial $\rho\propto r^{-2}$, the densities near $r=\rout$ are already very modest, and we simply continue the logarithmic grid for 20 additional cells, obtaining sufficient accuracy for this particular problem.

We compare the results of simulations using Astaroth to the analytical solution of the  ODE, and also to results of ZEUS-TW.
We have used a modestly sized logarithmic-spherical grid for these tests, with $560\times50\times8$ zones spanning the active zones located at
$\rin=6\times10^{13}\cm<r<\rout=1.5\times10^{16}\cm$, $0<\theta<\pi$, and $0<\phi<2\pi$. The low resolution used in this test for the $\phi$ coordinate is made possible because of the symmetry of the physics problem and its numerical implementation; a more general problem will need a higher resolution in the angular directions.
We have applied outflow boundary conditions at $r=\rin$ for both codes. The boundary $r=\rout$ has little influence because we stop the runs long before the expansion wave reaches $\rout$. It has been implemented as outflow in ZEUS and 
by setting
$v_r=0$ for the outer ghost zones in Astaroth, conditions essentially equivalent for this collapse problem. The azimuthal boundary condition is periodic, while the polar boundary condition is reflecting. 

Due to programming choices, differences persist between the nonuniform grids and central sink particle implementations of Astaroth and ZEUS. Both codes show convergence to self-similar analytical solutions, but they do it in different ways. Here we describe in detail these implementation differences, because they cause minor but noticeable effects in the respective paths to convergence.
Astaroth defines all its major quantities at the cell edges or vertices of its grid, including density, potential, and velocity. ZEUS on the other hand is grid-staggered: densities and potentials are defined at cell centers, with velocities defined at cell edges. The continuity equation of ZEUS is treated in a finite-volume manner, computing mass fluxes at all edges of its active computational domain, and applying them to evolve $\rho$. Mass is conserved except for the mass fluxes at the boundaries. The central sink particle as implemented in ZEUS receives the mass that leaves the computational domain at $r=\rin$, and has the initial value based on the radius of the inner edge of the first radial active zone. In this set of comparative runs, the different grids are accommodated as follows. 
Cell centers on the logarithmic grid of ZEUS are located at the arithmetic mean of their edges, by simple averaging. Those in Astaroth are located at the geometric means of their edges, by application of the logarithmic mapping function. This induces a tiny difference in the location of the densities by a factor of 1.0049, sufficiently close to unity.
A completely identical treatment of the central sink particle is not possible because of the different manner in which each code computes the mass fluxes.
Because of that, estimates of the central mass are important for comparing each code to the analytical solution, but as a method for comparing the codes to each other, they are sensitive to the different treatment of the initial central mass and mass fluxes.
At the time $t=1.5\times10^4\yr$ shown in Figure \ref{fig:sis}, these masses are $m_0=0.9523 (0.8944)$, $m_0'=0.9585 (0.9790)$, and $m_\mathrm{in}=0.9682 (0.9764)$ respectively for the Astaroth and ZEUS runs of that figure without modifications (red solid and magenta dashes). We can modify the runs to take the implementation differences into account.  We can change the central mass in ZEUS to the Astaroth value of $M_{\ast}$ and align the density grid locations.  To compare to the lower-order ZEUS result and avoid overshoot at the discontinuity in the solution, we can add a kinematic viscosity to Astaroth with $\nu=c_s\,\Delta r$.  With these modifications $m_0=0.9591 (0.9532)$, $m_0'=0.9677 (0.9788)$, and $m_\mathrm{in}=0.9775 (0.9858)$ for the modified runs of Astaroth, and ZEUS respectively.

Figure \ref{fig:sis} shows the $r$-dependence of the dimensionless infall velocity at time $t=1.5 \times 10^4\yr$.
Using the BICGSTAB solver, the number of steps required for convergence to a residual of $<10^{-4}$ immediately decreases in time from an initial value of 2381 to the range $\sim1$--$10$, and then increases back to $\sim30$.
While the overall behavior of this Astaroth run is more accurate than that of ZEUS-TW, the run using ZEUS-TW uses a Van Leer slope limiter, while no slope limiter is applied to the Astaroth run. This is plausibly the cause of the presence of an overshoot of the velocity curve of up to $\sim0.01 c_s$, further reduced by applying viscosity $\nu$ to resolve the discontinuous angle in the SIS self-similar solution.
Overall, the solutions approach the self-similar SIS with time, a behavior already observed for a magnetized setting in \citet{wang2025}, because for sufficiently long times the initial scale $\rin\ll c_s t$ becomes negligible. However, for sufficiently long times, self-similarity again suffers, as $c_s t$ approaches $\rout$: this last behavior is more pronounced in runs using uniform radial grids due to their smaller value of $\rout$.

\section{Summary} \label{sec:conclusions}
In this work, we 
  describe the implementation and testing of several efficient
solvers for self-gravity on GPUs, including novel combinations of numerical techniques in our MG solvers. This is timely because astrophysical simulations of self-gravitating systems across multiple scales rely critically on the speed and accuracy of the backbone numerical algorithms and hardware platforms.
The Astaroth platform's stencil-based acceleration excels at high-resolution, high-order HD and MHD simulations, making stencil-based iterative solvers a natural choice for the self-gravity implementation. We present the development, validation, and benchmarking of four GPU solvers for self-gravity within the Astaroth framework: SOR, CG, BICGSTAB, and MG\@. 
The open boundary conditions needed for most astrophysical applications are handled by multipolar expansions of user-adjustable order.

Cartesian implementations were developed for SOR, CG, and MG, while spherical implementations were developed for SOR and BICGSTAB. All Cartesian solvers were validated against the known analytical solution of a combination of spherical mass distributions, with the numerical residual drivable to near machine round-off precision (Figure \ref{fig:err1}). 
Mixed-precision storage, by keeping coarse-level data in single precision, reduces data transmission and memory bottlenecks and contributes to the speed of the MG implementation.
A key algorithmic novelty is the combination of numerically optimized SPAI smoothers with compact stencils for the MG solvers.

Solver speeds were measured in units of billions of zone updates per second (Gzups).
On one GPU of LUMI, MG achieves $1$--$1.5$ Gzups per V-cycle at $255^3$ resolution ($\sim1.5$--$3.5$ for $511^3$), with $3$--$7$ V-cycles sufficient for convergence to a relative residual of $10^{-6}$, comparable to speeds of other stencil-based algorithms in Astaroth, such as found in the 3D MHD benchmark on LUMI of $\sim0.3$ Gzups at $256^3$ zones on one GPU\@.

As a demonstration of the full coupling of self-gravity to hydrodynamics, we simulate the collapse of the singular isothermal sphere \citep{shu1977}, a classical problem in star-formation theory, using the BICGSTAB solver on a non-uniform spherical grid. We recover the self-similar infall velocity profile and central mass growth rate, in close agreement with independent ZEUS-TW runs.

The solvers described here provide a solid foundation for simulations involving self-gravity, with applicability to general elliptic problems beyond astrophysics. Future work will extend MG to non-uniform and spherical grids, optimize multi-GPU scaling, and couple the solvers fully to all parts of Astaroth.

{
\vspace{9mm}
\noindent
The authors acknowledge support for the CompAS Project from the Institute of Astronomy and Astrophysics, Academia Sinica (ASIAA), the Academia Sinica grant AS-IAIA-114-M01, and the National Science and Technology Council (NSTC) in Taiwan through grants 112-2112-M-001-030, 113-2112-M-001-008, and 114-2112-M-001-001-; the International Collaboration and Cooperation grant for COSMAGG that supports the exchanges between Taiwan and Finland: 113-2927-I-001-513-, 114-2927-I-001-506-, and Research Council of Finland project 359462. The authors thank the National Center for High-performance Computing (NCHC) of the National Applied Research Laboratories (NARLabs) in Taiwan for providing computational (nano5 GPU cluster) and storage resources and the ASIAA for in-house access to high-performance computing facilities. We also acknowledge the computing resources granted by the Finnish IT Center for Science (CSC; project number 2000403) and LUMI-G resources through the EuroHPC consortium (project number 462000986). M-MML acknowledges partial support from US NSF grant AST23-07950.
}

\vspace{5mm}
\facilities{NCHC (nano5 GPU cluster), and LUMI}

\software{Astaroth \citep{pekkila2022}, MATLAB \citep{MATLAB:2024}, Matplotlib \citep{Hunter:2007}}

\bibliography{poisson_paper}{}
\bibliographystyle{aasjournal}

\end{document}